\documentclass[12pt]{article}

\usepackage{graphicx}
\graphicspath{{./images-10/}{./images-8/}}

\let\prevc=\c

\let\plural=\relax

\makeatletter
\def\newleaf{\newpage
\newcount\tmp
\tmp=\c@page
\divide\tmp by 2
\multiply\tmp by 2
\ifnum\c@page=\tmp
~\newpage
\fi
}
\makeatother

\expandafter\ifx\csname proofnewpage\endcsname\relax

\fi

\def\color[#1]#2{}

\long\def\nop#1{}

\def\comment{\edef\cps{\the\parskip} \parskip=0.5cm \begingroup \tt}

\def\separator{\vskip 1cm\hrule\vskip 1cm}

\def\noproof#1{
\def\proof{{#1}\iffalse}
\let\qed=\fi
}

\begingroup
\makeatletter
\global\def\fakelabel#1#2{
\expandafter\ifx\csname fakelabelsome\endcsname\relax
\let\fakelabelsome\par
\AtEndDocument{\typeout{}}
\fi
\AtEndDocument{\typeout{NOTE: #1 is a fake label, marked #2}\typeout{}}
\@ifundefined {r@#1}
{\global\@namedef{r@#1}{#2}}
{}
}
\endgroup

\long\def\figurearrow#1#2{
\newbox\before
\setbox\before=\hbox{#1}
\newbox\after
\setbox\after=\hbox{#2}
\newdimen\vdim
\vdim=\ht\before
\ifdim\vdim<\the\ht\after
  \vdim=\ht\after
\fi
\hbox{
\vbox to \the\vdim{\vfill\box\before\vfill}
\vbox to \the\vdim{\vfill\hbox to 3cm{\hfill\Huge$\Rightarrow$\hfill}\vfill}
\vbox to \the\vdim{\vfill\box\after\vfill}
}
}

\expandafter\ifx\csname shortcite\endcsname\relax

\fi

\newbox\current

\long\def\plframebox#1{
\setbox\current\vbox{#1}		

\vbox to \ht\current {\hrule\vss
\hbox to \wd\current {%
\vrule \hss\box\current\hss \vrule}
\vss\hrule }
}

\long\def\eatpar#1{%
\ifx#1\par                      
\let\nextmove=\eatpar           
\else
\let\nextmove=#1
\fi
\noexpand\nextmove
}

\def\modifymargins#1#2{
\newdimen\addtoh
\newdimen\addtow
\addtoh=#1
\addtow=#2

\advance\topmargin by -\addtoh
\multiply\addtoh by 2
\advance\textheight by \addtoh

\advance\oddsidemargin by -\addtow
\advance\evensidemargin by -\addtow
\multiply\addtow by 2
\advance\textwidth by \addtow
}

\begingroup
\catcode`\~=11
\gdef\centertilde#1{\lower #1pt\hbox{~}}
\endgroup

\newcount\currenttime
\newcount\hour
\newcount\minute

\def\printtime{%
\currenttime=\time
\hour=\currenttime
\divide\hour by 60
\minute=-\hour
\multiply\minute by 60
\advance\minute by \currenttime
\the\hour:\ifnum\minute<10 0\fi\the\minute
}

\begingroup
\makeatletter
\global\let\@@date=\@date
\gdef\@date{\@@date\ --- \printtime}
\endgroup

\def\oggi{\number\day\space 
\ifcase\month\or
Gennaio\or Febbraio\or Marzo\or Aprile\or Maggio\or Giugno\or
Luglio\or Agosto\or Settembre\or Ottobre\or Novembre\or Dicembre\fi
\space \number\year}

\newcounter{rmexample}

\def\proof{\noindent {\sl Proof.\ \ }}

\def\qed{\hfill{\boxit{}}
  \ifdim\lastskip<\medskipamount \removelastskip\penalty55\medskip\fi}
\def\qedn#1{\hfill{\boxit{}$_#1$}
  \ifdim\lastskip<\medskipamount \removelastskip\penalty55\medskip\fi}
\long\def\boxit#1{\vbox{\hrule\hbox{\vrule\kern3pt
                  \vbox{\kern3pt#1\kern3pt}\kern3pt\vrule}\hrule}}

\def\var{V\!ar}

\def\true{{\sf true}}
\def\false{{\sf false}}

\def\p{{\rm P}}
\def\np{{\rm NP}}

\def\bh#1{\if#1{}{\rm BH}\else\mbox{BH$_{#1}$}\fi}

\def\pspace{{\rm PSPACE}}

\def\c{\mbox{$\leadsto$}}

\def\profont{\sf}

\def\x3c{{\profont x3c}}

\def\possnewtheorem#1#2{
\expandafter\ifx\csname #1\endcsname\relax
\newtheorem{#1}{#2}
\fi
}

\def\possnewtheoremthree#1[#2]#3{
\expandafter\ifx\csname #1\endcsname\relax
\newtheorem{#1}[#2]{#3}
\fi
}

\possnewtheorem{theorem}{Theorem}
\possnewtheorem{corollary}{Corollary}
\possnewtheorem{lemma}{Lemma}
\possnewtheoremthree{proposition}[theorem]{Proposition}
\possnewtheorem{definition}{Definition}
\possnewtheorem{question}{Question}
\possnewtheorem{example}{Example}
\possnewtheorem{nontheorem}{Counterexample}
\possnewtheorem{property}{Property}
\possnewtheorem{assumption}{Assumption}
\possnewtheorem{conjecture}{Conjecture}
\possnewtheorem{notation}{Notation}
\newenvironment{theorem*}[1]{{\noindent \bf Theorem~#1}\begin{it}}{\end{it}\

}

\def\after#1#2{#1~{\sf after}~#2}

\let\c=\prevc
\ifx\isproof\undefined\long\def\proofonly#1{}\relax\else\long\def\proofonly#1{#1}\fi
\expandafter\ifx\csname ttytty\endcsname\relax\modifymargins{60pt}{40pt}
\else\modifymargins{60pt}{0pt}\def\_{\tt\char 95}
\raggedright\parindent 40pt

\fi
\expandafter\ifx\csname ttytty\endcsname\relax
\def\ttytex#1{#1\nop}			
\else

\modifymargins{0pt}{-10pt}
\def\frac#1#2{#1/#2}
\def\_{\char 95}
\expandafter\ifx\csname makeatletter\endcsname\relax\makeatletter\fi
\def\@ttyfig#1#2{\def\specialtext{\special{txt:#2}}\specialtext\egroup}
\def\ttyfig{\relax\bgroup\catcode`\^^M\active\let^^M
\let\-\relax\catcode`\ \active\@ttyfig}
\let\ttytex\ttyfig
\expandafter\ifx\csname makeatother\endcsname\relax\makeatletter\fi
\fi

\possnewtheorem{algorithm}{Algorithm}

\title{Four algorithms for propositional forgetting}
\author{Paolo Liberatore\thanks{
DIIAG, Sapienza University of Rome.
{\tt liberato@diag.uniroma1.it}
}}
\date{}

\begin{document}

\maketitle

\begin{abstract}

Four algorithms for propositional forgetting are compared. The first performs
all possible resolutions and deletes the clauses containing a variable to
forget. The second forgets a variable at time by resolving and then deleting
all clauses that resolve on that variable. The third outputs the result of all
possible linear resolutions on the variables to forget. The fourth generates a
clause from the points of contradiction during a backtracking search. The
latter emerges as the winner, with the second and first having some role in
specific cases. The linear resolution algorithm performs poorly in this
implementation.

\end{abstract}

\section{Introduction}

Logical forgetting is reducing the amount of information by eliminating the
parts that are related to some kind of objects, such as variables, objects or
conditions~\cite{delg-17}. Propositional, modal, temporal, action, defeasible,
description and first-order logics are fields where it has been applied~\cite{%
bool-54,moin-07,delg-17,
vand-etal-09,
feng-etal-20,
erde-ferr-07,raja-etal-14,
anto-etal-12,
kone-etal-09,eite-06,
lin-reit-94,zhou-zhan-11
}. Others are answer set programming~\cite{wang-etal-14,gonc-etal-16},
argumentation theory~\cite{baum-etal-20}, belief
revision~\cite{naya-chen-lin-07} and circumscription~\cite{wang-etal-15}. In
the propositional case, it is also called variable
elimination~\cite{subb-prad-04} and is the core of one of the earliest
mechanisms for automated reasoning~\cite{davi-putn-60}. In modal logics, it is
often associated with its dual, uniform interpolation.


The applications of logical forgetting include relevance~\cite{lang-etal-03},
updating~\cite{naya-chen-lin-07}, automated
reasoning~\cite{davi-putn-60,dech-rish-94,subb-prad-04}, privacy
preserving~\cite{gonc-etal-17}, limited reasoning
agents~\cite{fagi-etal-95,raja-etal-14}, knowledge focusing~\cite{eite-kern-19}
and consistency restoring~\cite{lang-marq-10}.

Automated reasoning highlights the two sides of logical forgetting. The
algorithm by Davis and Putnam~\cite{davi-putn-60} establishes the
satisfiability of a propositional formula by repeatedly eliminating variables
while preserving the satisfiability of the formula. Bucket
elimination~\cite{dech-rish-94,raze-etal-21} follows the same principle. The
NiVER system~\cite{subb-prad-04} removes variables only if that does not take
too much time or produce too large a formula, which again would slow down the
overall process. Running time is central.

Yet, satisfiability of a single formula is not always the final goal of
automated reasoning. The same logical knowledge base may need to be queried
many times as to whether it entails or is consistent with other information. As
done by materialized views in databases~\cite{mist-etal-01}, if such queries
are often about some specific subset of information, it may be convenient to
isolate the relevant part of the knowledge base. The running time of forgetting
is counterbalanced by the speed-up it produces among many queries.

More generally, the aim of forgetting dictates its efficiency requirements.
Forgetting may be required for legal reasons~\cite{gonc-etal-17}, where
applications are usually not time constrained. Forgetting may be done to tell
whether some variables are related or influence some others; the result is
yes/no; even the memory the result of forgetting takes is not that important.

An algorithm performing a certain number of steps consumes at most a unit of
memory in each. The amount of required memory is bounded by that number. Not
the other way around. Memory can be reused. As an example, a trivially
memory-efficient algorithm for propositional satisfiability loops over all
possible interpretations, checking the satisfaction of the formula against
each. The only required memory is that needed for storing the interpretation,
the formula and the value of each subformula according to the interpretation.
Yet, the running time is linear in the number of possible interpretations,
which is exponential in the number of variables.

In practice, a computer may be run some more if time allows, but cannot
indefinitely be increased in memory. Hitting the memory limit prevents any
output to be generated. Ironically, this may even happen when the output is
very small. As an example, the satisfiability algorithm by Davis and
Putnam~\cite{davi-putn-60} may exponentially blow up the input formula by
iteratively eliminating variables, but the output is always a single bit:
either the formula is satisfiable or it is not. More generally, this is the
case when the variables to keep are few, as it is when focusing information on
a specific topic or querying is required only for a small subset of variables.

Depending on the goal of forgetting, its result may not even need to be stored
explicitly. For example, it may be discarded once a variable is found to be
influenced by certain others. It is also not needed if its final purpose is to
establish whether it is equivalent to a given other formula~\cite{libe-20-a}.


Memory requirements hit all three algorithms currently defined for forgetting
in propositional logics: double substitution, prime implicate selection and
variable elimination by resolution.

Double substitution was first introduced by Boole~\cite{bool-54} as variable
marginalization: the variable is first assumed true, then assumed false;
forgetting is the disjunction of the two cases. Prime implicate
selection~\cite[Proposition~19]{lang-etal-03} generates all prime implicates
and selects only the ones that do not contain any variable to forget. Variable
elimination~\cite{delg-wass-13,wang-15,delg-17} resolves and discards all
clauses that contain the variables to forget.

All three methods may take exponential space. This is particularly bad as
theory predicts that it can be avoided in all cases~\cite{libe-20-a}:
forgetting is at the second level of the polynomial hierarchy~\cite{stoc-76},
which implies its membership to \pspace. It can always be done in polynomial
space.

A naive but polynomial-space mechanism proves the claim: iterating over all
possible values of the variables to remember. Each one is tested for
inconsistency with the input formula. If so, the clause made by the negation of
the literals it satisfies is generated. Satisfiability only takes polynomial
space. Iterating over the values of the variables is linear in space as well.
Overall, forgetting this way only takes a linear amount of space. It is still
not a good way to forget, as it always takes exponential running time, even
when the other algorithms would be linear. A similar issue occurs in Answer Set
Programming: the semantical forms of forgetting require looping over all
possible answer sets the result may have, if implemented literally; for this
reason, syntactical methods are looked for~\cite{bert-etal-19}.


Whether forgetting can always be done in polynomial space is known: yes. Both
in theory and in practice. The open problem is to forget in polynomial space as
quickly as possible, not taking exponential time in all cases.

Four algorithms are compared:

\begin{description}

\item[Resolution closure] All clauses are resolved; the subsumed ones are
removed. The ones comprising variables to remember are output. It is correct
because only entailed clauses are produced this way. It is complete because the
prime implicates of forgetting are the minimally entailed clauses not
containing a variable to forget~\cite{lang-etal-03}. This mechanism is a
baseline for forgetting: it does not require anything else than generating
minimal clauses by resolution and selecting some by a simple rule. It is useful
to show how better the others fare.

\item[Variable elimination] All clauses containing a variable to forget are
iteratively resolved. The results are clauses not containing that variable.
These and the other clauses of the formula make the result of forgetting.
Variable marginalization (replacing a variable with true and false and
disjoining) gives the same result when applied to a set of clauses, if the
result is converted back to a set of clauses.

Variable elimination by resolution was introduced by Davis and
Putnam~\cite{davi-putn-60} to establish the satisfiability of a formula: when
all variables are removed, it produces no clauses if the formula is satisfiable
and an empty clause otherwise. It was later employed on some of the variables
to speed up the following call to a satisfiability checker~\cite{subb-prad-04}.
When applied on a given set of variables, it forgets them from the
formula~\cite{delg-wass-13,wang-15,delg-17}.

Research on propositional satisfiability concluded that this mechanism takes
exponential space on some formulae when forgetting all variables regardless of
the order of removal of variables~\cite{gali-77}. This is exactly one of the
cases where superpolynomial space should be avoided: the result is small, but
the memory necessary for producing it is large.

\item[Linear resolution] Variable elimination is the same as ordered
resolution. Proof complexity proves that it is surpassed on proof size by
linear resolution~\cite{goer-92,bone-etal-00,bure-pita-07}, which can also be
adapted to forgetting variables.


\item[Backtracking]

Partial models of a formula are found by backtracking over the possible values
of the variables. In reverse, backtracking finds partial models falsifying the
formula, which are dual to entailed clauses. Not all of them are allowed in the
result of forgetting because of the possible presence of variables to forget.
Delaying their assignment solves the problem. The resulting algorithm is
guaranteed to only take polynomial working space. Unit propagation speeds it
up, but needs to deal with variables to forget.

\end{description}

These four algorithms are described in details in a following section each.
Section~\ref{section-python} describes their implementation. A number of
experiments have been performed to assess their computational properties: time
and memory used and size of result. These experiments and their results are in
Section~\ref{section-comparison}. Section~\ref{section-every} presents some
considerations on how an arbitrary proof method could be adapted to forgetting,
in the same way as resolution and backtracking.
Section~\ref{section-conclusions} presents some conclusive remarks.

The backtracking algorithm proves superior or comparable to the others on all
considered measures. It is the quicker and takes comparable or less memory than
the best of the others, depending on how memory is measured. Its main drawback
is that it rebuilds its output from the semantics of the formula, disregarding
its syntax; this output may not be the most intuitive expression of forgetting
variables.

\section{Preliminaries}

The logic considered in this article is propositional Boolean logic over a
finite alphabet. It is based on a finite set of variables. A literal is a
variable or its negation. A clause is a set of literals, and is interpreted as
their disjunction: either one of them is true; it is written using the symbol
$\vee$ between literals. A formula is a set of clauses, and is interpreted as
their conjunction: all of them are true. An empty clause is a clause containing
no literals. It is denoted by $\bot$ and is always false by definition.

For every clause $C$, the notation $\neg C$ is used for the set of literals
that are exactly the opposite of each literal in $C$. For example, $\neg (a
\vee \neg b)$ is $\{\neg a, b\}$. In the other way around, if $I$ is a partial
interpretation, then $\neg I$ is the clause that is falsified exactly by $I$.
For example, $\neg \{\neg a, b\}$ is $a \vee \neg b$.

Propositional forgetting a variable is removing that variable from the formula
while maintaining the semantic of the formula on all other variables. The
typical definition is that forgetting a variable from a formula entails a
formula on the other variables if and only if the original formula does. This
is the same with mutual consistency instead of entailment.

Resolution~\cite{robi-65} is a rule that derives a clause $C \vee D$ from two
clauses $C \vee x$ and $D \vee \neg x$. The resulting clause is called the
resolvent of the two and is denoted by $C * D$. It logically follows from them.
This makes resolution a correct inference rule. Obtaining a clause $C$ from a
set of clauses $F$ by resolving clauses an arbitrary number of times is denoted
$F \vdash C$. Since resolution is correct, a consequence is $F \models C$.
Resolution is also refutationally complete: if $F \models \bot$ then $F \vdash
\bot$.

Resolution is entailment-complete~\cite{slag-etal-69}: if $F \models C$ then $F
\vdash C'$ where $C' \subseteq C$. This property is lost if resolution is
restricted in certain ways that are still refutationally complete, such as
resolving variables in a certain order~\cite{amir-mcil-05}.

\section{Resolution closure}
\label{section-close}

An equivalent definition of forgetting is the set of prime implicates of the
formula that comprise the variables to remember~\cite{lang-etal-03}. Resolution
generates all prime implicates of the formula, among
others~\cite{slag-etal-69}. Selecting the ones that do not contain the
variables to forget results in forgetting.

In order to reduce the number of generated clauses, the non-minimal ones are
removed multiple times. This is correct since a clause that contains another is
entailed by it. At each step, all clauses that resolve among them are resolved.
Their resolvents are added to the formula, which is then simplified by the
removal of non-minimal clauses.

\begin{algorithm}[Forget by Closure]
\label{algorithm-close}

\

\noindent forget\_close(Formula $F$, Variables $V$)

\begin{enumerate}

\item $R = \emptyset$

\item $N = F$

\item while $N \not= R$

\begin{enumerate}

\item $R = N$

\item for each $C \vee x ,~ D \vee \neg x \in R$ for some variable $x$

\begin{enumerate}

\item $N = N \cup \{C \vee x * D \vee \neg x\}$

\end{enumerate}

\item $N = \{C \in N \mid \not\exists C' \in N ,~ C' \subset C\}$

\end{enumerate}

\item return
$\{C \in R \mid \not\exists x \in V .~ x \in C \mbox{ or } \neg x \in C\}$

\end{enumerate}

\end{algorithm}

At each step, the current formula is copied into a new set $R$. All pairs of
clauses of $R$ that resolve are resolved: $C \vee x * D \vee \neg x$ denotes
the resolvent of $C \vee x$ and $D \vee \neg x$, which is added to $N$. This
set is then minimized. If nothing changes, the loop ends. The return value is
the set of clauses of $R$ not containing any variable to forget.

Minimizing at every step takes time but limits the inflation of the formula.
Not doing it would unnecessarily enlarge $N$ by redundant clauses, which may
produce even more clauses in the following steps.

The algorithm is very simple. Apart from resolution and minimization, it only
takes few lines of code. A couple of additional data structures would improve
it.

The first is an index of the clauses that contain a given literal. For each
literal, a set stores all clauses that contain it. This index speeds up finding
the pairs of clauses that resolve. It negates the need of checking each pair of
clauses of the formula for opposite literals. If a clause is in the set of $x$
and another in the set of $\neg x$, they are guaranteed to resolve. The only
required check is whether their resolvent is tautology.

The second is an index of the clauses by their size. This would simplify
minimizing. A clause no longer needs to be checked against all others, only
against the smaller ones. Starting from the smallest might increase the chances
of detecting redundant clauses earlier.

\section{Variable elimination}
\label{section-eliminate}

A way to forget a variable from a formula is to remove all clauses containing
that variable after resolving them in all possible ways. This is called
variable elimination or resolving that variable out. It was introduced by Davis
and Putnam~\cite{davi-putn-60} to establish the satisfiability of a formula.
Eliminating a variable from a formula preserves all consequences of the formula
that do not contain the variable. This includes contradiction: a formula
entails contradiction if and only it does so after eliminating all variables.

The method was applied to forgetting in the Horn case by Delgrande and
Wasserman~\cite{delg-wass-13} and in the general case by Wang~\cite{wang-15}
and Delgrande~\cite{delg-17}. Only a change is required: not all variables are
removed, only the ones to forget. 

\begin{algorithm}[Forget by Elimination]
\label{algorithm-eliminate}

\

\noindent forget\_eliminate(Formula $F$, Variables $V$)

\begin{enumerate}

\item for each $x \in V$:

\begin{enumerate}

\item $P = \{C \in F \mid x \in C\}$

\item $N = \{C \in F \mid \neg x \in C\}$

\item $F = F \backslash (P \cup N) \cup \bigcup_{C \in P, D \in N} C * D$

\end{enumerate}

\item return F

\end{enumerate}

\end{algorithm}

Proof complexity classes variable elimination as an exponential-size method:
for some formulae, the size of a minimal resolution proof is
exponential~\cite{gali-77}. This is unconditional: it does not depend on any
complexity theory assumption like the non-equality of the complexity classes
\p{} and \np. A proof being exponential means that the clauses become
exponentially many while eliminating all variables.


The algorithm has a long history, starting more than half a century
ago~\cite{davi-putn-60,rish-dech-00,osam-etal-21}. Even its version that
removes only a subset of the variables~\cite{subb-prad-04} predates its
application to propositional forgetting. The original algorithm by Davis and
Putnam also included two optimizations (unit propagation and pure literals)
that are neglected by proof complexity and the adaptations to forgetting.

The program is implemented using the same resolution subroutine of the others.
It could be sped up by indexing the clauses by the literals they contain: for
each variable, two sets store the clauses that contain the variable
respectively negated and unnegated. These sets would make finding the clauses
to resolve trivial: each clause of the first set is resolved with each of the
second. The non-tautological results are kept, the original clauses discarded.
To save time, clauses may not be removed immediately from the sets but only
marked as removed. Alternatively, the order of elimination of variables is
chosen beforehand, and a clause is not added in the sets of a variable if it
also contains a variable that is removed before it; this variant is called
bucket elimination~\cite{dech-rish-94,dech-99}.

This optimization is not implemented, but kept into account when counting the
running time of the algorithm: the time spent to find the clauses to resolve is
neglected. Only the number of resolutions is added. While the program checks
every pair of clauses in the formula, only the pairs that resolve are counted.
From the time accounting point of view, this is the same as implementing the
optimization.

\section{Linear resolution}
\label{section-linear}

\proofonly{

\separator

note that the experimental results in dech-rish-94 tell that some directional
resolution is good in some cases

{\bf why linear resolution, what it is}

\

}

Variable elimination is a form of resolution, called directional
resolution~\cite{dech-rish-94}. Adapting it to forgetting is straightforward,
by resolving on variables to forget only. On the downside, it is not the best
resolution strategy, both from the theoretical and practical point of
view~\cite{gali-77,goer-92,bone-etal-00,dech-rish-94}. Better resolution
policies exists. A commonly used one is linear
resolution~\cite{love-70,luck-70,zamo-shar-71,inou-92,buss-joha-16}. Adapting
it to forgetting is equally simple, but requires a separate proof of
correctness.

Linear resolution hinges around a center clause. This is initially a clause of
the input formula. At each step, the policy is to always resolve the center
clause with a side clause, which is either a clause of the input formula or a
previous center clause. The result is the new center clause. This mechanism is
refutationally complete: if a formula is unsatisfiable, the empty clause
follows from a linear resolution.

\proofonly{

\separator

{\bf How to adapt linear resolution to forgetting and proof.}

\

}

This way of forgetting requires all clauses of the original formula to be
linearly resolved in all possible ways. Running time decreases by reducing the
number of lines to explore. Forgetting allows for two reductions:

\begin{itemize}

\item only resolving on variables to forget;

\item stopping when the center clause does not contain any variable to forget.

\end{itemize}

Like all restrictions of resolution, linear resolution only generates clauses
entailed by the original formula. The variant for forgetting only generates
clauses comprising variables to remember. The result is a formula on the
variables to remember that is entailed by the original formula. This is not
enough. A further requirement is that every formula on the variables to
remember that is entailed by the original formula is also entailed by the
result of forgetting. Proving this condition only on the clauses comprising all
variables to remember is sufficient~\cite{libe-20-a}.

A clause that contains all variables is a full clause~\cite{abba-kull-18}. A
clause that contains all variables to remember is a full remembrance clause.
Linear resolution with the two additional restrictions for forgetting is proved
to generate a subset of every full remembrance clause that is entailed by the
formula.

A full remembrance clause $C$ is entailed by $F$ if and only if $F \cup \neg C
\models \bot$, where $\neg C$ is the set comprising the negation of the
literals of $C$:

\[
\neg C = \{\neg l \mid l \in C\}
\]

Clauses entailed by others can be removed from a formula without altering its
semantics. This is the case for clauses that contains others. Therefore, $F
\cup \neg C \models \bot$ still holds when removing from $F$ all clauses that
contain a literal in $\neg C$, the opposite of a literal of $C$.

\[
G = \{D \in F \mid \not\exists l \in C ~.~ \neg l \in D\}
\]

The entailment $G \cup \neg C \models \bot$ holds since $G$ is equivalent to
$F$. Therefore, the empty clause follows from $G \cup \neg C$ by linear
resolution. This is denoted $G \cup \neg C \vdash \bot$. The claim is that such
a linear resolution can be altered to derive a subset of $C$ from $G$ while
obeying the two above restriction.

\begin{center}
\setlength{\unitlength}{3947sp}%
\begingroup\makeatletter\ifx\SetFigFont\undefined%
\gdef\SetFigFont#1#2#3#4#5{%
  \reset@font\fontsize{#1}{#2pt}%
  \fontfamily{#3}\fontseries{#4}\fontshape{#5}%
  \selectfont}%
\fi\endgroup%
\begin{picture}(7080,630)(2311,-3751)
\thinlines
{\color[rgb]{0,0,0}\put(5476,-3661){\line( 0, 1){300}}
}%
{\color[rgb]{0,0,0}\put(5251,-3661){\line( 1, 0){450}}
}%
{\color[rgb]{0,0,0}\put(6151,-3661){\line( 1, 0){450}}
}%
{\color[rgb]{0,0,0}\put(6376,-3661){\line( 0, 1){300}}
}%
{\color[rgb]{0,0,0}\put(7051,-3661){\line( 1, 0){450}}
}%
{\color[rgb]{0,0,0}\put(7276,-3661){\line( 0, 1){300}}
}%
{\color[rgb]{0,0,0}\put(7951,-3661){\line( 1, 0){450}}
}%
{\color[rgb]{0,0,0}\put(8176,-3661){\line( 0, 1){300}}
}%
{\color[rgb]{0,0,0}\put(8851,-3661){\line( 1, 0){450}}
}%
{\color[rgb]{0,0,0}\put(9076,-3661){\line( 0, 1){300}}
}%
{\color[rgb]{0,0,0}\put(4351,-3661){\line( 1, 0){450}}
}%
{\color[rgb]{0,0,0}\put(4576,-3361){\line( 0,-1){300}}
}%
\put(9376,-3736){\makebox(0,0)[lb]{\smash{{\SetFigFont{12}{14.4}
{\rmdefault}{\mddefault}{\updefault}{\color[rgb]{0,0,0}$\bot$}%
}}}}
\put(3226,-3736){\makebox(0,0)[b]{\smash{{\SetFigFont{12}{14.4}
{\rmdefault}{\mddefault}{\updefault}{\color[rgb]{0,0,0}$?$}%
}}}}
\put(3676,-3286){\makebox(0,0)[b]{\smash{{\SetFigFont{12}{14.4}
{\rmdefault}{\mddefault}{\updefault}{\color[rgb]{0,0,0}$?$}%
}}}}
\put(2776,-3286){\makebox(0,0)[b]{\smash{{\SetFigFont{12}{14.4}
{\rmdefault}{\mddefault}{\updefault}{\color[rgb]{0,0,0}$?$}%
}}}}
\put(2326,-3736){\makebox(0,0)[b]{\smash{{\SetFigFont{12}{14.4}
{\rmdefault}{\mddefault}{\updefault}{\color[rgb]{0,0,0}$?$}%
}}}}
\put(5026,-3736){\makebox(0,0)[b]{\smash{{\SetFigFont{12}{14.4}
{\rmdefault}{\mddefault}{\updefault}{\color[rgb]{0,0,0}$?$}%
}}}}
\put(5476,-3286){\makebox(0,0)[b]{\smash{{\SetFigFont{12}{14.4}
{\rmdefault}{\mddefault}{\updefault}{\color[rgb]{0,0,0}$?$}%
}}}}
\put(5926,-3736){\makebox(0,0)[b]{\smash{{\SetFigFont{12}{14.4}
{\rmdefault}{\mddefault}{\updefault}{\color[rgb]{0,0,0}$?$}%
}}}}
\put(6376,-3286){\makebox(0,0)[b]{\smash{{\SetFigFont{12}{14.4}
{\rmdefault}{\mddefault}{\updefault}{\color[rgb]{0,0,0}$?$}%
}}}}
\put(6826,-3736){\makebox(0,0)[b]{\smash{{\SetFigFont{12}{14.4}
{\rmdefault}{\mddefault}{\updefault}{\color[rgb]{0,0,0}$?$}%
}}}}
\put(7276,-3286){\makebox(0,0)[b]{\smash{{\SetFigFont{12}{14.4}
{\rmdefault}{\mddefault}{\updefault}{\color[rgb]{0,0,0}$?$}%
}}}}
\put(7726,-3736){\makebox(0,0)[b]{\smash{{\SetFigFont{12}{14.4}
{\rmdefault}{\mddefault}{\updefault}{\color[rgb]{0,0,0}$?$}%
}}}}
\put(8176,-3286){\makebox(0,0)[b]{\smash{{\SetFigFont{12}{14.4}
{\rmdefault}{\mddefault}{\updefault}{\color[rgb]{0,0,0}$?$}%
}}}}
\put(8626,-3736){\makebox(0,0)[b]{\smash{{\SetFigFont{12}{14.4}
{\rmdefault}{\mddefault}{\updefault}{\color[rgb]{0,0,0}$?$}%
}}}}
\put(9076,-3286){\makebox(0,0)[b]{\smash{{\SetFigFont{12}{14.4}
{\rmdefault}{\mddefault}{\updefault}{\color[rgb]{0,0,0}$?$}%
}}}}
\put(4126,-3736){\makebox(0,0)[b]{\smash{{\SetFigFont{12}{14.4}
{\rmdefault}{\mddefault}{\updefault}{\color[rgb]{0,0,0}$?$}%
}}}}
\put(4576,-3286){\makebox(0,0)[b]{\smash{{\SetFigFont{12}{14.4}
{\rmdefault}{\mddefault}{\updefault}{\color[rgb]{0,0,0}$?$}%
}}}}
{\color[rgb]{0,0,0}\put(3451,-3661){\line( 1, 0){450}}
}%
{\color[rgb]{0,0,0}\put(3676,-3661){\line( 0, 1){300}}
}%
{\color[rgb]{0,0,0}\put(2551,-3661){\line( 1, 0){450}}
}%
{\color[rgb]{0,0,0}\put(2776,-3661){\line( 0, 1){300}}
}%
\end{picture}%
\nop{
   |   |   |   |   |   |   |
  -+- -+- -+- -+- -+- -+- -+- false
}
\end{center}

The derivation starts from a clause of $G \cup \neg C$ and iteratively resolves
it until it is empty. These unknown clauses and their resolved are denoted as
question marks in the figure. If none of them contain $l$ or $\neg l$, then $G
\vdash \false$, which means that $\false$ expresses forgetting.

Otherwise, some clauses in the derivation contain $l$ or $\neg l$. These
literals are removed by some resolution steps since the final clause $\bot$
does not contain them. By construction, $l$ is negative only in the clause
$\neg l$. This clause only resolves with a clause containing $l$, and the
result does not contain these literals. The conclusion is that the only clause
containing $\neg l$ in the derivation is the clause $\neg l$ itself. No other
clause contains $\neg l$ together with other literals.

At some point, resolution removes $l$ since the final clause is empty. Since
$\neg l$ is the only clause containing $\neg l$ that occurs in the derivation,
this step resolves it with a clause $D \vee l$.

The resolution of $\neg l$ with a clause $D \vee l$ may occur several times in
the derivation. The first occurrence is considered. The center clause may be
$\neg l$ or $D \vee \neg l$. The first case can be reduced to the second: since
$\neg l$ is also a clause of $G \cup \neg C$, everything before it can be cut
out, and this clause swapped with the other.

\begin{center}
\setlength{\unitlength}{3947sp}%
\begingroup\makeatletter\ifx\SetFigFont\undefined%
\gdef\SetFigFont#1#2#3#4#5{%
  \reset@font\fontsize{#1}{#2pt}%
  \fontfamily{#3}\fontseries{#4}\fontshape{#5}%
  \selectfont}%
\fi\endgroup%
\begin{picture}(7080,630)(2311,-3751)
\thinlines
{\color[rgb]{0,0,0}\put(5476,-3661){\line( 0, 1){300}}
}%
{\color[rgb]{0,0,0}\put(5251,-3661){\line( 1, 0){450}}
}%
{\color[rgb]{0,0,0}\put(6151,-3661){\line( 1, 0){450}}
}%
{\color[rgb]{0,0,0}\put(6376,-3661){\line( 0, 1){300}}
}%
{\color[rgb]{0,0,0}\put(7051,-3661){\line( 1, 0){450}}
}%
{\color[rgb]{0,0,0}\put(7276,-3661){\line( 0, 1){300}}
}%
{\color[rgb]{0,0,0}\put(7951,-3661){\line( 1, 0){450}}
}%
{\color[rgb]{0,0,0}\put(8176,-3661){\line( 0, 1){300}}
}%
{\color[rgb]{0,0,0}\put(8851,-3661){\line( 1, 0){450}}
}%
{\color[rgb]{0,0,0}\put(9076,-3661){\line( 0, 1){300}}
}%
{\color[rgb]{0,0,0}\put(4351,-3661){\line( 1, 0){450}}
}%
{\color[rgb]{0,0,0}\put(4576,-3361){\line( 0,-1){300}}
}%
\put(9376,-3736){\makebox(0,0)[lb]{\smash{{\SetFigFont{12}{14.4}
{\rmdefault}{\mddefault}{\updefault}{\color[rgb]{0,0,0}$\bot$}%
}}}}
\put(3226,-3736){\makebox(0,0)[b]{\smash{{\SetFigFont{12}{14.4}
{\rmdefault}{\mddefault}{\updefault}{\color[rgb]{0,0,0}$?$}%
}}}}
\put(3676,-3286){\makebox(0,0)[b]{\smash{{\SetFigFont{12}{14.4}
{\rmdefault}{\mddefault}{\updefault}{\color[rgb]{0,0,0}$?$}%
}}}}
\put(2776,-3286){\makebox(0,0)[b]{\smash{{\SetFigFont{12}{14.4}
{\rmdefault}{\mddefault}{\updefault}{\color[rgb]{0,0,0}$?$}%
}}}}
\put(2326,-3736){\makebox(0,0)[b]{\smash{{\SetFigFont{12}{14.4}
{\rmdefault}{\mddefault}{\updefault}{\color[rgb]{0,0,0}$?$}%
}}}}
\put(5026,-3736){\makebox(0,0)[b]{\smash{{\SetFigFont{12}{14.4}
{\rmdefault}{\mddefault}{\updefault}{\color[rgb]{0,0,0}$?$}%
}}}}
\put(5476,-3286){\makebox(0,0)[b]{\smash{{\SetFigFont{12}{14.4}
{\rmdefault}{\mddefault}{\updefault}{\color[rgb]{0,0,0}$?$}%
}}}}
\put(5926,-3736){\makebox(0,0)[b]{\smash{{\SetFigFont{12}{14.4}
{\rmdefault}{\mddefault}{\updefault}{\color[rgb]{0,0,0}$?$}%
}}}}
\put(6376,-3286){\makebox(0,0)[b]{\smash{{\SetFigFont{12}{14.4}
{\rmdefault}{\mddefault}{\updefault}{\color[rgb]{0,0,0}$?$}%
}}}}
\put(6826,-3736){\makebox(0,0)[b]{\smash{{\SetFigFont{12}{14.4}
{\rmdefault}{\mddefault}{\updefault}{\color[rgb]{0,0,0}$?$}%
}}}}
\put(7276,-3286){\makebox(0,0)[b]{\smash{{\SetFigFont{12}{14.4}
{\rmdefault}{\mddefault}{\updefault}{\color[rgb]{0,0,0}$?$}%
}}}}
\put(7726,-3736){\makebox(0,0)[b]{\smash{{\SetFigFont{12}{14.4}
{\rmdefault}{\mddefault}{\updefault}{\color[rgb]{0,0,0}$?$}%
}}}}
\put(8176,-3286){\makebox(0,0)[b]{\smash{{\SetFigFont{12}{14.4}
{\rmdefault}{\mddefault}{\updefault}{\color[rgb]{0,0,0}$?$}%
}}}}
\put(8626,-3736){\makebox(0,0)[b]{\smash{{\SetFigFont{12}{14.4}
{\rmdefault}{\mddefault}{\updefault}{\color[rgb]{0,0,0}$?$}%
}}}}
\put(9076,-3286){\makebox(0,0)[b]{\smash{{\SetFigFont{12}{14.4}
{\rmdefault}{\mddefault}{\updefault}{\color[rgb]{0,0,0}$?$}%
}}}}
{\color[rgb]{0,0,0}\put(3451,-3661){\line( 1, 0){450}}
}%
{\color[rgb]{0,0,0}\put(3676,-3661){\line( 0, 1){300}}
}%
{\color[rgb]{0,0,0}\put(2551,-3661){\line( 1, 0){450}}
}%
{\color[rgb]{0,0,0}\put(2776,-3661){\line( 0, 1){300}}
}%
\put(4576,-3286){\makebox(0,0)[b]{\smash{{\SetFigFont{12}{14.4}
{\rmdefault}{\mddefault}{\updefault}{\color[rgb]{0,0,0}$D \vee l$}%
}}}}
\put(4126,-3736){\makebox(0,0)[b]{\smash{{\SetFigFont{12}{14.4}
{\rmdefault}{\mddefault}{\updefault}{\color[rgb]{0,0,0}$\neg l$}%
}}}}
\end{picture}%
\nop{
                D v l
   |   |   |      |   |   |   |
  -+- -+- -+- -l -+- -+- -+- -+- false
}
\end{center}

This change turns $\neg l$ from a center clause to a side clause.

\begin{center}
\setlength{\unitlength}{3947sp}%
\begingroup\makeatletter\ifx\SetFigFont\undefined%
\gdef\SetFigFont#1#2#3#4#5{%
  \reset@font\fontsize{#1}{#2pt}%
  \fontfamily{#3}\fontseries{#4}\fontshape{#5}%
  \selectfont}%
\fi\endgroup%
\begin{picture}(7080,630)(2311,-3751)
\thinlines
{\color[rgb]{0,0,0}\put(5476,-3661){\line( 0, 1){300}}
}%
{\color[rgb]{0,0,0}\put(5251,-3661){\line( 1, 0){450}}
}%
{\color[rgb]{0,0,0}\put(6151,-3661){\line( 1, 0){450}}
}%
{\color[rgb]{0,0,0}\put(6376,-3661){\line( 0, 1){300}}
}%
{\color[rgb]{0,0,0}\put(7051,-3661){\line( 1, 0){450}}
}%
{\color[rgb]{0,0,0}\put(7276,-3661){\line( 0, 1){300}}
}%
{\color[rgb]{0,0,0}\put(7951,-3661){\line( 1, 0){450}}
}%
{\color[rgb]{0,0,0}\put(8176,-3661){\line( 0, 1){300}}
}%
{\color[rgb]{0,0,0}\put(8851,-3661){\line( 1, 0){450}}
}%
{\color[rgb]{0,0,0}\put(9076,-3661){\line( 0, 1){300}}
}%
{\color[rgb]{0,0,0}\put(4351,-3661){\line( 1, 0){450}}
}%
{\color[rgb]{0,0,0}\put(4576,-3361){\line( 0,-1){300}}
}%
\put(9376,-3736){\makebox(0,0)[lb]{\smash{{\SetFigFont{12}{14.4}
{\rmdefault}{\mddefault}{\updefault}{\color[rgb]{0,0,0}$\bot$}%
}}}}
\put(5026,-3736){\makebox(0,0)[b]{\smash{{\SetFigFont{12}{14.4}
{\rmdefault}{\mddefault}{\updefault}{\color[rgb]{0,0,0}$?$}%
}}}}
\put(5476,-3286){\makebox(0,0)[b]{\smash{{\SetFigFont{12}{14.4}
{\rmdefault}{\mddefault}{\updefault}{\color[rgb]{0,0,0}$?$}%
}}}}
\put(5926,-3736){\makebox(0,0)[b]{\smash{{\SetFigFont{12}{14.4}
{\rmdefault}{\mddefault}{\updefault}{\color[rgb]{0,0,0}$?$}%
}}}}
\put(6376,-3286){\makebox(0,0)[b]{\smash{{\SetFigFont{12}{14.4}
{\rmdefault}{\mddefault}{\updefault}{\color[rgb]{0,0,0}$?$}%
}}}}
\put(6826,-3736){\makebox(0,0)[b]{\smash{{\SetFigFont{12}{14.4}
{\rmdefault}{\mddefault}{\updefault}{\color[rgb]{0,0,0}$?$}%
}}}}
\put(7276,-3286){\makebox(0,0)[b]{\smash{{\SetFigFont{12}{14.4}
{\rmdefault}{\mddefault}{\updefault}{\color[rgb]{0,0,0}$?$}%
}}}}
\put(7726,-3736){\makebox(0,0)[b]{\smash{{\SetFigFont{12}{14.4}
{\rmdefault}{\mddefault}{\updefault}{\color[rgb]{0,0,0}$?$}%
}}}}
\put(8176,-3286){\makebox(0,0)[b]{\smash{{\SetFigFont{12}{14.4}
{\rmdefault}{\mddefault}{\updefault}{\color[rgb]{0,0,0}$?$}%
}}}}
\put(8626,-3736){\makebox(0,0)[b]{\smash{{\SetFigFont{12}{14.4}
{\rmdefault}{\mddefault}{\updefault}{\color[rgb]{0,0,0}$?$}%
}}}}
\put(9076,-3286){\makebox(0,0)[b]{\smash{{\SetFigFont{12}{14.4}
{\rmdefault}{\mddefault}{\updefault}{\color[rgb]{0,0,0}$?$}%
}}}}
\put(4126,-3736){\makebox(0,0)[b]{\smash{{\SetFigFont{12}{14.4}
{\rmdefault}{\mddefault}{\updefault}{\color[rgb]{0,0,0}$D \vee l$}%
}}}}
\put(4576,-3286){\makebox(0,0)[b]{\smash{{\SetFigFont{12}{14.4}
{\rmdefault}{\mddefault}{\updefault}{\color[rgb]{0,0,0}$\neg l$}%
}}}}
\end{picture}%
\nop{
                 -l
                  |   |   |   |
           D v l -+- -+- -+- -+- false
}
\end{center}

The general situation is that $\neg l$ is a side clause and $D \vee l$ is a
center clause. Resolving them gives $D$. The following clauses are denoted
$E$, $F$, and so on.

\begin{center}
\setlength{\unitlength}{3947sp}%
\begingroup\makeatletter\ifx\SetFigFont\undefined%
\gdef\SetFigFont#1#2#3#4#5{%
  \reset@font\fontsize{#1}{#2pt}%
  \fontfamily{#3}\fontseries{#4}\fontshape{#5}%
  \selectfont}%
\fi\endgroup%
\begin{picture}(7080,688)(2761,-3809)
\thinlines
{\color[rgb]{0,0,0}\put(3901,-3661){\line( 1, 0){450}}
}%
{\color[rgb]{0,0,0}\put(4126,-3661){\line( 0, 1){300}}
}%
{\color[rgb]{0,0,0}\put(5926,-3661){\line( 0, 1){300}}
}%
{\color[rgb]{0,0,0}\put(5701,-3661){\line( 1, 0){450}}
}%
{\color[rgb]{0,0,0}\put(6601,-3661){\line( 1, 0){450}}
}%
{\color[rgb]{0,0,0}\put(6826,-3661){\line( 0, 1){300}}
}%
{\color[rgb]{0,0,0}\put(7501,-3661){\line( 1, 0){450}}
}%
{\color[rgb]{0,0,0}\put(7726,-3661){\line( 0, 1){300}}
}%
{\color[rgb]{0,0,0}\put(3001,-3661){\line( 1, 0){450}}
}%
{\color[rgb]{0,0,0}\put(3226,-3661){\line( 0, 1){300}}
}%
{\color[rgb]{0,0,0}\put(8401,-3661){\line( 1, 0){450}}
}%
{\color[rgb]{0,0,0}\put(8626,-3661){\line( 0, 1){300}}
}%
{\color[rgb]{0,0,0}\put(9301,-3661){\line( 1, 0){450}}
}%
{\color[rgb]{0,0,0}\put(9526,-3661){\line( 0, 1){300}}
}%
\put(4576,-3736){\makebox(0,0)[b]{\smash{{\SetFigFont{12}{14.4}
{\rmdefault}{\mddefault}{\updefault}{\color[rgb]{0,0,0}$D \vee l$}%
}}}}
\put(2776,-3736){\makebox(0,0)[b]{\smash{{\SetFigFont{12}{14.4}
{\rmdefault}{\mddefault}{\updefault}{\color[rgb]{0,0,0}$?$}%
}}}}
\put(3676,-3736){\makebox(0,0)[b]{\smash{{\SetFigFont{12}{14.4}
{\rmdefault}{\mddefault}{\updefault}{\color[rgb]{0,0,0}$?$}%
}}}}
\put(3226,-3286){\makebox(0,0)[b]{\smash{{\SetFigFont{12}{14.4}
{\rmdefault}{\mddefault}{\updefault}{\color[rgb]{0,0,0}$?$}%
}}}}
\put(4126,-3286){\makebox(0,0)[b]{\smash{{\SetFigFont{12}{14.4}
{\rmdefault}{\mddefault}{\updefault}{\color[rgb]{0,0,0}$?$}%
}}}}
\put(5926,-3286){\makebox(0,0)[b]{\smash{{\SetFigFont{12}{14.4}
{\rmdefault}{\mddefault}{\updefault}{\color[rgb]{0,0,0}$?$}%
}}}}
\put(6826,-3286){\makebox(0,0)[b]{\smash{{\SetFigFont{12}{14.4}
{\rmdefault}{\mddefault}{\updefault}{\color[rgb]{0,0,0}$?$}%
}}}}
\put(7726,-3286){\makebox(0,0)[b]{\smash{{\SetFigFont{12}{14.4}
{\rmdefault}{\mddefault}{\updefault}{\color[rgb]{0,0,0}$?$}%
}}}}
\put(8176,-3736){\makebox(0,0)[b]{\smash{{\SetFigFont{12}{14.4}
{\rmdefault}{\mddefault}{\updefault}{\color[rgb]{0,0,0}$?$}%
}}}}
\put(8626,-3286){\makebox(0,0)[b]{\smash{{\SetFigFont{12}{14.4}
{\rmdefault}{\mddefault}{\updefault}{\color[rgb]{0,0,0}$?$}%
}}}}
\put(9076,-3736){\makebox(0,0)[b]{\smash{{\SetFigFont{12}{14.4}
{\rmdefault}{\mddefault}{\updefault}{\color[rgb]{0,0,0}$?$}%
}}}}
\put(9526,-3286){\makebox(0,0)[b]{\smash{{\SetFigFont{12}{14.4}
{\rmdefault}{\mddefault}{\updefault}{\color[rgb]{0,0,0}$?$}%
}}}}
{\color[rgb]{0,0,0}\put(4801,-3661){\line( 1, 0){450}}
}%
{\color[rgb]{0,0,0}\put(5026,-3361){\line( 0,-1){300}}
}%
\put(5476,-3736){\makebox(0,0)[b]{\smash{{\SetFigFont{12}{14.4}
{\rmdefault}{\mddefault}{\updefault}{\color[rgb]{0,0,0}$D$}%
}}}}
\put(6376,-3736){\makebox(0,0)[b]{\smash{{\SetFigFont{12}{14.4}
{\rmdefault}{\mddefault}{\updefault}{\color[rgb]{0,0,0}$E$}%
}}}}
\put(7276,-3736){\makebox(0,0)[b]{\smash{{\SetFigFont{12}{14.4}
{\rmdefault}{\mddefault}{\updefault}{\color[rgb]{0,0,0}$F$}%
}}}}
\put(5026,-3286){\makebox(0,0)[b]{\smash{{\SetFigFont{12}{14.4}
{\rmdefault}{\mddefault}{\updefault}{\color[rgb]{0,0,0}$\neg l$}%
}}}}
\put(9826,-3736){\makebox(0,0)[lb]{\smash{{\SetFigFont{12}{14.4}
{\rmdefault}{\mddefault}{\updefault}{\color[rgb]{0,0,0}$\bot$}%
}}}}
\end{picture}%
\nop{
                    -l
   |   |   |         |     |     |     |   |
  -+- -+- -+- D v l -+- D -+- E -+- F -+- -+- false
}
\end{center}

Removing this step still leaves this line a valid resolution derivation, with
the addition of $l$ to the clauses following $D \vee l$.

\begin{center}
\setlength{\unitlength}{3947sp}%
\begingroup\makeatletter\ifx\SetFigFont\undefined%
\gdef\SetFigFont#1#2#3#4#5{%
  \reset@font\fontsize{#1}{#2pt}%
  \fontfamily{#3}\fontseries{#4}\fontshape{#5}%
  \selectfont}%
\fi\endgroup%
\begin{picture}(7080,688)(2761,-3809)
\thinlines
{\color[rgb]{0,0,0}\put(3901,-3661){\line( 1, 0){450}}
}%
{\color[rgb]{0,0,0}\put(4126,-3661){\line( 0, 1){300}}
}%
{\color[rgb]{0,0,0}\put(5926,-3661){\line( 0, 1){300}}
}%
{\color[rgb]{0,0,0}\put(5701,-3661){\line( 1, 0){450}}
}%
{\color[rgb]{0,0,0}\put(6601,-3661){\line( 1, 0){450}}
}%
{\color[rgb]{0,0,0}\put(6826,-3661){\line( 0, 1){300}}
}%
{\color[rgb]{0,0,0}\put(7501,-3661){\line( 1, 0){450}}
}%
{\color[rgb]{0,0,0}\put(7726,-3661){\line( 0, 1){300}}
}%
{\color[rgb]{0,0,0}\put(3001,-3661){\line( 1, 0){450}}
}%
{\color[rgb]{0,0,0}\put(3226,-3661){\line( 0, 1){300}}
}%
{\color[rgb]{0,0,0}\put(8401,-3661){\line( 1, 0){450}}
}%
{\color[rgb]{0,0,0}\put(8626,-3661){\line( 0, 1){300}}
}%
{\color[rgb]{0,0,0}\put(9301,-3661){\line( 1, 0){450}}
}%
{\color[rgb]{0,0,0}\put(9526,-3661){\line( 0, 1){300}}
}%
\put(4576,-3736){\makebox(0,0)[b]{\smash{{\SetFigFont{12}{14.4}
{\rmdefault}{\mddefault}{\updefault}{\color[rgb]{0,0,0}$D \vee l$}%
}}}}
\put(2776,-3736){\makebox(0,0)[b]{\smash{{\SetFigFont{12}{14.4}
{\rmdefault}{\mddefault}{\updefault}{\color[rgb]{0,0,0}$?$}%
}}}}
\put(3676,-3736){\makebox(0,0)[b]{\smash{{\SetFigFont{12}{14.4}
{\rmdefault}{\mddefault}{\updefault}{\color[rgb]{0,0,0}$?$}%
}}}}
\put(3226,-3286){\makebox(0,0)[b]{\smash{{\SetFigFont{12}{14.4}
{\rmdefault}{\mddefault}{\updefault}{\color[rgb]{0,0,0}$?$}%
}}}}
\put(4126,-3286){\makebox(0,0)[b]{\smash{{\SetFigFont{12}{14.4}
{\rmdefault}{\mddefault}{\updefault}{\color[rgb]{0,0,0}$?$}%
}}}}
\put(5926,-3286){\makebox(0,0)[b]{\smash{{\SetFigFont{12}{14.4}
{\rmdefault}{\mddefault}{\updefault}{\color[rgb]{0,0,0}$?$}%
}}}}
\put(6826,-3286){\makebox(0,0)[b]{\smash{{\SetFigFont{12}{14.4}
{\rmdefault}{\mddefault}{\updefault}{\color[rgb]{0,0,0}$?$}%
}}}}
\put(7726,-3286){\makebox(0,0)[b]{\smash{{\SetFigFont{12}{14.4}
{\rmdefault}{\mddefault}{\updefault}{\color[rgb]{0,0,0}$?$}%
}}}}
\put(8176,-3736){\makebox(0,0)[b]{\smash{{\SetFigFont{12}{14.4}
{\rmdefault}{\mddefault}{\updefault}{\color[rgb]{0,0,0}$?$}%
}}}}
\put(8626,-3286){\makebox(0,0)[b]{\smash{{\SetFigFont{12}{14.4}
{\rmdefault}{\mddefault}{\updefault}{\color[rgb]{0,0,0}$?$}%
}}}}
\put(9076,-3736){\makebox(0,0)[b]{\smash{{\SetFigFont{12}{14.4}
{\rmdefault}{\mddefault}{\updefault}{\color[rgb]{0,0,0}$?$}%
}}}}
\put(9526,-3286){\makebox(0,0)[b]{\smash{{\SetFigFont{12}{14.4}
{\rmdefault}{\mddefault}{\updefault}{\color[rgb]{0,0,0}$?$}%
}}}}
{\color[rgb]{0,0,0}\put(4801,-3661){\line( 1, 0){900}}
}%
\put(6376,-3736){\makebox(0,0)[b]{\smash{{\SetFigFont{12}{14.4}
{\rmdefault}{\mddefault}{\updefault}{\color[rgb]{0,0,0}$E \vee l$}%
}}}}
\put(7276,-3736){\makebox(0,0)[b]{\smash{{\SetFigFont{12}{14.4}
{\rmdefault}{\mddefault}{\updefault}{\color[rgb]{0,0,0}$F \vee l$}%
}}}}
\end{picture}%
\nop{
   |   |   |               |         |         |   |
  -+- -+- -+- D v l ----- -+- E v l -+- F v l -+- -+-
}
\end{center}

A following step may still resolve $l$ with $\neg l$, but the same procedure
applies again. This resolution is removed and $l$ added to the following
clauses. The final clause is added $l$.

\begin{center}
\setlength{\unitlength}{3947sp}%
\begingroup\makeatletter\ifx\SetFigFont\undefined%
\gdef\SetFigFont#1#2#3#4#5{%
  \reset@font\fontsize{#1}{#2pt}%
  \fontfamily{#3}\fontseries{#4}\fontshape{#5}%
  \selectfont}%
\fi\endgroup%
\begin{picture}(7080,688)(2761,-3809)
\thinlines
{\color[rgb]{0,0,0}\put(3901,-3661){\line( 1, 0){450}}
}%
{\color[rgb]{0,0,0}\put(4126,-3661){\line( 0, 1){300}}
}%
{\color[rgb]{0,0,0}\put(5926,-3661){\line( 0, 1){300}}
}%
{\color[rgb]{0,0,0}\put(5701,-3661){\line( 1, 0){450}}
}%
{\color[rgb]{0,0,0}\put(6601,-3661){\line( 1, 0){450}}
}%
{\color[rgb]{0,0,0}\put(6826,-3661){\line( 0, 1){300}}
}%
{\color[rgb]{0,0,0}\put(7501,-3661){\line( 1, 0){450}}
}%
{\color[rgb]{0,0,0}\put(7726,-3661){\line( 0, 1){300}}
}%
{\color[rgb]{0,0,0}\put(3001,-3661){\line( 1, 0){450}}
}%
{\color[rgb]{0,0,0}\put(3226,-3661){\line( 0, 1){300}}
}%
{\color[rgb]{0,0,0}\put(8401,-3661){\line( 1, 0){450}}
}%
{\color[rgb]{0,0,0}\put(8626,-3661){\line( 0, 1){300}}
}%
{\color[rgb]{0,0,0}\put(9301,-3661){\line( 1, 0){450}}
}%
{\color[rgb]{0,0,0}\put(9526,-3661){\line( 0, 1){300}}
}%
\put(4576,-3736){\makebox(0,0)[b]{\smash{{\SetFigFont{12}{14.4}
{\rmdefault}{\mddefault}{\updefault}{\color[rgb]{0,0,0}$D \vee l$}%
}}}}
\put(2776,-3736){\makebox(0,0)[b]{\smash{{\SetFigFont{12}{14.4}
{\rmdefault}{\mddefault}{\updefault}{\color[rgb]{0,0,0}$?$}%
}}}}
\put(3676,-3736){\makebox(0,0)[b]{\smash{{\SetFigFont{12}{14.4}
{\rmdefault}{\mddefault}{\updefault}{\color[rgb]{0,0,0}$?$}%
}}}}
\put(3226,-3286){\makebox(0,0)[b]{\smash{{\SetFigFont{12}{14.4}
{\rmdefault}{\mddefault}{\updefault}{\color[rgb]{0,0,0}$?$}%
}}}}
\put(4126,-3286){\makebox(0,0)[b]{\smash{{\SetFigFont{12}{14.4}
{\rmdefault}{\mddefault}{\updefault}{\color[rgb]{0,0,0}$?$}%
}}}}
\put(5926,-3286){\makebox(0,0)[b]{\smash{{\SetFigFont{12}{14.4}
{\rmdefault}{\mddefault}{\updefault}{\color[rgb]{0,0,0}$?$}%
}}}}
\put(6826,-3286){\makebox(0,0)[b]{\smash{{\SetFigFont{12}{14.4}
{\rmdefault}{\mddefault}{\updefault}{\color[rgb]{0,0,0}$?$}%
}}}}
\put(7726,-3286){\makebox(0,0)[b]{\smash{{\SetFigFont{12}{14.4}
{\rmdefault}{\mddefault}{\updefault}{\color[rgb]{0,0,0}$?$}%
}}}}
\put(8176,-3736){\makebox(0,0)[b]{\smash{{\SetFigFont{12}{14.4}
{\rmdefault}{\mddefault}{\updefault}{\color[rgb]{0,0,0}$?$}%
}}}}
\put(8626,-3286){\makebox(0,0)[b]{\smash{{\SetFigFont{12}{14.4}
{\rmdefault}{\mddefault}{\updefault}{\color[rgb]{0,0,0}$?$}%
}}}}
\put(9076,-3736){\makebox(0,0)[b]{\smash{{\SetFigFont{12}{14.4}
{\rmdefault}{\mddefault}{\updefault}{\color[rgb]{0,0,0}$?$}%
}}}}
\put(9526,-3286){\makebox(0,0)[b]{\smash{{\SetFigFont{12}{14.4}
{\rmdefault}{\mddefault}{\updefault}{\color[rgb]{0,0,0}$?$}%
}}}}
{\color[rgb]{0,0,0}\put(4801,-3661){\line( 1, 0){900}}
}%
\put(6376,-3736){\makebox(0,0)[b]{\smash{{\SetFigFont{12}{14.4}
{\rmdefault}{\mddefault}{\updefault}{\color[rgb]{0,0,0}$E \vee l$}%
}}}}
\put(7276,-3736){\makebox(0,0)[b]{\smash{{\SetFigFont{12}{14.4}
{\rmdefault}{\mddefault}{\updefault}{\color[rgb]{0,0,0}$F \vee l$}%
}}}}
\put(9826,-3736){\makebox(0,0)[lb]{\smash{{\SetFigFont{12}{14.4}
{\rmdefault}{\mddefault}{\updefault}{\color[rgb]{0,0,0}$l \vee \bot$}%
}}}}
\end{picture}%
\nop{
   |   |   |               |         |         |   |
  -+- -+- -+- D v l ----- -+- E v l -+- F v l -+- -+- false v l
}
\end{center}

Since this resolution does not resolve on $\neg l$, this clause can be removed
from the formula $G \cup \neg C$. The result is a resolution from
{} $G \cup \neg C \backslash \{\neg l\}$ to $\bot \vee l$.
It can be iterated over all literals of $\neg C$. For each, either the
resolution does not involve $\neg l$, or can be removed $\neg l$ at the cost of
adding $l$ to the final clause. The result is a derivation from $G$ to a subset
of $C$.

\begin{center}
\setlength{\unitlength}{3947sp}%
\begingroup\makeatletter\ifx\SetFigFont\undefined%
\gdef\SetFigFont#1#2#3#4#5{%
  \reset@font\fontsize{#1}{#2pt}%
  \fontfamily{#3}\fontseries{#4}\fontshape{#5}%
  \selectfont}%
\fi\endgroup%
\begin{picture}(7080,688)(2761,-3809)
\thinlines
{\color[rgb]{0,0,0}\put(3901,-3661){\line( 1, 0){450}}
}%
{\color[rgb]{0,0,0}\put(4126,-3661){\line( 0, 1){300}}
}%
{\color[rgb]{0,0,0}\put(5926,-3661){\line( 0, 1){300}}
}%
{\color[rgb]{0,0,0}\put(5701,-3661){\line( 1, 0){450}}
}%
{\color[rgb]{0,0,0}\put(6601,-3661){\line( 1, 0){450}}
}%
{\color[rgb]{0,0,0}\put(6826,-3661){\line( 0, 1){300}}
}%
{\color[rgb]{0,0,0}\put(7501,-3661){\line( 1, 0){450}}
}%
{\color[rgb]{0,0,0}\put(7726,-3661){\line( 0, 1){300}}
}%
{\color[rgb]{0,0,0}\put(3001,-3661){\line( 1, 0){450}}
}%
{\color[rgb]{0,0,0}\put(3226,-3661){\line( 0, 1){300}}
}%
{\color[rgb]{0,0,0}\put(8401,-3661){\line( 1, 0){450}}
}%
{\color[rgb]{0,0,0}\put(8626,-3661){\line( 0, 1){300}}
}%
{\color[rgb]{0,0,0}\put(9301,-3661){\line( 1, 0){450}}
}%
{\color[rgb]{0,0,0}\put(9526,-3661){\line( 0, 1){300}}
}%
\put(4576,-3736){\makebox(0,0)[b]{\smash{{\SetFigFont{12}{14.4}
{\rmdefault}{\mddefault}{\updefault}{\color[rgb]{0,0,0}$D \vee l$}%
}}}}
\put(2776,-3736){\makebox(0,0)[b]{\smash{{\SetFigFont{12}{14.4}
{\rmdefault}{\mddefault}{\updefault}{\color[rgb]{0,0,0}$?$}%
}}}}
\put(3676,-3736){\makebox(0,0)[b]{\smash{{\SetFigFont{12}{14.4}
{\rmdefault}{\mddefault}{\updefault}{\color[rgb]{0,0,0}$?$}%
}}}}
\put(3226,-3286){\makebox(0,0)[b]{\smash{{\SetFigFont{12}{14.4}
{\rmdefault}{\mddefault}{\updefault}{\color[rgb]{0,0,0}$?$}%
}}}}
\put(4126,-3286){\makebox(0,0)[b]{\smash{{\SetFigFont{12}{14.4}
{\rmdefault}{\mddefault}{\updefault}{\color[rgb]{0,0,0}$?$}%
}}}}
\put(5926,-3286){\makebox(0,0)[b]{\smash{{\SetFigFont{12}{14.4}
{\rmdefault}{\mddefault}{\updefault}{\color[rgb]{0,0,0}$?$}%
}}}}
\put(6826,-3286){\makebox(0,0)[b]{\smash{{\SetFigFont{12}{14.4}
{\rmdefault}{\mddefault}{\updefault}{\color[rgb]{0,0,0}$?$}%
}}}}
\put(7726,-3286){\makebox(0,0)[b]{\smash{{\SetFigFont{12}{14.4}
{\rmdefault}{\mddefault}{\updefault}{\color[rgb]{0,0,0}$?$}%
}}}}
\put(8176,-3736){\makebox(0,0)[b]{\smash{{\SetFigFont{12}{14.4}
{\rmdefault}{\mddefault}{\updefault}{\color[rgb]{0,0,0}$?$}%
}}}}
\put(8626,-3286){\makebox(0,0)[b]{\smash{{\SetFigFont{12}{14.4}
{\rmdefault}{\mddefault}{\updefault}{\color[rgb]{0,0,0}$?$}%
}}}}
\put(9076,-3736){\makebox(0,0)[b]{\smash{{\SetFigFont{12}{14.4}
{\rmdefault}{\mddefault}{\updefault}{\color[rgb]{0,0,0}$?$}%
}}}}
\put(9526,-3286){\makebox(0,0)[b]{\smash{{\SetFigFont{12}{14.4}
{\rmdefault}{\mddefault}{\updefault}{\color[rgb]{0,0,0}$?$}%
}}}}
{\color[rgb]{0,0,0}\put(4801,-3661){\line( 1, 0){900}}
}%
\put(6376,-3736){\makebox(0,0)[b]{\smash{{\SetFigFont{12}{14.4}
{\rmdefault}{\mddefault}{\updefault}{\color[rgb]{0,0,0}$E \vee l$}%
}}}}
\put(7276,-3736){\makebox(0,0)[b]{\smash{{\SetFigFont{12}{14.4}
{\rmdefault}{\mddefault}{\updefault}{\color[rgb]{0,0,0}$F \vee l$}%
}}}}
\put(9826,-3736){\makebox(0,0)[lb]{\smash{{\SetFigFont{12}{14.4}
{\rmdefault}{\mddefault}{\updefault}{\color[rgb]{0,0,0}$C' \subseteq C$}%
}}}}
\end{picture}%
\nop{
   |   |   |   |   |   |   |
  -+- -+- -+- -+- -+- -+- -+- C' c C
}
\end{center}

Since all resolutions on a clause of $C$ have been removed and $C$ contains all
variables to remember, this derivation only resolves on the variables to
forget. It satisfies the first condition: only resolve on variables to forget.

The second condition is that resolution stops on clauses comprising variables
to remember. The contrary is that a center clause $D$ contains only variables
to remember but is not the final one in the derivation. It resolves with a side
clause $E$.

\begin{center}
\setlength{\unitlength}{3947sp}%
\begingroup\makeatletter\ifx\SetFigFont\undefined%
\gdef\SetFigFont#1#2#3#4#5{%
  \reset@font\fontsize{#1}{#2pt}%
  \fontfamily{#3}\fontseries{#4}\fontshape{#5}%
  \selectfont}%
\fi\endgroup%
\begin{picture}(7080,630)(2311,-3751)
\thinlines
{\color[rgb]{0,0,0}\put(5476,-3661){\line( 0, 1){300}}
}%
{\color[rgb]{0,0,0}\put(5251,-3661){\line( 1, 0){450}}
}%
{\color[rgb]{0,0,0}\put(6151,-3661){\line( 1, 0){450}}
}%
{\color[rgb]{0,0,0}\put(6376,-3661){\line( 0, 1){300}}
}%
{\color[rgb]{0,0,0}\put(7051,-3661){\line( 1, 0){450}}
}%
{\color[rgb]{0,0,0}\put(7276,-3661){\line( 0, 1){300}}
}%
{\color[rgb]{0,0,0}\put(7951,-3661){\line( 1, 0){450}}
}%
{\color[rgb]{0,0,0}\put(8176,-3661){\line( 0, 1){300}}
}%
{\color[rgb]{0,0,0}\put(8851,-3661){\line( 1, 0){450}}
}%
{\color[rgb]{0,0,0}\put(9076,-3661){\line( 0, 1){300}}
}%
{\color[rgb]{0,0,0}\put(4351,-3661){\line( 1, 0){450}}
}%
{\color[rgb]{0,0,0}\put(4576,-3361){\line( 0,-1){300}}
}%
{\color[rgb]{0,0,0}\put(2551,-3661){\line( 1, 0){450}}
}%
{\color[rgb]{0,0,0}\put(2776,-3661){\line( 0, 1){300}}
}%
{\color[rgb]{0,0,0}\put(3451,-3661){\line( 1, 0){450}}
}%
{\color[rgb]{0,0,0}\put(3676,-3661){\line( 0, 1){300}}
}%
\put(9376,-3736){\makebox(0,0)[lb]{\smash{{\SetFigFont{12}{14.4}
{\rmdefault}{\mddefault}{\updefault}{\color[rgb]{0,0,0}$\bot$}%
}}}}
\put(5026,-3736){\makebox(0,0)[b]{\smash{{\SetFigFont{12}{14.4}
{\rmdefault}{\mddefault}{\updefault}{\color[rgb]{0,0,0}$D$}%
}}}}
\put(5476,-3286){\makebox(0,0)[b]{\smash{{\SetFigFont{12}{14.4}
{\rmdefault}{\mddefault}{\updefault}{\color[rgb]{0,0,0}$E$}%
}}}}
\put(2326,-3736){\makebox(0,0)[b]{\smash{{\SetFigFont{12}{14.4}
{\rmdefault}{\mddefault}{\updefault}{\color[rgb]{0,0,0}$?$}%
}}}}
\put(3226,-3736){\makebox(0,0)[b]{\smash{{\SetFigFont{12}{14.4}
{\rmdefault}{\mddefault}{\updefault}{\color[rgb]{0,0,0}$?$}%
}}}}
\put(4126,-3736){\makebox(0,0)[b]{\smash{{\SetFigFont{12}{14.4}
{\rmdefault}{\mddefault}{\updefault}{\color[rgb]{0,0,0}$?$}%
}}}}
\put(5926,-3736){\makebox(0,0)[b]{\smash{{\SetFigFont{12}{14.4}
{\rmdefault}{\mddefault}{\updefault}{\color[rgb]{0,0,0}$?$}%
}}}}
\put(2776,-3286){\makebox(0,0)[b]{\smash{{\SetFigFont{12}{14.4}
{\rmdefault}{\mddefault}{\updefault}{\color[rgb]{0,0,0}$?$}%
}}}}
\put(3676,-3286){\makebox(0,0)[b]{\smash{{\SetFigFont{12}{14.4}
{\rmdefault}{\mddefault}{\updefault}{\color[rgb]{0,0,0}$?$}%
}}}}
\put(4576,-3286){\makebox(0,0)[b]{\smash{{\SetFigFont{12}{14.4}
{\rmdefault}{\mddefault}{\updefault}{\color[rgb]{0,0,0}$?$}%
}}}}
\put(6376,-3286){\makebox(0,0)[b]{\smash{{\SetFigFont{12}{14.4}
{\rmdefault}{\mddefault}{\updefault}{\color[rgb]{0,0,0}$?$}%
}}}}
\put(6826,-3736){\makebox(0,0)[b]{\smash{{\SetFigFont{12}{14.4}
{\rmdefault}{\mddefault}{\updefault}{\color[rgb]{0,0,0}$?$}%
}}}}
\put(7276,-3286){\makebox(0,0)[b]{\smash{{\SetFigFont{12}{14.4}
{\rmdefault}{\mddefault}{\updefault}{\color[rgb]{0,0,0}$?$}%
}}}}
\put(7726,-3736){\makebox(0,0)[b]{\smash{{\SetFigFont{12}{14.4}
{\rmdefault}{\mddefault}{\updefault}{\color[rgb]{0,0,0}$?$}%
}}}}
\put(8176,-3286){\makebox(0,0)[b]{\smash{{\SetFigFont{12}{14.4}
{\rmdefault}{\mddefault}{\updefault}{\color[rgb]{0,0,0}$?$}%
}}}}
\put(8626,-3736){\makebox(0,0)[b]{\smash{{\SetFigFont{12}{14.4}
{\rmdefault}{\mddefault}{\updefault}{\color[rgb]{0,0,0}$?$}%
}}}}
\put(9076,-3286){\makebox(0,0)[b]{\smash{{\SetFigFont{12}{14.4}
{\rmdefault}{\mddefault}{\updefault}{\color[rgb]{0,0,0}$?$}%
}}}}
\end{picture}%
\nop{
                     E
   |   |   |   |     |   |   |   |
  -+- -+- -+- -+- D -+- -+- -+- -+- C' c C
}
\end{center}

Since $D$ only contains variables to remember, the resolution with $E$ is on
one of them. But all resolutions on variables to remember have been removed
from the construction. This is a contradiction.

This proves that if $C$ is an entailed clause comprising all variables to
remember, a subset $C'$ of its is at the end of a derivation that only resolves
on variables to forget and stops at the first clause comprising variables to
remember.

\

\proofonly{

\separator

{\bf A-ordering resolution for forgetting.}

\

\

}

The center clause of a linear resolution may be resolved with an arbitrary
clause of the formula or any previous clause in the line. Each choice is a new
derivation to explore. Efficiency increases by reducing their number while
still maintaining refutation completeness. This is achieved by either
restricting the clauses in the line or the resolving literal. The first
possibility is for example exploited by s-linear resolution and merge
resolution, the second by A-ordering resolution or the various kinds of
SL-resolution.

An A-ordering is just an arbitrary linear order of the variables. A center
clause is always resolved on its maximal variable.

Variables are forgotten by resolving only on them and stopping when the center
clause no longer contains any. The ordering compares only these variables,
since the others are not resolved anyway.

If $F$ entails a full remembrance clause $C$, then $F \cup \neg C$ entails a
contradiction. Since A-ordering resolution is refutationally complete,
contradiction follows by linear resolution from $F \cup \neg C$ using an
arbitrary order, including the order that is the same on the variables to
remember and compares the others arbitrarily.

As for general linear resolution, the clauses that contain the literals of
$\neg C$ are removed from $F$. The resolutions on these literals are removed
from the derivation. The result is a linear resolution that follows the given
order and ends with a clause that is contained in $C$.

This proves that A-ordering resolution with an arbitrary order of the variables
to forget generates enough clauses to express forgetting.

\proofonly{

\separator

{\bf The algorithm.}

\

}

The complete algorithm for forgetting by A-ordering linear resolution follows.

\begin{algorithm}[Forget by A-Ordering Linear Resolution]
\label{algorithm-linear}

\

\noindent forget\_linear(Clause $C$, Formula $S$, Variables $V$)

\begin{enumerate}

\item $X = \var(C) \cap V$

\item if $X = \emptyset$

\begin{enumerate}

\item return $\{C\}$

\end{enumerate}

\item $x = \max(X)$

\item let $l \in C$ be such that $\var(l) = x$

\item $R = \emptyset$

\item for each $D \in S$ such that $\neg l \in D$:

\begin{enumerate}

\item $E = C * D$

\item $R = R \cup \mbox{forget\_linear}(E, S, V)$

\end{enumerate}

\item return $R$

\end{enumerate}

\

\noindent forget\_linear(Formula $F$, Variables $V$)

\begin{enumerate}

\item $R = \emptyset$

\item for each $C \in F$:

\begin{enumerate}

\item $R = R \cup \mbox{forget\_linear}(C, F, V)$

\end{enumerate}

\item return $R$

\end{enumerate}

\end{algorithm}

\section{Backtracking}
\label{section-backtracking}

The satisfiability of a propositional formula can be established by
backtracking: each variable is first set to true and then to false; for each
value, a recursive call establishes whether the formula is consistent with that
value.

The parameters of the recursive calls are the formula and the current values of
some variables. The base case is when the formula evaluates to true or false,
even if some variables are currently not assigned.

What matters for forgetting are the points where the formula is false. The two
recursive subcalls are performed with two incompatible partial models, one
setting a variable to true and one to false. Because of this, every satisfying
partial assignment is incompatible with each other, and they cover all
interpretations of the variables.

A partial assignment $I$ contradicts a formula $F$ if and only if
{} $F \cup \{l \mid l \in I\} \models \bot$.
This condition is the same as
{} $F \models \neg (\wedge \{l \mid l \in I\})$.
A further reformulation is
{} $F \models \vee \{\neg l \mid l \in I\}$.
The consequent is denoted $\neg I$, the clause comprising the negation of all
literals satisfied by $I$. The conclusion is therefore $F \models \neg I$. For
example,
{} $\{x_1=\true, x_2=\false, x_3=\true\}$
contradicting $F$ is the same as
{} $F \models \neg x_1 \vee x_2 \vee \neg x_3$.

The collection of the clauses obtained this way during backtracking is
equivalent to the original formula. Forgetting requires not only equivalence
but also the ab sense of the variables to forget. Backtracking needs to be
adapted to obey this condition.

The simplest solution is to first assign the variables to remember and then the
variables to forget. Clauses are output not in the base case of recursion, but
in the highest point of unsatisfiability in the recursive tree, and only if the
partial assignment does not include any variable to forget.

The following figure shows an example of backtracking when the variables to
remember are $x$ and $y$ and the ones to forget $z$ and $w$. The nodes of the
tree are labeled with the literals that are true according to the partial
assignment. Mark $X$ is for formula false and $=$ for formula true.

\begin{center}
\setlength{\unitlength}{3947sp}%
\begingroup\makeatletter\ifx\SetFigFont\undefined%
\gdef\SetFigFont#1#2#3#4#5{%
  \reset@font\fontsize{#1}{#2pt}%
  \fontfamily{#3}\fontseries{#4}\fontshape{#5}%
  \selectfont}%
\fi\endgroup%
\begin{picture}(2602,2962)(3725,-5548)
{\color[rgb]{0,0,0}\thinlines
\put(4651,-3511){\circle{336}}
}%
{\color[rgb]{0,0,0}\put(6151,-3511){\circle{336}}
}%
{\color[rgb]{0,0,0}\put(3901,-4261){\circle{336}}
}%
{\color[rgb]{0,0,0}\put(5401,-4261){\circle{336}}
}%
{\color[rgb]{0,0,0}\put(6151,-5011){\circle{336}}
}%
{\color[rgb]{0,0,0}\put(4651,-5011){\circle{336}}
}%
{\color[rgb]{0,0,0}\put(5401,-2761){\circle{336}}
}%
{\color[rgb]{0,0,0}\put(5326,-2911){\line(-1,-1){525}}
}%
{\color[rgb]{0,0,0}\put(5476,-2911){\line( 1,-1){525}}
}%
{\color[rgb]{0,0,0}\put(4726,-3661){\line( 1,-1){525}}
}%
{\color[rgb]{0,0,0}\put(4576,-3661){\line(-1,-1){525}}
}%
{\color[rgb]{0,0,0}\put(5326,-4411){\line(-1,-1){525}}
}%
{\color[rgb]{0,0,0}\put(5476,-4411){\line( 1,-1){525}}
}%
{\color[rgb]{0,0,0}\put(3751,-4486){\line( 1,-1){300}}
}%
{\color[rgb]{0,0,0}\put(3751,-4786){\line( 1, 1){300}}
}%
{\color[rgb]{0,0,0}\put(6001,-5236){\line( 1,-1){300}}
}%
{\color[rgb]{0,0,0}\put(6001,-5536){\line( 1, 1){300}}
}%
{\color[rgb]{0,0,0}\put(4501,-5311){\line( 1, 0){300}}
}%
{\color[rgb]{0,0,0}\put(4501,-5461){\line( 1, 0){300}}
}%
\put(4651,-3511){\makebox(0,0)[b]{\smash{{\SetFigFont{12}{14.4}
{\rmdefault}{\mddefault}{\updefault}{\color[rgb]{0,0,0}$y$}%
}}}}
\put(5401,-2761){\makebox(0,0)[b]{\smash{{\SetFigFont{12}{14.4}
{\rmdefault}{\mddefault}{\updefault}{\color[rgb]{0,0,0}$x$}%
}}}}
\put(4951,-3211){\makebox(0,0)[b]{\smash{{\SetFigFont{12}{14.4}
{\rmdefault}{\mddefault}{\updefault}{\color[rgb]{0,0,0}$T$}%
}}}}
\put(5851,-3211){\makebox(0,0)[b]{\smash{{\SetFigFont{12}{14.4}
{\rmdefault}{\mddefault}{\updefault}{\color[rgb]{0,0,0}$F$}%
}}}}
\put(4276,-3886){\makebox(0,0)[b]{\smash{{\SetFigFont{12}{14.4}
{\rmdefault}{\mddefault}{\updefault}{\color[rgb]{0,0,0}$T$}%
}}}}
\put(5101,-3886){\makebox(0,0)[b]{\smash{{\SetFigFont{12}{14.4}
{\rmdefault}{\mddefault}{\updefault}{\color[rgb]{0,0,0}$F$}%
}}}}
\put(5401,-4261){\makebox(0,0)[b]{\smash{{\SetFigFont{12}{14.4}
{\rmdefault}{\mddefault}{\updefault}{\color[rgb]{0,0,0}$z$}%
}}}}
\put(4951,-4636){\makebox(0,0)[b]{\smash{{\SetFigFont{12}{14.4}
{\rmdefault}{\mddefault}{\updefault}{\color[rgb]{0,0,0}$T$}%
}}}}
\put(5851,-4636){\makebox(0,0)[b]{\smash{{\SetFigFont{12}{14.4}
{\rmdefault}{\mddefault}{\updefault}{\color[rgb]{0,0,0}$F$}%
}}}}
\put(4501,-5161){\makebox(0,0)[rb]{\smash{{\SetFigFont{12}{14.4}
{\rmdefault}{\mddefault}{\updefault}{\color[rgb]{0,0,0}$\{x,\neg y,z\}$}%
}}}}
\put(6301,-5161){\makebox(0,0)[lb]{\smash{{\SetFigFont{12}{14.4}
{\rmdefault}{\mddefault}{\updefault}{\color[rgb]{0,0,0}$\{x,\neg y,\neg z\}$}%
}}}}
\put(5551,-4186){\makebox(0,0)[lb]{\smash{{\SetFigFont{12}{14.4}
{\rmdefault}{\mddefault}{\updefault}{\color[rgb]{0,0,0}$\{x,\neg y\}$}%
}}}}
\put(3751,-4186){\makebox(0,0)[rb]{\smash{{\SetFigFont{12}{14.4}
{\rmdefault}{\mddefault}{\updefault}{\color[rgb]{0,0,0}$\{x,y\}$}%
}}}}
\put(4801,-3661){\makebox(0,0)[lb]{\smash{{\SetFigFont{12}{14.4}
{\rmdefault}{\mddefault}{\updefault}{\color[rgb]{0,0,0}$\{x\}$}%
}}}}
\put(5626,-2836){\makebox(0,0)[lb]{\smash{{\SetFigFont{12}{14.4}
{\rmdefault}{\mddefault}{\updefault}{\color[rgb]{0,0,0}$0$}%
}}}}
\end{picture}%
\nop{
                           0
           +---------------+------------+
           x                           -x
   +-------+-------+
  xy              x-y
   X        +------+-----+
          x-yz         x-y-z
            =            X
}
\end{center}

The literals $x,\neg y,\neg z${\plural} correspond to the clause $\neg x \vee y
\vee z$, which is not generated since it contains $z$. The formula being true
in its sibling $x,\neg y,z$ makes it true in the parent $x,\neg y$. Therefore,
the formula is false in $x,y$ and true in its sibling $x,\neg y$. This is the
highest point of unsatisfaction. A clause is generated. It is $\neg x \vee \neg
y$, the negation of $x,y$.

This example shows the main points of the algorithm:

\begin{itemize}

\item first assign the variables to remember, then the variables to forget;

\item a clause is generated when the formula is false in a node and true in the
sibling; in the example: false in $x,y$, true in $x,\neg y$;

\item it is generated only if the partial model assigns no variable to forget;
in the example: it is not generated in $x, \neg y, z$, although it satisfies
the formula and its sibling does not;

\item the clause comprises the literals the partial assignment in the node
falsifies: $x,y$ produces $\neg x \vee \neg y$.

\end{itemize}

The second and last point alone generate implicates of a
formula~\cite{cast-96,schr-96}. Forgetting requires the addition of the first
and third point: assign variables in a certain order and do not generate some
clauses.

\

\proofonly{

actually, the two above article generate implicants rather than implicates;
each is generated by a node where the formula is true; it is the same as
generating the implicates by changing the signs and working from the nodes
where the formula is false

not implemented: if the first recursive subcall returns true and the branching
variable is already one to be forgotten, there is no need to perform the other
subcall, since the return value will be "sat" anyway and no clause would be
generated anyway

}

The main contributor to efficiency of modern backtracking-based algorithms is
unit propagation. If all literals of a clause are falsified by the current
values but one, that one has to be true for the formula to be satisfied. This
evaluation is added to the others.

The problem with forgetting is when the algorithm is still branching on the
variables to remember and the new value is on a variable to forget.

An alternative view of unit propagation explains why. Setting a variable to a
certain value corresponds to evaluating that variable in the next step; the
wrong value falsifies the formula immediately. An example is forgetting $y$
from a formula that contains $\neg x \vee y$ and the first branching variable
is $x$.

\begin{center}
\setlength{\unitlength}{3947sp}%
\begingroup\makeatletter\ifx\SetFigFont\undefined%
\gdef\SetFigFont#1#2#3#4#5{%
  \reset@font\fontsize{#1}{#2pt}%
  \fontfamily{#3}\fontseries{#4}\fontshape{#5}%
  \selectfont}%
\fi\endgroup%
\begin{picture}(3341,4012)(2986,-6598)
{\color[rgb]{0,0,0}\thinlines
\put(4801,-6061){\circle{336}}
}%
{\color[rgb]{0,0,0}\put(3901,-4261){\circle{336}}
}%
{\color[rgb]{0,0,0}\put(4651,-3511){\circle{336}}
}%
{\color[rgb]{0,0,0}\put(5401,-4261){\circle{336}}
}%
{\color[rgb]{0,0,0}\put(6151,-3511){\circle{336}}
}%
{\color[rgb]{0,0,0}\put(5401,-2761){\circle{336}}
}%
{\color[rgb]{0,0,0}\multiput(3901,-4411)(-4.68750,-7.03125){33}{\makebox(1.6667,11.6667){\tiny.}}
\multiput(3751,-4636)(4.68750,-7.03125){33}{\makebox(1.6667,11.6667){\tiny.}}
\multiput(3901,-4861)(-4.68750,-7.03125){33}{\makebox(1.6667,11.6667){\tiny.}}
\multiput(3751,-5086)(6.00000,-6.00000){26}{\makebox(1.6667,11.6667){\tiny.}}
}%
{\color[rgb]{0,0,0}\put(4576,-3661){\line(-1,-1){525}}
}%
{\color[rgb]{0,0,0}\put(5326,-2911){\line(-1,-1){525}}
}%
{\color[rgb]{0,0,0}\put(4726,-3661){\line( 1,-1){525}}
}%
{\color[rgb]{0,0,0}\put(5476,-2911){\line( 1,-1){525}}
}%
{\color[rgb]{0,0,0}\put(4651,-6361){\line( 1, 0){300}}
}%
{\color[rgb]{0,0,0}\put(4651,-6511){\line( 1, 0){300}}
}%
{\color[rgb]{0,0,0}\put(5251,-4486){\line( 1,-1){300}}
}%
{\color[rgb]{0,0,0}\put(5251,-4786){\line( 1, 1){300}}
}%
\put(4276,-3886){\makebox(0,0)[b]{\smash{{\SetFigFont{12}{14.4}
{\rmdefault}{\mddefault}{\updefault}{\color[rgb]{0,0,0}$T$}%
}}}}
\put(5101,-3886){\makebox(0,0)[b]{\smash{{\SetFigFont{12}{14.4}
{\rmdefault}{\mddefault}{\updefault}{\color[rgb]{0,0,0}$F$}%
}}}}
\put(4651,-3511){\makebox(0,0)[b]{\smash{{\SetFigFont{12}{14.4}
{\rmdefault}{\mddefault}{\updefault}{\color[rgb]{0,0,0}$y$}%
}}}}
\put(5701,-4336){\makebox(0,0)[lb]{\smash{{\SetFigFont{12}{14.4}
{\rmdefault}{\mddefault}{\updefault}{\color[rgb]{0,0,0}$\{x,\neg y\}$}%
}}}}
\put(5851,-3211){\makebox(0,0)[b]{\smash{{\SetFigFont{12}{14.4}
{\rmdefault}{\mddefault}{\updefault}{\color[rgb]{0,0,0}$F$}%
}}}}
\put(4951,-3211){\makebox(0,0)[b]{\smash{{\SetFigFont{12}{14.4}
{\rmdefault}{\mddefault}{\updefault}{\color[rgb]{0,0,0}$T$}%
}}}}
\put(5401,-2761){\makebox(0,0)[b]{\smash{{\SetFigFont{12}{14.4}
{\rmdefault}{\mddefault}{\updefault}{\color[rgb]{0,0,0}$x$}%
}}}}
\put(3676,-4261){\makebox(0,0)[rb]{\smash{{\SetFigFont{12}{14.4}
{\rmdefault}{\mddefault}{\updefault}{\color[rgb]{0,0,0}$\{x,y\}$}%
}}}}
\put(4426,-3511){\makebox(0,0)[rb]{\smash{{\SetFigFont{12}{14.4}
{\rmdefault}{\mddefault}{\updefault}{\color[rgb]{0,0,0}$\{x\}$}%
}}}}
{\color[rgb]{0,0,0}\put(3751,-5086){\line( 1,-1){900}}
}%
\end{picture}%
\nop{
                       0
                 +-----+-----+
                 x           -x
             +---+---+      ...
            xy      x-y 
           ...       X
           ...
            =
}
\end{center}

The branch $x,\neg y$ closes immediately because it falsifies $\neg x \vee y$.
No further branching is needed. Search is only necessary in the other branch
$x,y$. If it ends up falsifying the formula as well, the formula is also false
in the parent $x$. The clause $\neg x$ is generated if it satisfies the
formula. The problem is with the other case, when some descendant of $x,y$
satisfies the formula. The result is that the formula is true in $x,y$ and
false in $x,\neg y$. The latter is a top point of unsatisfaction. The clause
$\neg x \vee y$ would be generated, but does not because it contains $y$, a
variable to forget.

The search in the subtree rooted at $x$ ends, and the algorithm moves to $\neg
x$. All its nodes contain $\neg x$. All clauses that are generated contain its
negation $x$. This excludes a clause like $\neg x \vee z$ where $z$ is a
variable to remember. Even if this exact clause is in the formula, it is not
generated.

Without forgetting, such a clause is output in the subtree when
unsatisfiability is detected in a node and satisfiability in the other.

\begin{center}
\setlength{\unitlength}{3947sp}%
\begingroup\makeatletter\ifx\SetFigFont\undefined%
\gdef\SetFigFont#1#2#3#4#5{%
  \reset@font\fontsize{#1}{#2pt}%
  \fontfamily{#3}\fontseries{#4}\fontshape{#5}%
  \selectfont}%
\fi\endgroup%
\begin{picture}(3341,4012)(2986,-6598)
{\color[rgb]{0,0,0}\thinlines
\put(4801,-6061){\circle{336}}
}%
{\color[rgb]{0,0,0}\put(3901,-4261){\circle{336}}
}%
{\color[rgb]{0,0,0}\put(4651,-3511){\circle{336}}
}%
{\color[rgb]{0,0,0}\put(5401,-4261){\circle{336}}
}%
{\color[rgb]{0,0,0}\put(6151,-3511){\circle{336}}
}%
{\color[rgb]{0,0,0}\put(5401,-2761){\circle{336}}
}%
{\color[rgb]{0,0,0}\multiput(3901,-4411)(-4.68750,-7.03125){33}{\makebox(1.6667,11.6667){\tiny.}}
\multiput(3751,-4636)(4.68750,-7.03125){33}{\makebox(1.6667,11.6667){\tiny.}}
\multiput(3901,-4861)(-4.68750,-7.03125){33}{\makebox(1.6667,11.6667){\tiny.}}
\multiput(3751,-5086)(6.00000,-6.00000){26}{\makebox(1.6667,11.6667){\tiny.}}
}%
{\color[rgb]{0,0,0}\put(4576,-3661){\line(-1,-1){525}}
}%
{\color[rgb]{0,0,0}\put(5326,-2911){\line(-1,-1){525}}
}%
{\color[rgb]{0,0,0}\put(4726,-3661){\line( 1,-1){525}}
}%
{\color[rgb]{0,0,0}\put(5476,-2911){\line( 1,-1){525}}
}%
{\color[rgb]{0,0,0}\put(4651,-6361){\line( 1, 0){300}}
}%
{\color[rgb]{0,0,0}\put(4651,-6511){\line( 1, 0){300}}
}%
{\color[rgb]{0,0,0}\put(5251,-4486){\line( 1,-1){300}}
}%
{\color[rgb]{0,0,0}\put(5251,-4786){\line( 1, 1){300}}
}%
\put(4276,-3886){\makebox(0,0)[b]{\smash{{\SetFigFont{12}{14.4}
{\rmdefault}{\mddefault}{\updefault}{\color[rgb]{0,0,0}$T$}%
}}}}
\put(5101,-3886){\makebox(0,0)[b]{\smash{{\SetFigFont{12}{14.4}
{\rmdefault}{\mddefault}{\updefault}{\color[rgb]{0,0,0}$F$}%
}}}}
\put(4651,-3511){\makebox(0,0)[b]{\smash{{\SetFigFont{12}{14.4}
{\rmdefault}{\mddefault}{\updefault}{\color[rgb]{0,0,0}$y$}%
}}}}
\put(5701,-4336){\makebox(0,0)[lb]{\smash{{\SetFigFont{12}{14.4}
{\rmdefault}{\mddefault}{\updefault}{\color[rgb]{0,0,0}$\{x,\neg y\}$}%
}}}}
\put(5851,-3211){\makebox(0,0)[b]{\smash{{\SetFigFont{12}{14.4}
{\rmdefault}{\mddefault}{\updefault}{\color[rgb]{0,0,0}$F$}%
}}}}
\put(4951,-3211){\makebox(0,0)[b]{\smash{{\SetFigFont{12}{14.4}
{\rmdefault}{\mddefault}{\updefault}{\color[rgb]{0,0,0}$T$}%
}}}}
\put(5401,-2761){\makebox(0,0)[b]{\smash{{\SetFigFont{12}{14.4}
{\rmdefault}{\mddefault}{\updefault}{\color[rgb]{0,0,0}$x$}%
}}}}
\put(3676,-4261){\makebox(0,0)[rb]{\smash{{\SetFigFont{12}{14.4}
{\rmdefault}{\mddefault}{\updefault}{\color[rgb]{0,0,0}$\{x,y\}$}%
}}}}
\put(4426,-3511){\makebox(0,0)[rb]{\smash{{\SetFigFont{12}{14.4}
{\rmdefault}{\mddefault}{\updefault}{\color[rgb]{0,0,0}$\{x\}$}%
}}}}
{\color[rgb]{0,0,0}\put(4051,-5311){\circle{336}}
}%
{\color[rgb]{0,0,0}\put(3301,-6061){\circle{336}}
}%
{\color[rgb]{0,0,0}\put(4126,-5461){\line( 1,-1){525}}
}%
{\color[rgb]{0,0,0}\put(3976,-5461){\line(-1,-1){525}}
}%
{\color[rgb]{0,0,0}\put(3151,-6286){\line( 1,-1){300}}
}%
{\color[rgb]{0,0,0}\put(3151,-6586){\line( 1, 1){300}}
}%
\put(4426,-5686){\makebox(0,0)[b]{\smash{{\SetFigFont{12}{14.4}
{\rmdefault}{\mddefault}{\updefault}{\color[rgb]{0,0,0}$F$}%
}}}}
\put(3676,-5686){\makebox(0,0)[b]{\smash{{\SetFigFont{12}{14.4}
{\rmdefault}{\mddefault}{\updefault}{\color[rgb]{0,0,0}$T$}%
}}}}
\put(4051,-5311){\makebox(0,0)[b]{\smash{{\SetFigFont{12}{14.4}
{\rmdefault}{\mddefault}{\updefault}{\color[rgb]{0,0,0}$z$}%
}}}}
\put(3001,-6136){\makebox(0,0)[rb]{\smash{{\SetFigFont{12}{14.4}
{\rmdefault}{\mddefault}{\updefault}{\color[rgb]{0,0,0}$\{x,y,\ldots,z\}$}%
}}}}
\end{picture}%
\nop{
                       0
                 +-----+-----+
                 x           -x
             +---+---+      ...
            xy      x-y 
           ...       X
           ...
      +-----+-----+
   xy...z      xy...-z
      X           =
}
\end{center}

The clause that would be generated in $x,y,\ldots,z$ is
{} $\neg x \vee \neg y \vee \cdots \vee \neg z$,
but is not generated because it contains $y$, a variable to forget. However,
this variable is not necessary to unsatisfiability. The interpretation $x,
\ldots, z$ still falsifies the formula since the missing variable $y$ is
entailed by $x$ anyway. In terms of clauses,
{} $\neg x \vee \neg y \vee \cdots \vee \neg z$
and
{} $\neg x \vee y$
resolve into
{} $\neg x \vee \cdots \vee \neg z$.
This clause is entailed and does contain the variable to forget $y$. It is a
clause to generate.

The same outcome is obtained graphically by delaying the evaluation of the
variables to forget to the leaves. Instead of branching on $y$ right after $x$,
it is done at the very end.

\begin{center}
\setlength{\unitlength}{3947sp}%
\begingroup\makeatletter\ifx\SetFigFont\undefined%
\gdef\SetFigFont#1#2#3#4#5{%
  \reset@font\fontsize{#1}{#2pt}%
  \fontfamily{#3}\fontseries{#4}\fontshape{#5}%
  \selectfont}%
\fi\endgroup%
\begin{picture}(3652,4762)(2675,-7348)
{\color[rgb]{0,0,0}\thinlines
\put(6151,-3511){\circle{336}}
}%
{\color[rgb]{0,0,0}\put(3901,-4261){\circle{336}}
}%
{\color[rgb]{0,0,0}\put(5401,-2761){\circle{336}}
}%
{\color[rgb]{0,0,0}\put(3301,-6061){\circle{336}}
}%
{\color[rgb]{0,0,0}\put(4801,-6061){\circle{336}}
}%
{\color[rgb]{0,0,0}\put(4051,-5311){\circle{336}}
}%
{\color[rgb]{0,0,0}\put(2851,-6811){\circle{336}}
}%
{\color[rgb]{0,0,0}\put(3751,-6811){\circle{300}}
}%
{\color[rgb]{0,0,0}\put(4351,-6811){\circle{336}}
}%
{\color[rgb]{0,0,0}\put(5251,-6811){\circle{336}}
}%
{\color[rgb]{0,0,0}\put(5476,-2911){\line( 1,-1){525}}
}%
{\color[rgb]{0,0,0}\multiput(3901,-4411)(-4.68750,-7.03125){33}{\makebox(1.6667,11.6667){\tiny.}}
\multiput(3751,-4636)(4.68750,-7.03125){33}{\makebox(1.6667,11.6667){\tiny.}}
\multiput(3901,-4861)(-4.68750,-7.03125){33}{\makebox(1.6667,11.6667){\tiny.}}
\multiput(3751,-5086)(6.00000,-6.00000){26}{\makebox(1.6667,11.6667){\tiny.}}
}%
{\color[rgb]{0,0,0}\put(3976,-5461){\line(-1,-1){525}}
}%
{\color[rgb]{0,0,0}\put(4126,-5461){\line( 1,-1){525}}
}%
{\color[rgb]{0,0,0}\put(5326,-2911){\line(-1,-1){1275}}
}%
{\color[rgb]{0,0,0}\multiput(3226,-6211)(-3.11866,-7.79665){69}{\makebox(1.6667,11.6667){\tiny.}}
}%
{\color[rgb]{0,0,0}\multiput(3376,-6211)(3.11866,-7.79665){69}{\makebox(1.6667,11.6667){\tiny.}}
}%
{\color[rgb]{0,0,0}\multiput(4726,-6211)(-3.11866,-7.79665){69}{\makebox(1.6667,11.6667){\tiny.}}
}%
{\color[rgb]{0,0,0}\multiput(4876,-6211)(3.11866,-7.79665){69}{\makebox(1.6667,11.6667){\tiny.}}
}%
{\color[rgb]{0,0,0}\put(2701,-7036){\line( 1,-1){300}}
}%
{\color[rgb]{0,0,0}\put(2701,-7336){\line( 1, 1){300}}
}%
{\color[rgb]{0,0,0}\put(3601,-7036){\line( 1,-1){300}}
}%
{\color[rgb]{0,0,0}\put(3601,-7336){\line( 1, 1){300}}
}%
{\color[rgb]{0,0,0}\put(4201,-7111){\line( 1, 0){300}}
}%
{\color[rgb]{0,0,0}\put(4201,-7261){\line( 1, 0){300}}
}%
{\color[rgb]{0,0,0}\put(5101,-7036){\line( 1,-1){300}}
}%
{\color[rgb]{0,0,0}\put(5101,-7336){\line( 1, 1){300}}
}%
\put(5401,-2761){\makebox(0,0)[b]{\smash{{\SetFigFont{12}{14.4}
{\rmdefault}{\mddefault}{\updefault}{\color[rgb]{0,0,0}$x$}%
}}}}
\put(4951,-3211){\makebox(0,0)[b]{\smash{{\SetFigFont{12}{14.4}
{\rmdefault}{\mddefault}{\updefault}{\color[rgb]{0,0,0}$T$}%
}}}}
\put(5851,-3211){\makebox(0,0)[b]{\smash{{\SetFigFont{12}{14.4}
{\rmdefault}{\mddefault}{\updefault}{\color[rgb]{0,0,0}$F$}%
}}}}
\put(4051,-5311){\makebox(0,0)[b]{\smash{{\SetFigFont{12}{14.4}
{\rmdefault}{\mddefault}{\updefault}{\color[rgb]{0,0,0}$z$}%
}}}}
\put(3676,-5686){\makebox(0,0)[b]{\smash{{\SetFigFont{12}{14.4}
{\rmdefault}{\mddefault}{\updefault}{\color[rgb]{0,0,0}$T$}%
}}}}
\put(4426,-5686){\makebox(0,0)[b]{\smash{{\SetFigFont{12}{14.4}
{\rmdefault}{\mddefault}{\updefault}{\color[rgb]{0,0,0}$F$}%
}}}}
\put(3301,-6061){\makebox(0,0)[b]{\smash{{\SetFigFont{12}{14.4}
{\rmdefault}{\mddefault}{\updefault}{\color[rgb]{0,0,0}$y$}%
}}}}
\put(3076,-6436){\makebox(0,0)[b]{\smash{{\SetFigFont{12}{14.4}
{\rmdefault}{\mddefault}{\updefault}{\color[rgb]{0,0,0}$T$}%
}}}}
\put(3526,-6436){\makebox(0,0)[b]{\smash{{\SetFigFont{12}{14.4}
{\rmdefault}{\mddefault}{\updefault}{\color[rgb]{0,0,0}$F$}%
}}}}
\put(4576,-6436){\makebox(0,0)[b]{\smash{{\SetFigFont{12}{14.4}
{\rmdefault}{\mddefault}{\updefault}{\color[rgb]{0,0,0}$T$}%
}}}}
\put(5101,-6436){\makebox(0,0)[b]{\smash{{\SetFigFont{12}{14.4}
{\rmdefault}{\mddefault}{\updefault}{\color[rgb]{0,0,0}$F$}%
}}}}
\put(3001,-6136){\makebox(0,0)[rb]{\smash{{\SetFigFont{12}{14.4}
{\rmdefault}{\mddefault}{\updefault}{\color[rgb]{0,0,0}$\{x, \ldots, z\}$}%
}}}}
\put(5551,-6886){\makebox(0,0)[lb]{\smash{{\SetFigFont{12}{14.4}
{\rmdefault}{\mddefault}{\updefault}{\color[rgb]{0,0,0}$\{x, \ldots, \neg z,\neg y\}$}%
}}}}
\end{picture}%
\nop{
                      0
                 +----+----+
                 x        -x
                ...       ...
                ...
        +--------+--------+
      x...z            x...-z
     +--+--+          +---+---+
 xy...z  x-y...z  xy...-z  x-y...-z
     X      X         =       X
}
\end{center}

The highest node where the formula is unsatisfiable and is satisfiable in the
sibling is $x,\ldots,z$. The generated clause is
{} $\neg x \vee \cdots \vee \neg z$.

While delaying branching on $y$ solves the problem in theory, it is
inconvenient in practice, as evaluating $y=\true$ might generate other
variables by unit propagation, with a possibly exponential reduction of the
recursion tree.

Rather, $y$ is generated as usual. When the formula is unsatisfiable in a node
and satisfiable in the other, the value of $y$ is discarded. The generated
clause does not contain this variable.

The treatment is different when $y$ is a variable to remember. Unit propagation
is the same as branching on $y$ right after $x$. This is possible this time
since $y$ is variable to remember. Still better, an algorithm can be
implemented exactly this way: by branching first on the variables to remember
that occurs in unit clauses. This way, no special treatment is required to
ensure that a clause is generated if unsatisfiability results in a child and
satisfiability in the other.

Operatively, when a variable $y$ is in a unit clause:

\begin{itemize}

\item if $y$ is a variable to forget, its value is added to the partial model;

\item otherwise, $y$ (or another variable occurring in a unit clause) is chosen
as the next branching variable;

\item the generated clauses only include the literals on the variables to
remember.

\end{itemize}

\

The complete algorithm follows. Its starting point is the empty partial model,
which assigns no variable. The $\mbox{forget\_backtracking}(I,F,V)$ subroutine
returns either $\true$, $\false$ or a formula. The first two values are only
returned if branching took place on a variable to forget. They tell the formula
$F$ satisfiable or not according to the partial model $I$. The third value is
returned when branching took place on a variable to remember, and is the result
of forgetting $V$ from $F \cup I$.

The propagate$(I,F,V)$ subroutine performs unit propagation on the variables in
$V$ only. It returns three values. The first is either $\true$, $\false$ or
$undef$, depending on whether propagation ended up respectively satisfying all
clauses, falsifying all clauses or neither. The second and the third return
values are respectively the updated model and simplified formula.

\begin{algorithm}[Forget by Backtracking]
implies
\label{algorithm-backtracking}

\

\noindent forget\_backtracking(PartialModel $I$, Formula $F$, Variables $V$)

\begin{enumerate}

\item $R,I,F = \mbox{propagate}(I, F, V)$

\item if $R = \true$ or $R = \false$
\begin{enumerate}
\item return $R$
\end{enumerate}

\item if $\exists x \in \var(F) \backslash V$
such that $\var(C) = \{x\}$ for
some $C \in F$
\begin{enumerate}
\item $B = x$
\end{enumerate}

\item else if $\var(F) \backslash \var(I) \backslash V \not= \emptyset$
\begin{enumerate}
\item $B = $ a variable in $\var(F) \backslash \var(I) \backslash V$
\end{enumerate}

\item else
\begin{enumerate}
\item $B = $ a variable in $\var(F) \backslash \var(I)$
\end{enumerate}

\item $T = \mbox{forget\_backtrack}(I \cup \{B\}, F, V)$

\item $F = \mbox{forget\_backtrack}(I \cup \{\neg B\}, F, V)$

\item if $T = F$
\begin{enumerate}
\item return $T$
\end{enumerate}

\item if $B \in V$

\begin{enumerate}
\item return $\true$
\end{enumerate}

\item if $T$ then
\begin{enumerate}
\item $I = I \cup \{B\}$
\end{enumerate}

\item else
\begin{enumerate}
\item $I = I \cup \{\neg B\}$
\end{enumerate}

\item return $\{\bigvee \{\neg l \mid l \in I ,~ \var(l) \not\in V\}\}$

\end{enumerate}

\

\noindent forget\_backtracking(Formula $F$, Variables $V$)

\begin{enumerate}

\item return forget\_backtracking$(\emptyset, F, V)$

\end{enumerate}

\end{algorithm}

\section{Implementation}
\label{section-python}

The four algorithms are implemented in Python~\cite{vanr-drak-11}. They are
currently available in the repository {\tt
https://github.com/paololiberatore/four}.

Variables are single characters or html-like entities like {\tt \&name;}. A
sequence {\tt abc} stands for the clause $a \vee b \vee c$. The forms {\tt
ab->cd} and {\tt ef=gh} are allowed.

The programs output the result of forgetting, self-reported time and memory and
tracing information. Every output line but the result of forgetting starts with
the character {\tt \#}. The second character in the line is a space for tracing
information. Time and memory are reported as {\tt \#T=number} and {\tt
\#M=number}.

The numbers are not the exact time and memory consumption. They do not include
constant factors and are about the best way to perform all operations rather
than the way they are implemented. For example, if a certain part of the
algorithm could be performed in linear time, the printed time may be the size
of the formula. It neglects all implementation details.

The programs print multiple resource usage information lines {\tt \#T=number}
and {\tt \#M=number}. The first kind is the ideal duration of a certain
operation; total time is their sum. The second kind is the ideal memory usage
at a certain time; the overall memory requirement is their maximum.

\proofonly{

The orderings followed by the algorithms:

- the elimination algorithm extracts the variables to resolve out from a set
  with a for loop; the ordering is therefore not given; in practice, it changes
  every time

- the linear resolution algorithm chooses the variable in the center clause
  according to the alphabetic order; since the algorithm as written resolve on
  the maximal variable, the ordering is the reverse alphabetic ordering

- the backtracking algorithm makes a pool of variables to choose from; when no
  variable to remember is in a unit clause, it extracts one with
  next(iter(pool)); it is therefore an arbitrary variable in the pool; in
  practice, it changes every time

}

\section{Comparison}
\label{section-comparison}

\def\image#1{%
	\includegraphics%
		[width=0.6\textwidth,natwidth=640,natheight=480,scale=0.4]{#1}}

The four algorithms are compared on random formulae of three clauses each. Each
literal in each clause is randomly chosen from a fixed alphabet. A time limit
of 10 seconds is set for each formula. Time and memory as reported by the
programs is stored, as well as actual time in seconds and maximal memory usage
in kilobytes from the operating system. Ten random formulae are generated for
every number of total variables, of variables to forget and clauses.

\proofonly{

The random formula generator generate.py can generate three different kinds of
formulae: random 3-cnf, random 2, 3, and sqrt(n+1) clauses, and random long
negative clauses with implications among them. The third kind is inspired by
the hard formulae for variable elimination according to proof complexity. The
experiments have been performed only on the first kind of formulae.

}

Real time and memory usage tell which programs are better than which, but
neglect unimplemented optimizations. Self-reported time and memory keep them
into account. Being asymptotic, these measures are better suited for
determining how the resources used by a single algorithm change among the
experiments rather than comparing different algorithms on a single experiments.
Real time and memory usage are for comparison between different programs; self
reporting time and memory usage is for comparing each algorithm with itself on
different inputs.

The experiments have been performed on all combinations of three to ten total
variables, zero to all variables to forget, and clauses ranging from one to
five times the total number of variables. All following diagrams are summaries
of the results of these tests.



\begin{figure}
\begin{center}

\ttytex{
\noindent
\begin{tabular}{cc}
\image{plot-variables-secs-close.ps}		&
\image{plot-variables-secs-eliminate.ps}	\\
\image{plot-variables-secs-linear.ps}		&
\image{plot-variables-secs-backtrack.ps}
\end{tabular}
}{
[variables -> secs (real time)]
}

\caption{Real time, depending on the total number of variables}
\label{variables-secs}
\end{center}
\end{figure}

The real time spent by the programs depends on the number of total variables as
shown in Figure~\ref{variables-secs}. Each point is the result of a test,
positioned according to its total number of variables and running time. As
expected, all programs take more on larger formulae. The close and elimination
algorithms show an exponential-like increase, which is not so marked for the
backtracking algorithm. The close algorithm hits the timeout of ten seconds for
some formulae of nine variables, while the elimination and backtracking
algorithms do not. The behavior of the linear resolution algorithm is
confounded by the timeouts reached already for a very small number of
variables. On absolute values, the backtracking algorithm is the clear winner.


\begin{figure}
\begin{center}

\ttytex{
\noindent
\begin{tabular}{cc}
\image{plot-variables-kb-close.ps}		&
\image{plot-variables-kb-eliminate.ps}		\\
\image{plot-variables-kb-linear.ps}		&
\image{plot-variables-kb-backtrack.ps}
\end{tabular}
}{
[variables -> kb (real memory)]
}

\caption{Actual memory used, depending on the total number of variables}
\label{variables-kb}
\end{center}
\end{figure}

Figure~\ref{variables-kb} depicts the real memory the programs use, depending
on the total number of variables. Timeouts are shown as 14000, a value larger
than any other in these experiments. The close algorithm is exponential as long
as it does not time out. The elimination algorithm takes less memory, but the
increase is still exponential. The backtracking algorithm is almost unaffected
by the increase in the number of variables. The linear algorithm is hard to
evaluate due to the timeouts; when it does not time out, it is even better than
backtracking.


\begin{figure}
\begin{center}

\ttytex{
\noindent
\begin{tabular}{cc}
\image{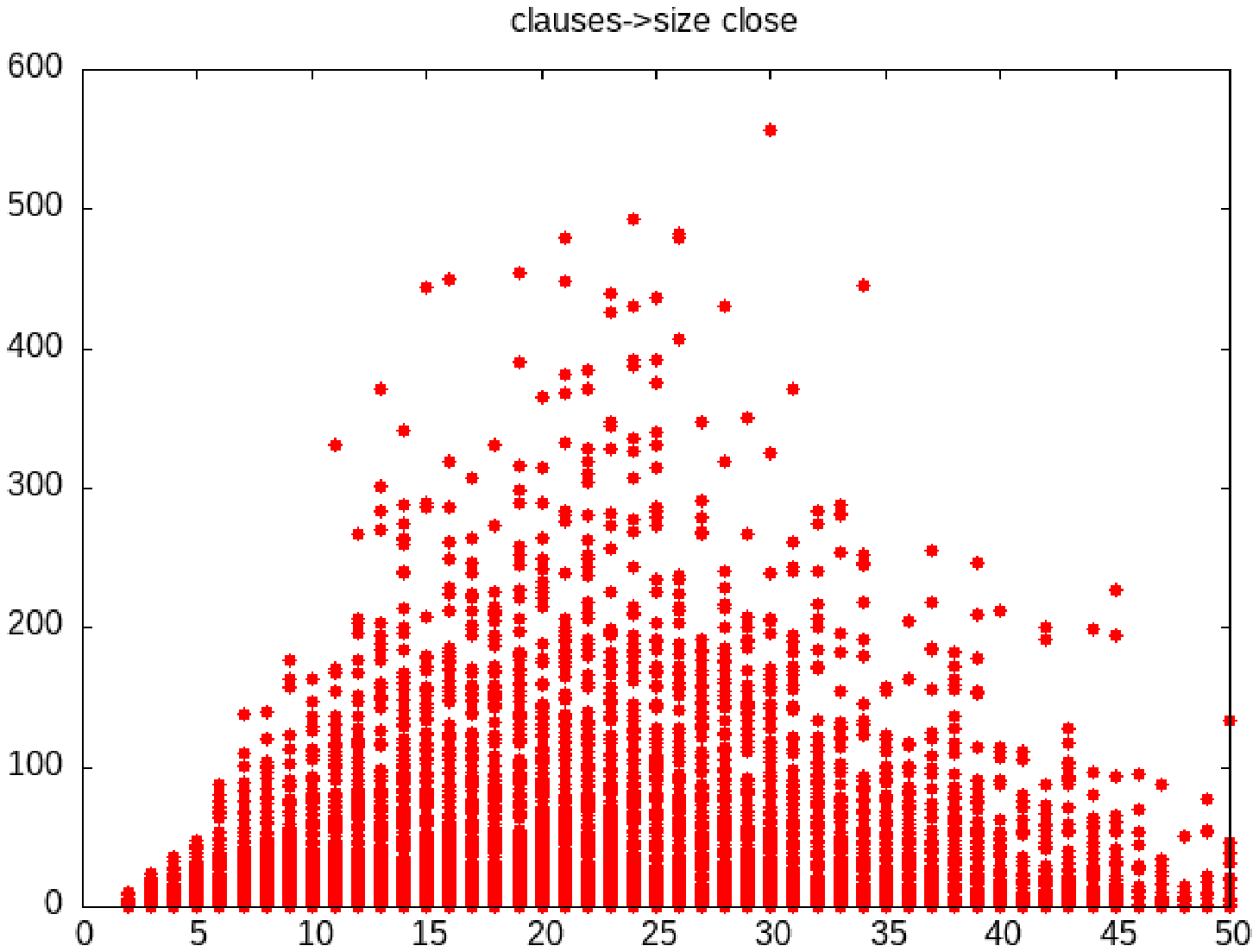}			&
\image{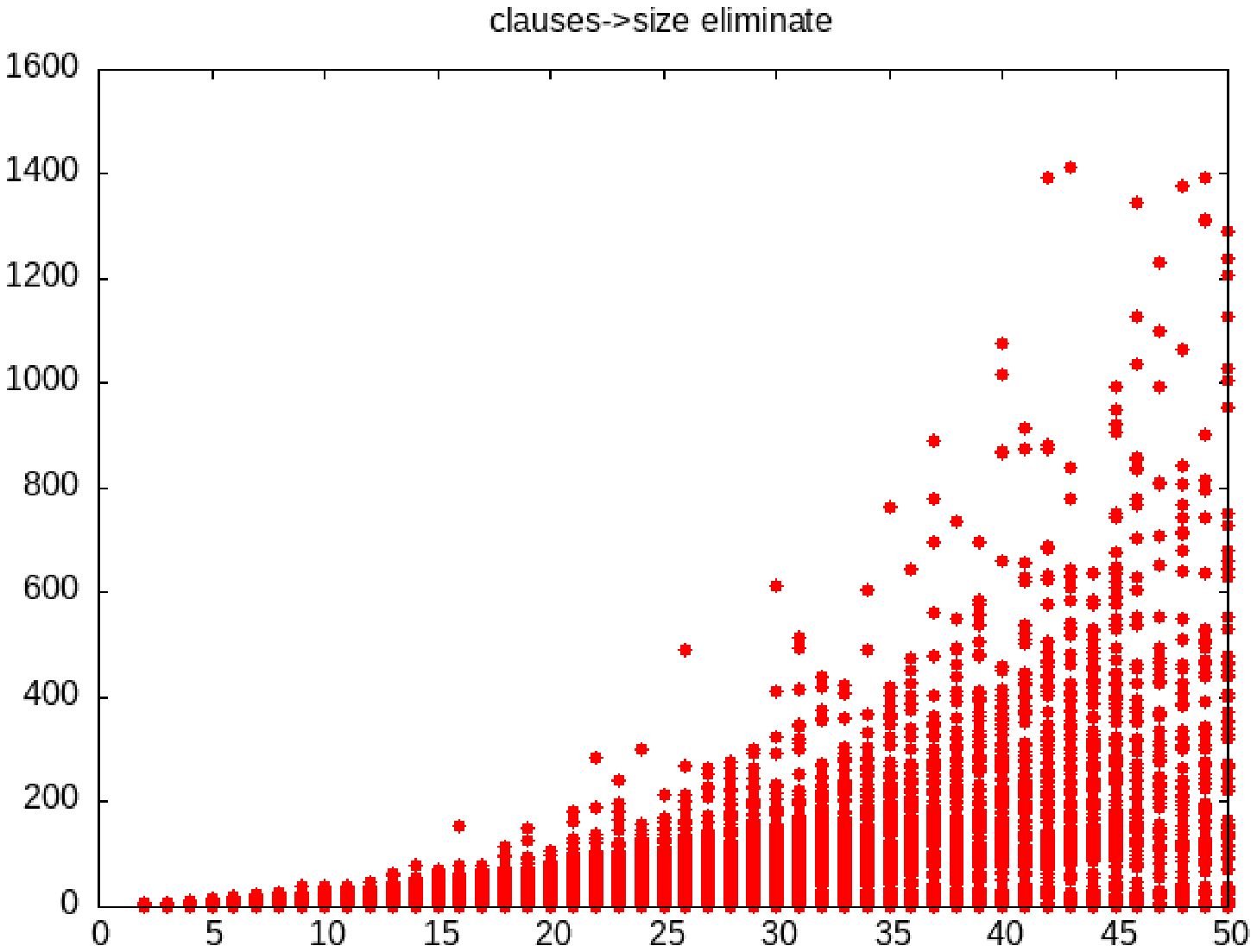}			\\
\image{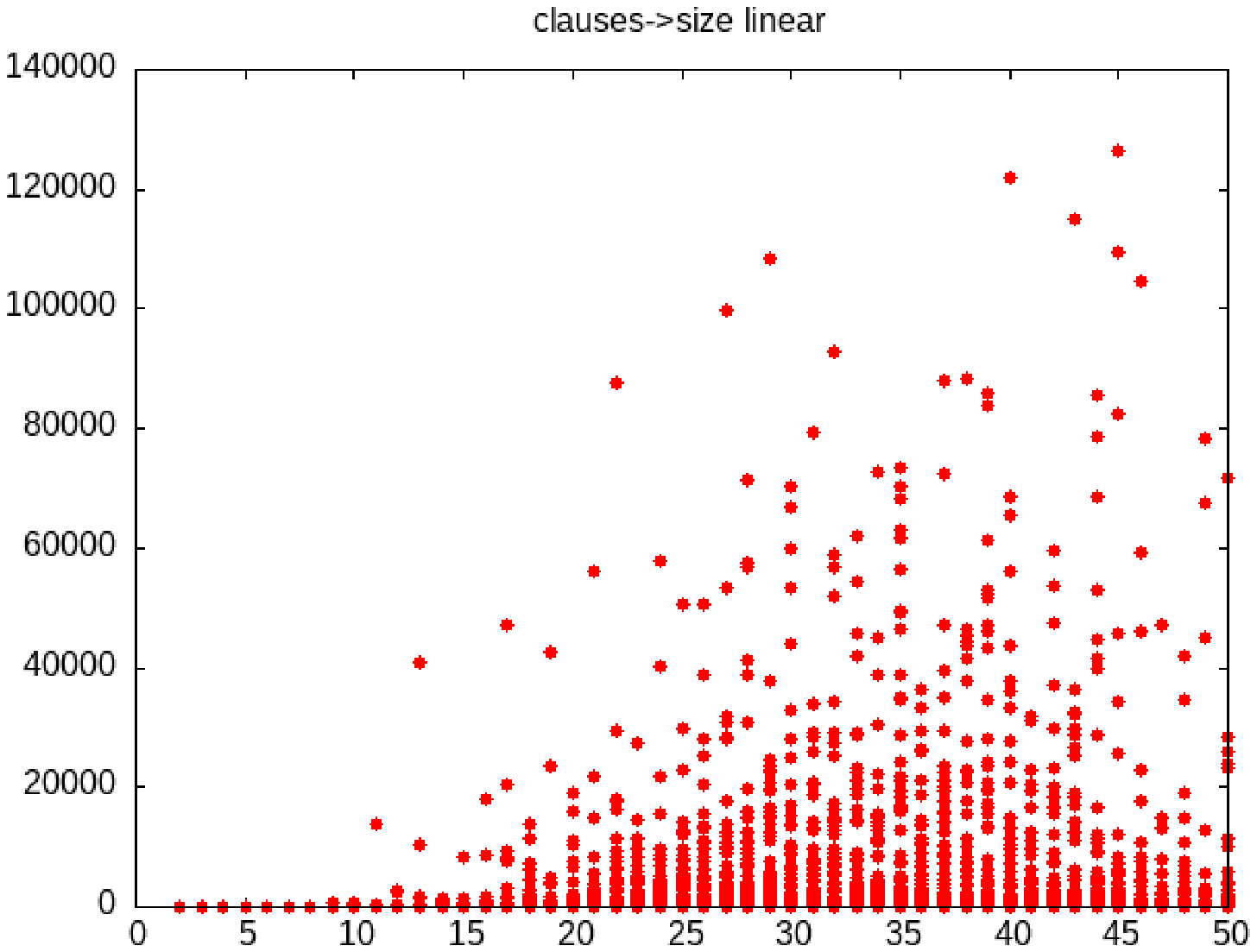}			&
\image{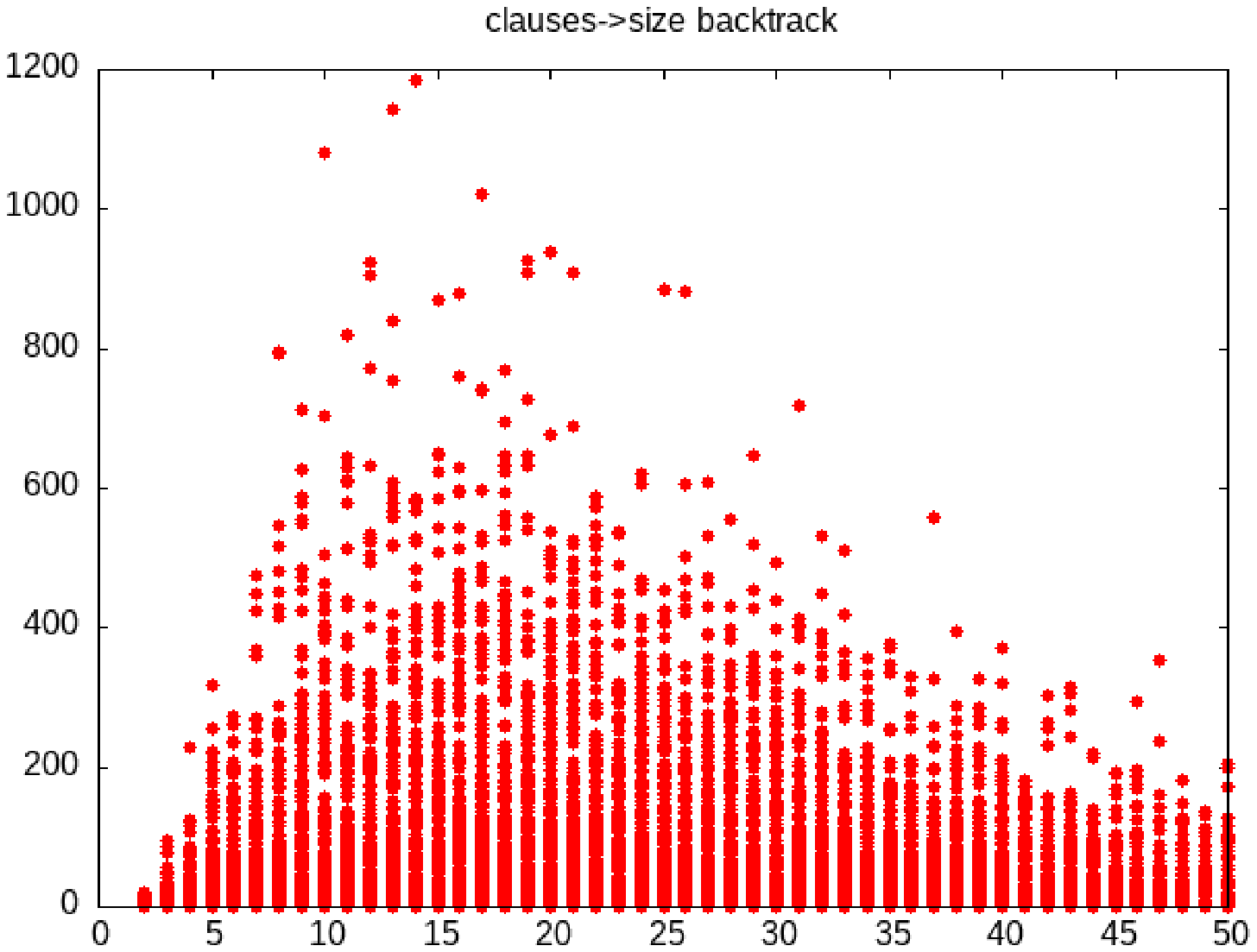}
\end{tabular}
}{
[clauses -> size]
}

\caption{Size of output, depending on the size of input}
\label{clauses-size}
\end{center}
\end{figure}

Figure~\ref{clauses-size} shows the size the output produced by the algorithms
in relation to the number of clauses of the input. They all follow a bell
curve, but the shape and the position of their peak depends on the algorithm.
Since the output is a simplification of the input, few input clauses are
expected to produce few output clauses. The following decrease is not so
obvious. The limit case shows an explanation: the formula that contains all
possible clauses is inconsistent, and all inconsistent formulae are equivalent
to very short ones, like $\{x,\neg x\}$. When the formula is not inconsistent
but contains many clauses, an algorithm may be able to combine many of them to
form shorter clauses, such as when $a \vee b \vee c$ and $a \vee b \vee \neg c$
produce $a \vee b$ by resolution. The ability to do so depends on the
algorithm, which explains the differences in the bell curves. The size of the
output is not plotted if the algorithm timeouts since the output would be
incomplete anyway.


\begin{figure}
\begin{center}

\ttytex{
\noindent
\begin{tabular}{cc}
\image{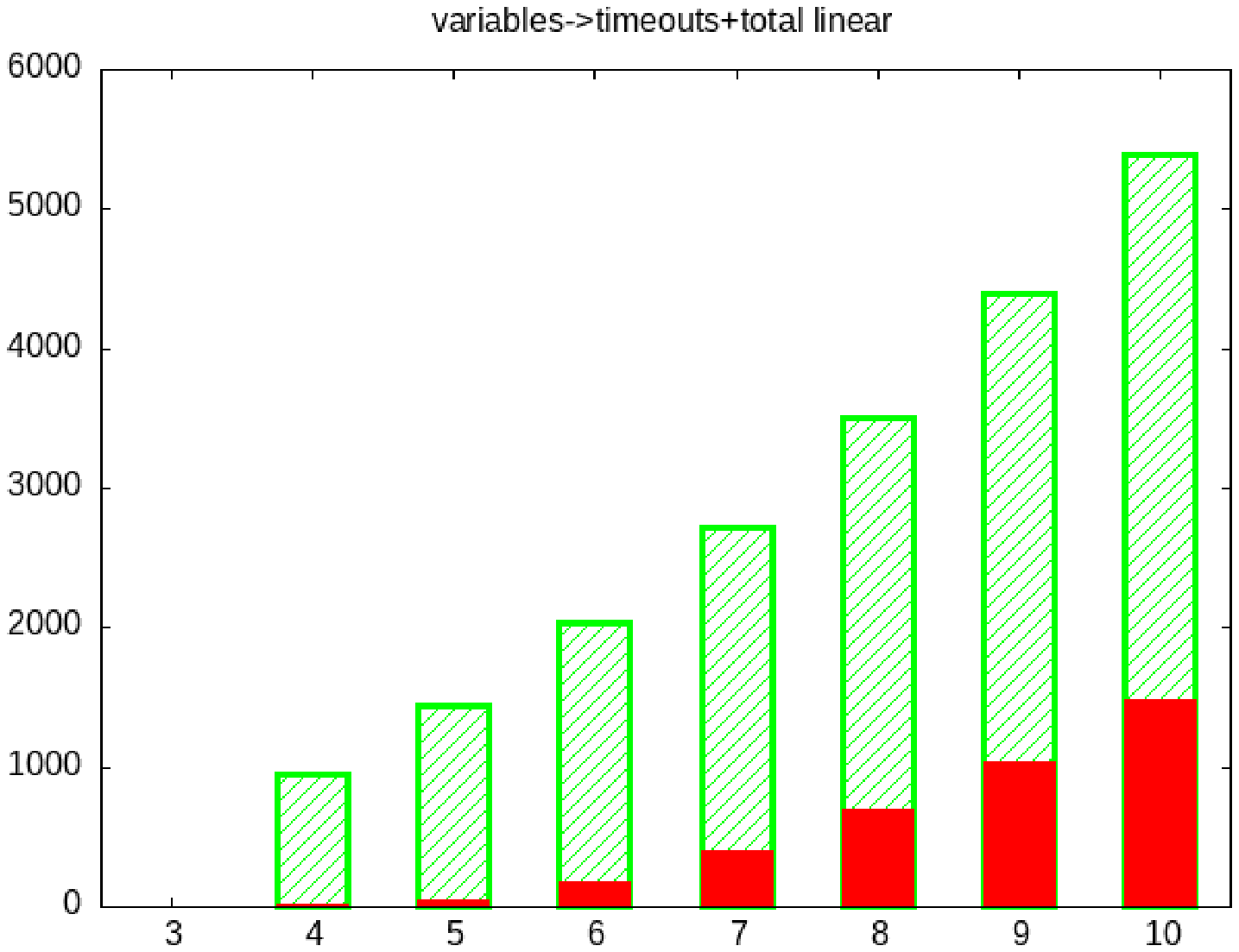}		&
\image{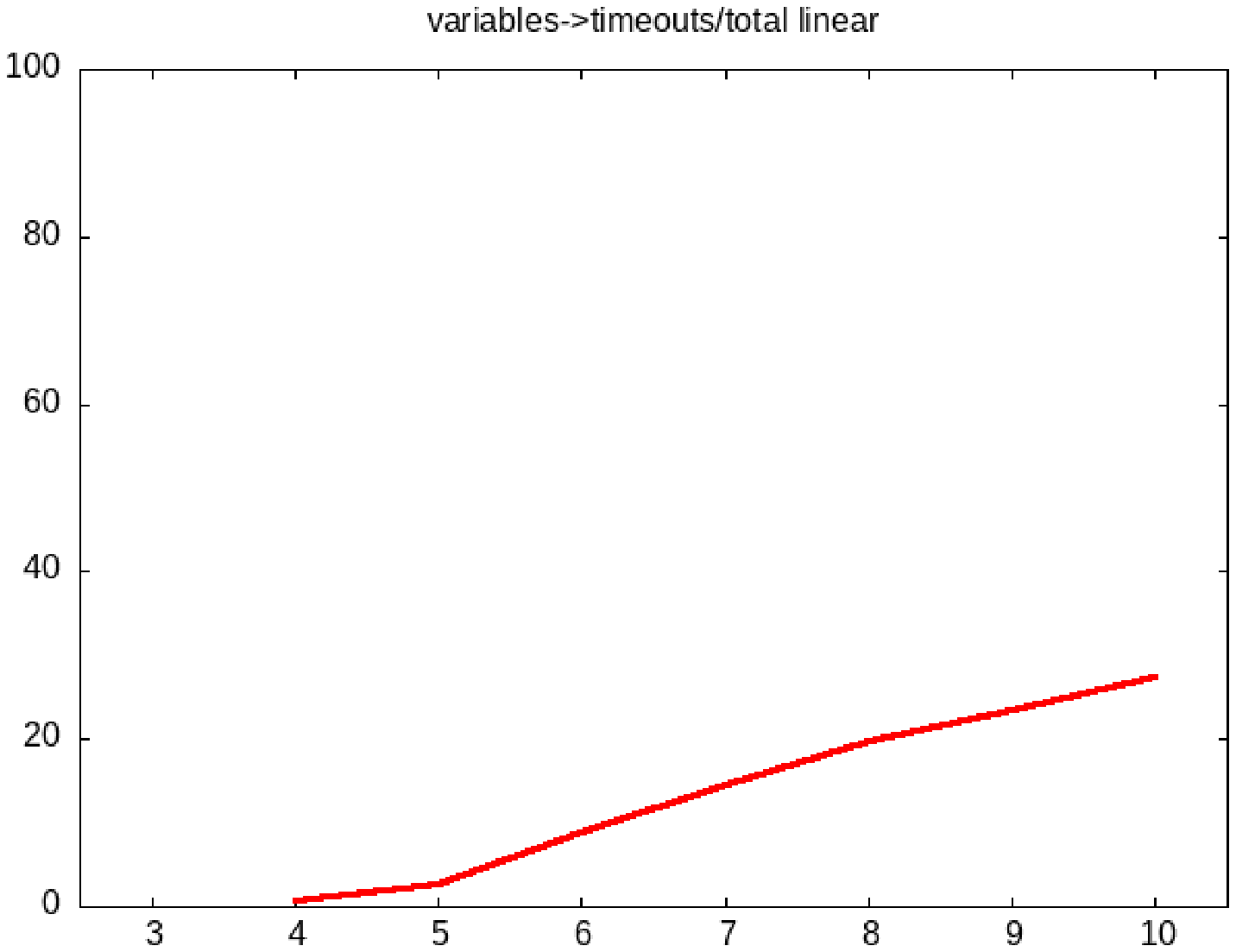}		\\
\image{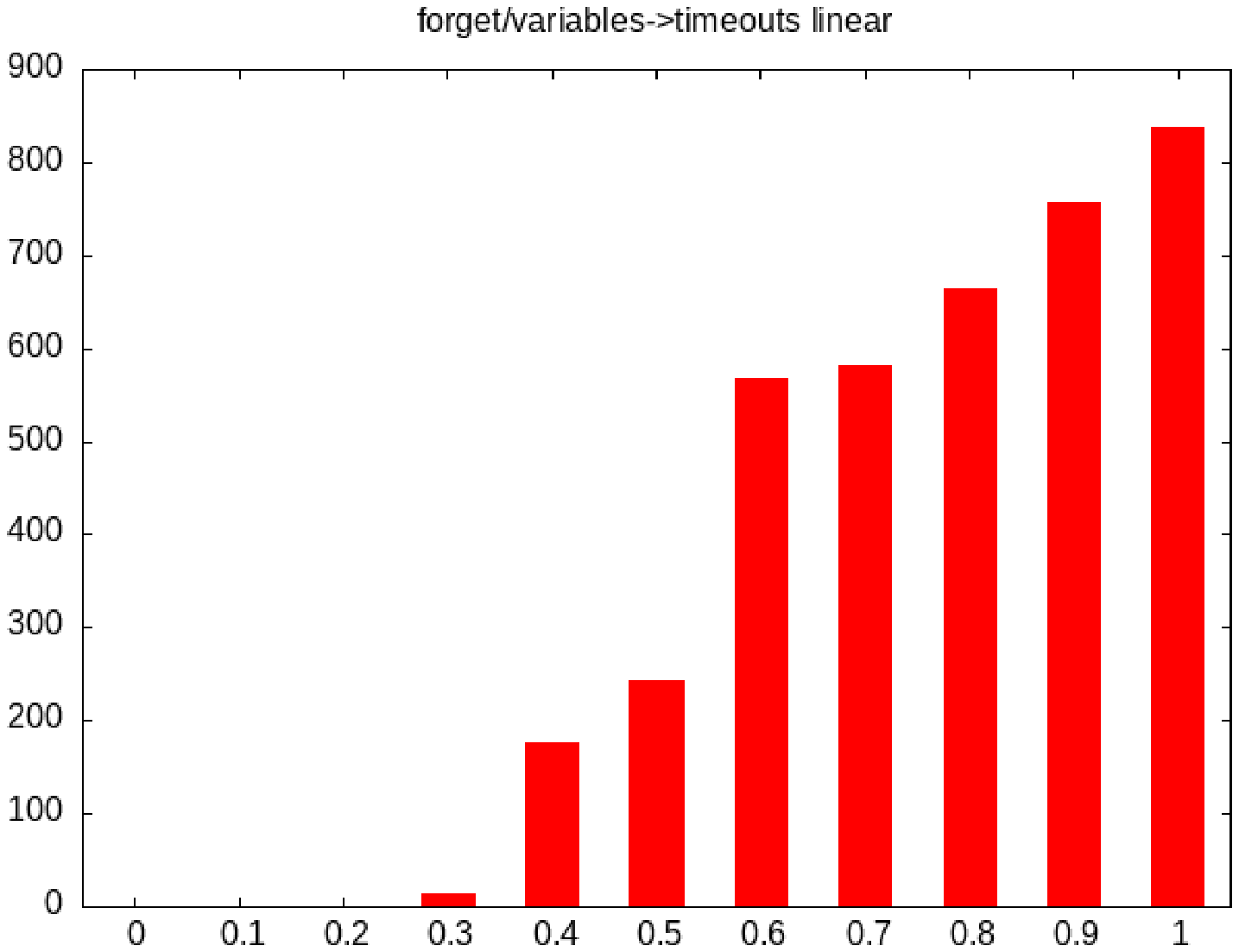}		&
\end{tabular}
}{
[timeouts for the linear algorithm]
}

\caption{Timeouts of the linear algorithm}
\label{timeouts}
\end{center}
\end{figure}

Figure~\ref{timeouts} shows the number of timeouts of the linear algorithm
depending on the number of variables and the fraction of variables to forget.
The first shows the number of timeouts and total tests for each number of
variables. The second is the percentage of timeouts over the total tests. The
increase looks almost linear, meaning that the percentage of failed tests
increased linearly with the number of variables. The graph showing the number
of timeouts in relation to the fraction of variables to forget groups them in
blocks: for example, the height of the bar at $0.5$ is the number of timeouts
when the fraction of variables to forget is between $0.45$ included and $0.55$
excluded. The graph exhibits an irregular increase. The irregularity may be
done to an uneven distribution of the fractions on the abscissa. More
significant is the increase: it suggests that the linear algorithm looses
efficiency as the fraction of variables to forget increases.


The above are direct comparisons between the algorithms. They are based on the
actual time and memory usage and size of the output. They are quantitative
measure on a common scale. The individual numbers can be directly compared: an
algorithm took 0.390 seconds, another took 5.660 seconds.

Below is a comparison of the ideal efficiency of the algorithms. It is based on
the self-reported time and memory usage, which disregards multiplicative
constants and keeps into account unimplemented optimizations. They do not allow
for a number-to-number comparison between the algorithms. Yet, they show how
the time and efficiency of the algorithms relates to the input parameters
(total number of variables, number of variables to forget, number of clauses).

All plots are for eight variables, the maximal number when at least three of
the four algorithms did not time out. The timeouts of the linear algorithm are
omitted since self-reported time and memory are incomplete measures anyway in
such cases.


\begin{figure}
\begin{center}

\ttytex{
\noindent
\begin{tabular}{cc}
\image{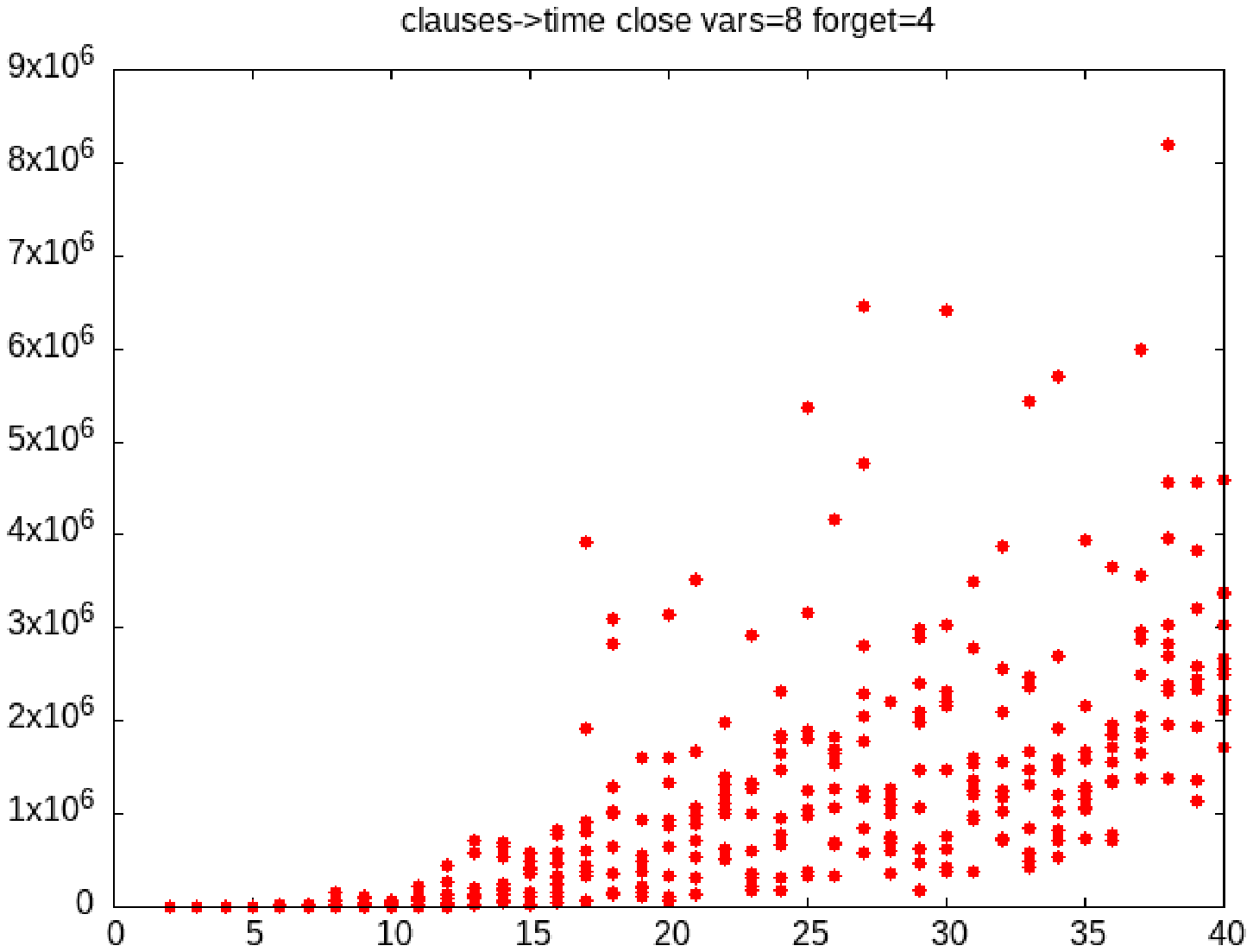}		&
\image{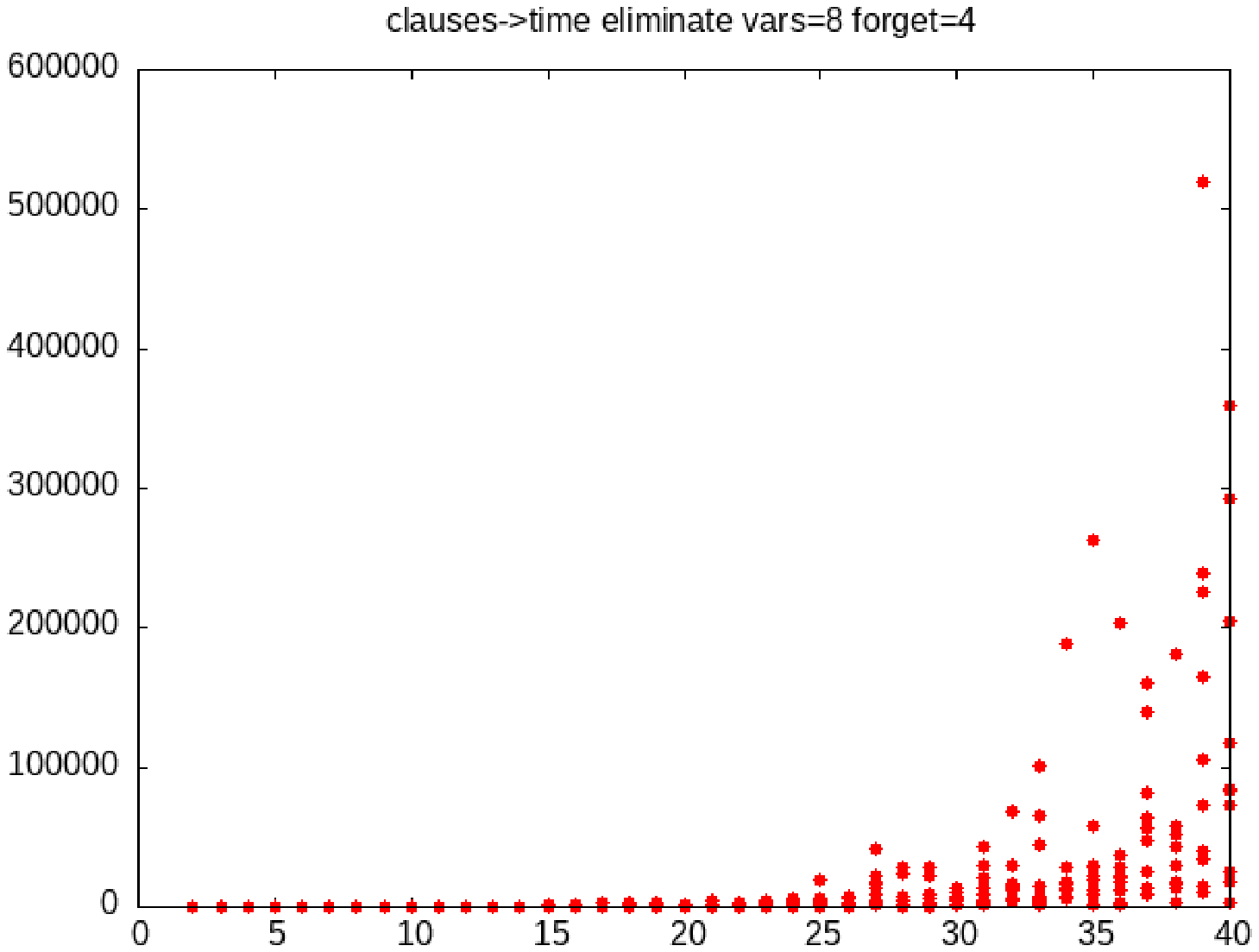}		\\
\image{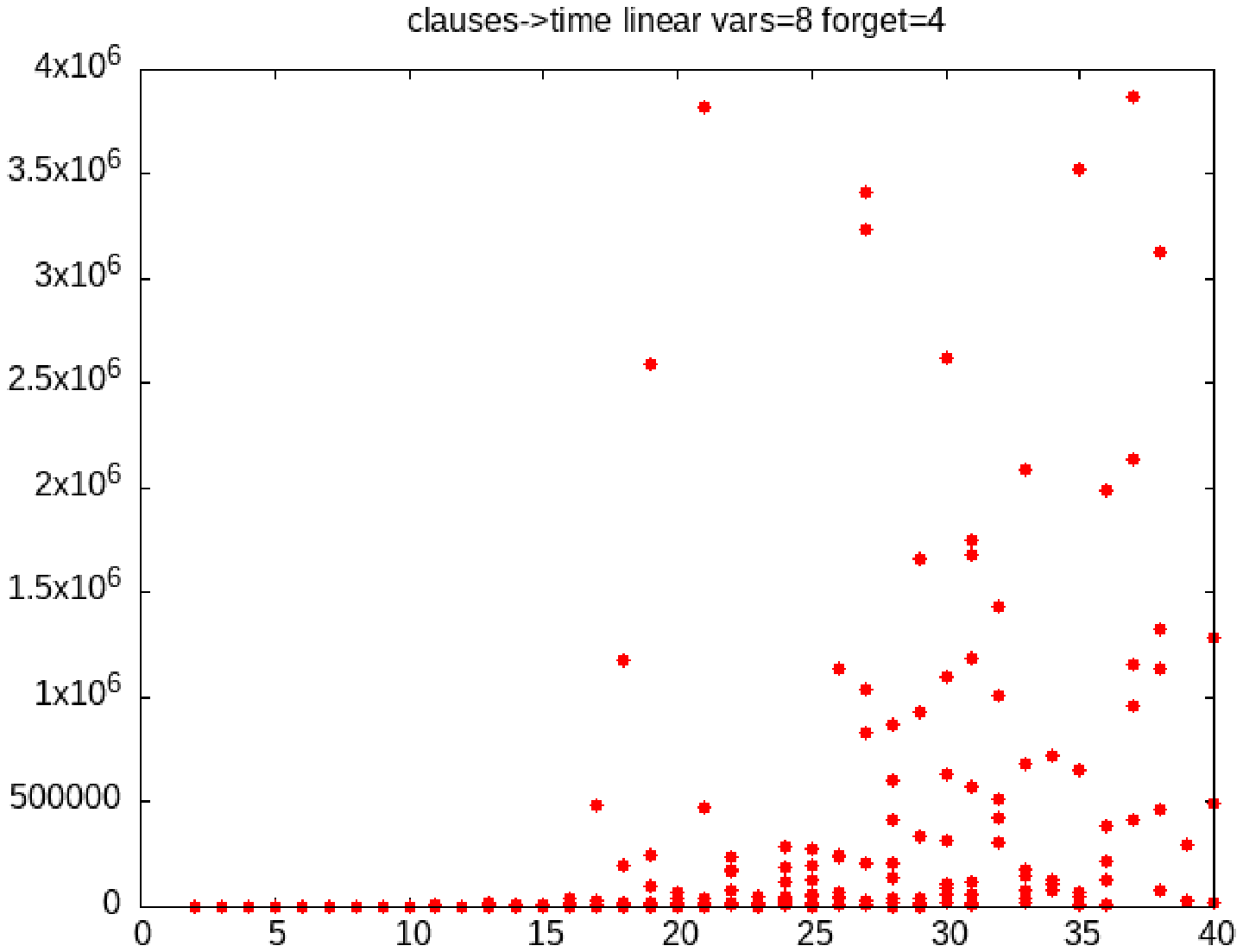}		&
\image{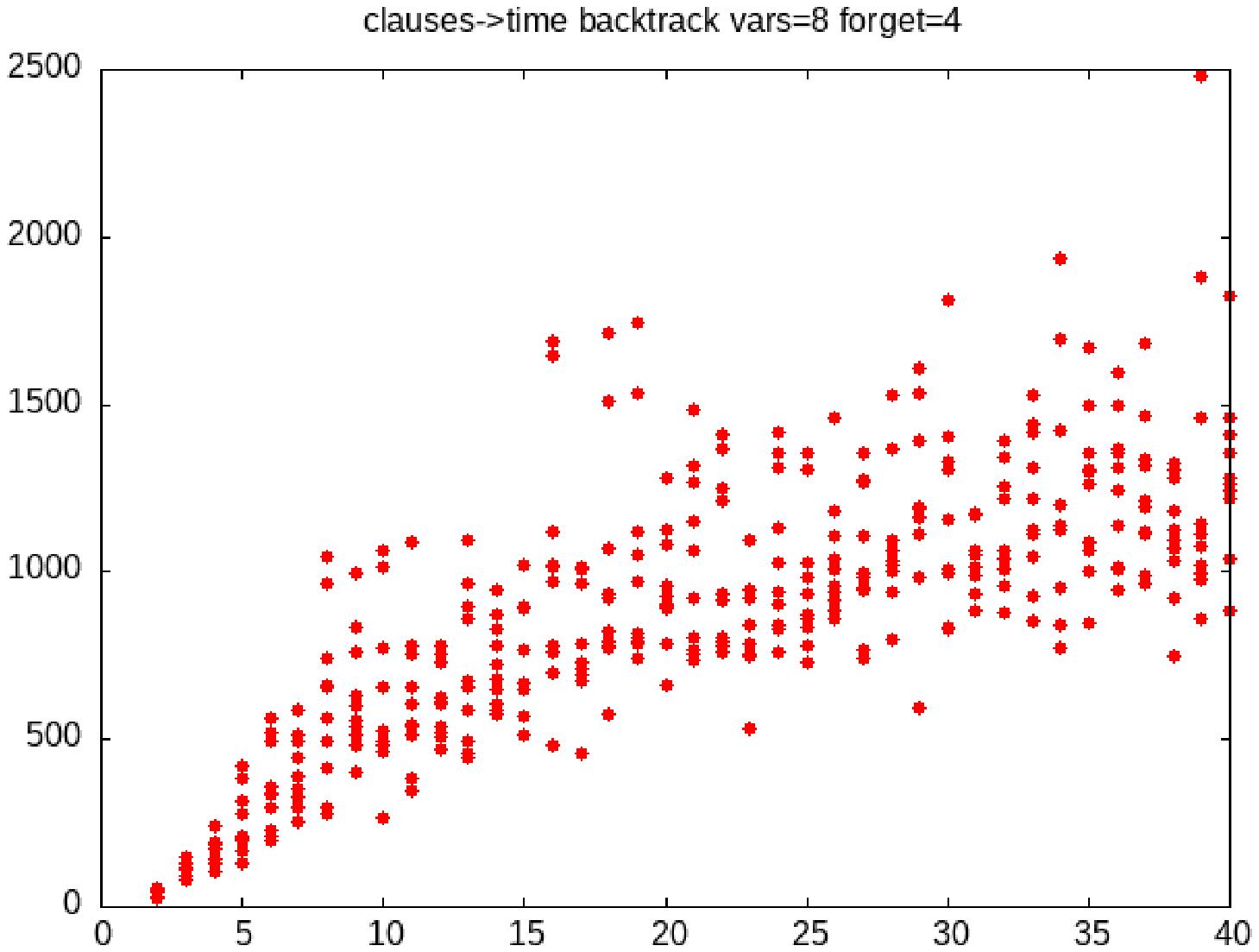}
\end{tabular}
}{
[clauses -> time]
}

\caption{Self-reported time, depending on the number of input clauses}
\label{clauses-time}
\end{center}
\end{figure}

Figure~\ref{clauses-time} shows how the number of input clauses affects the
running time. The number of total variables and variables to forget are fixed
to respectively eight and four. Different numbers lead to similar plots. A
common trait is that all algorithms spend more time on larger formulae, as
expected: a large formula takes more time to be processed. Yet, how time
depends on size varies a lot between algorithms. The curve of the close
algorithm steadily increases. The curve of the elimination algorithm keeps
increasing; it also exhibits a concentration of points at the bottom of the
diagram, with a few outliers. The linear resolution algorithm suffers from many
timeouts as the number of clauses increases; yet, its behavior appears
qualitatively similar to the elimination algorithm. The backtracking algorithm
is the least affected by the increase in the number of clauses; the
distribution looks like a logarithmic one, with an initial raise that slows
down quickly.


\begin{figure}
\begin{center}

\ttytex{
\noindent
\begin{tabular}{cc}
\image{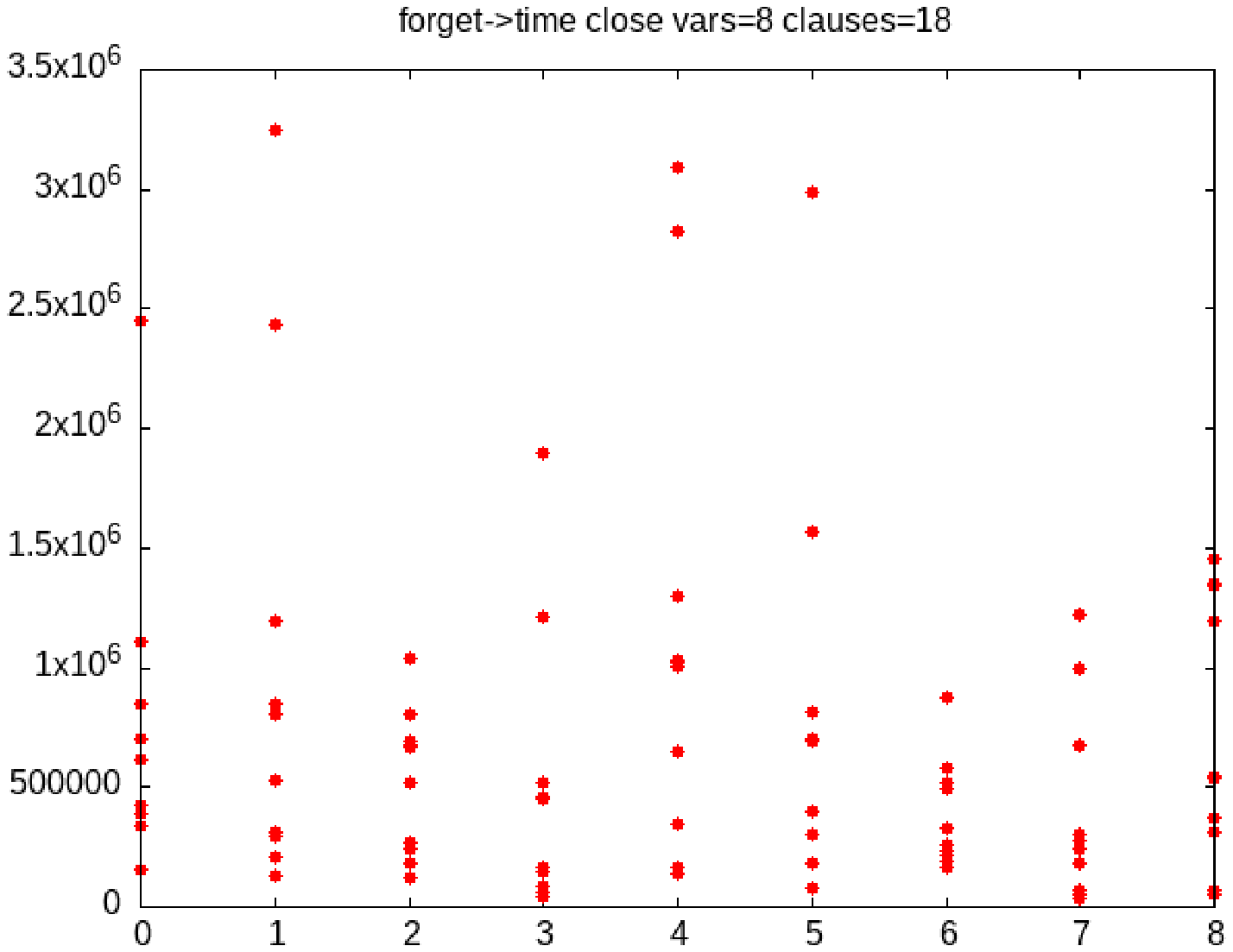}		&
\image{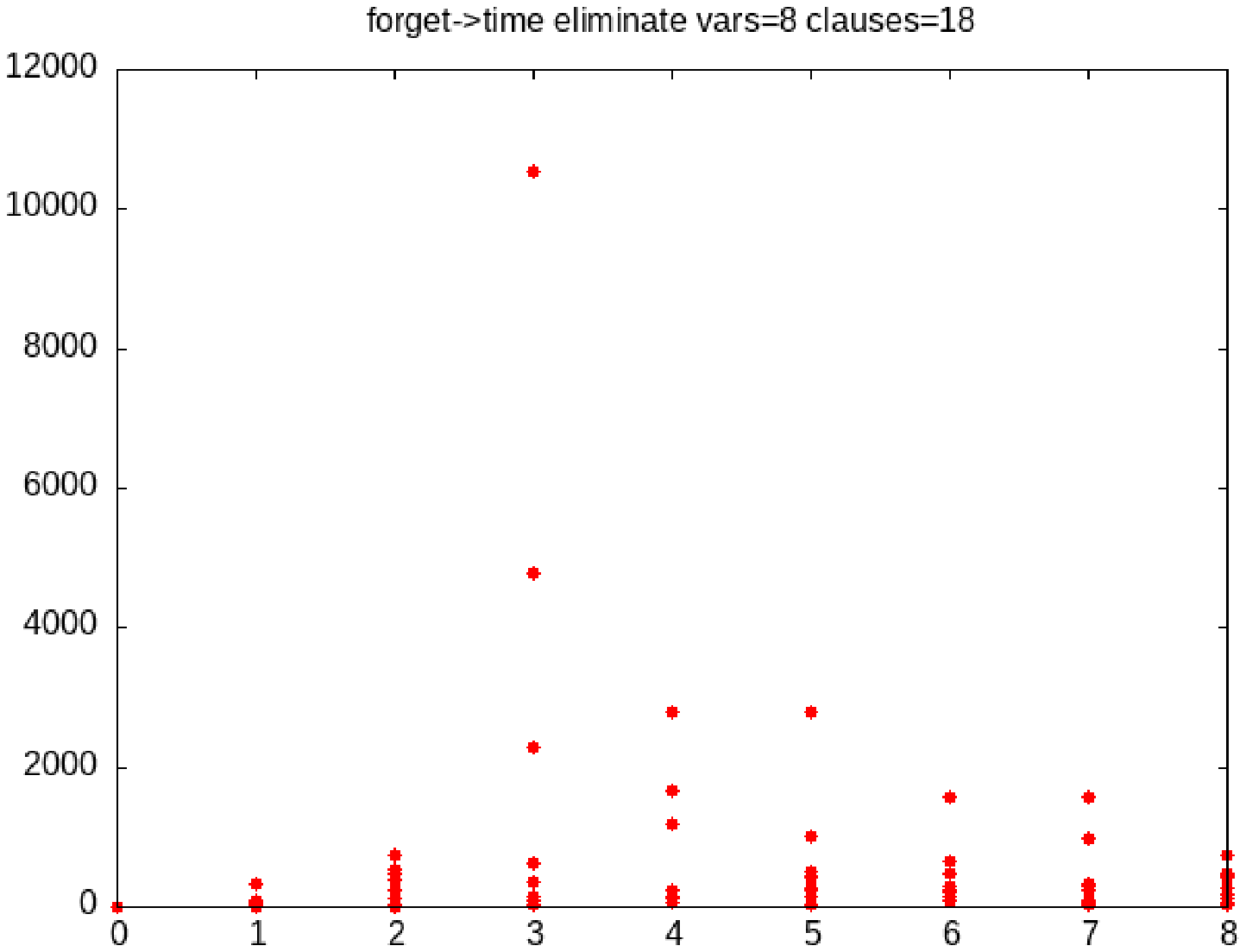}		\\
\image{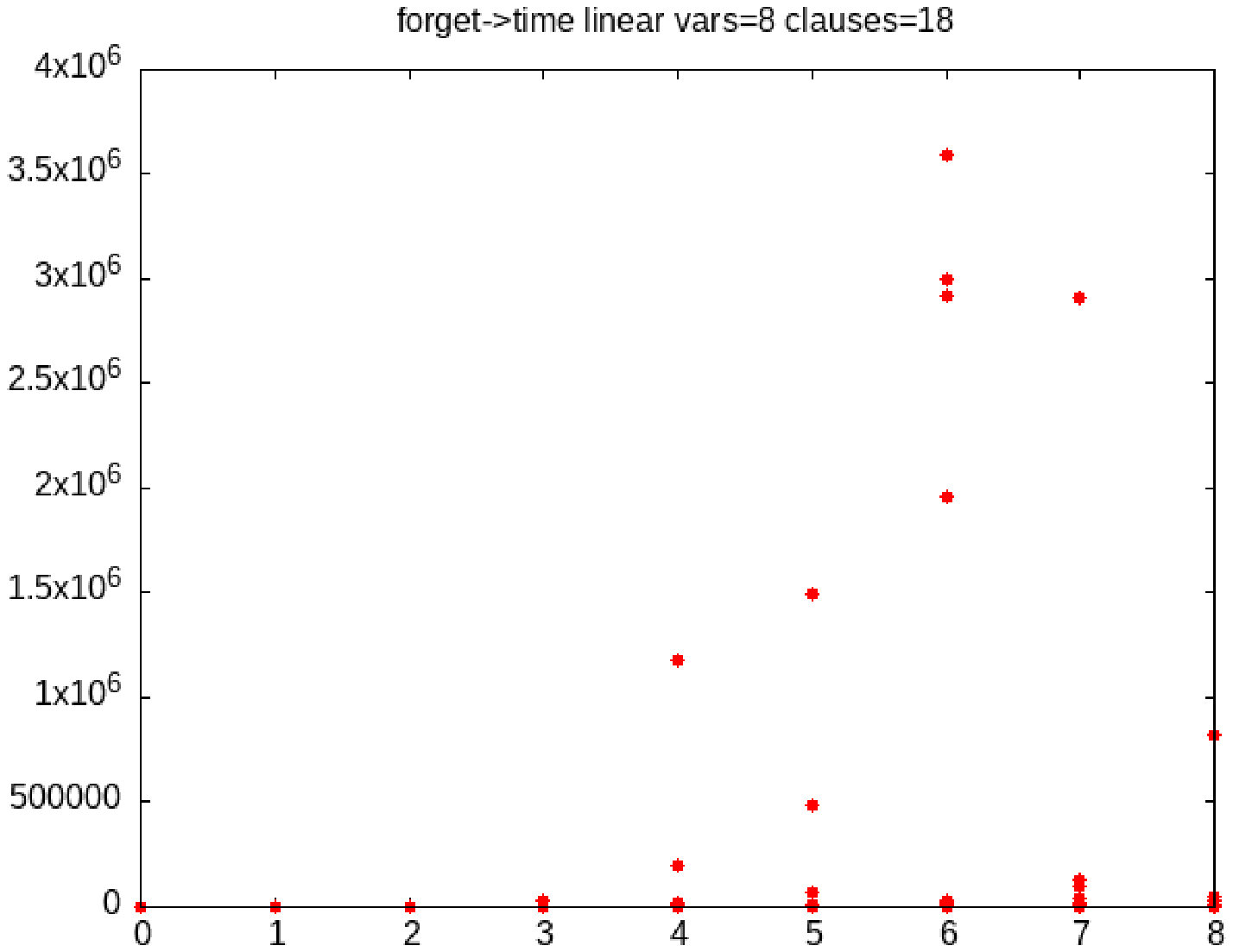}		&
\image{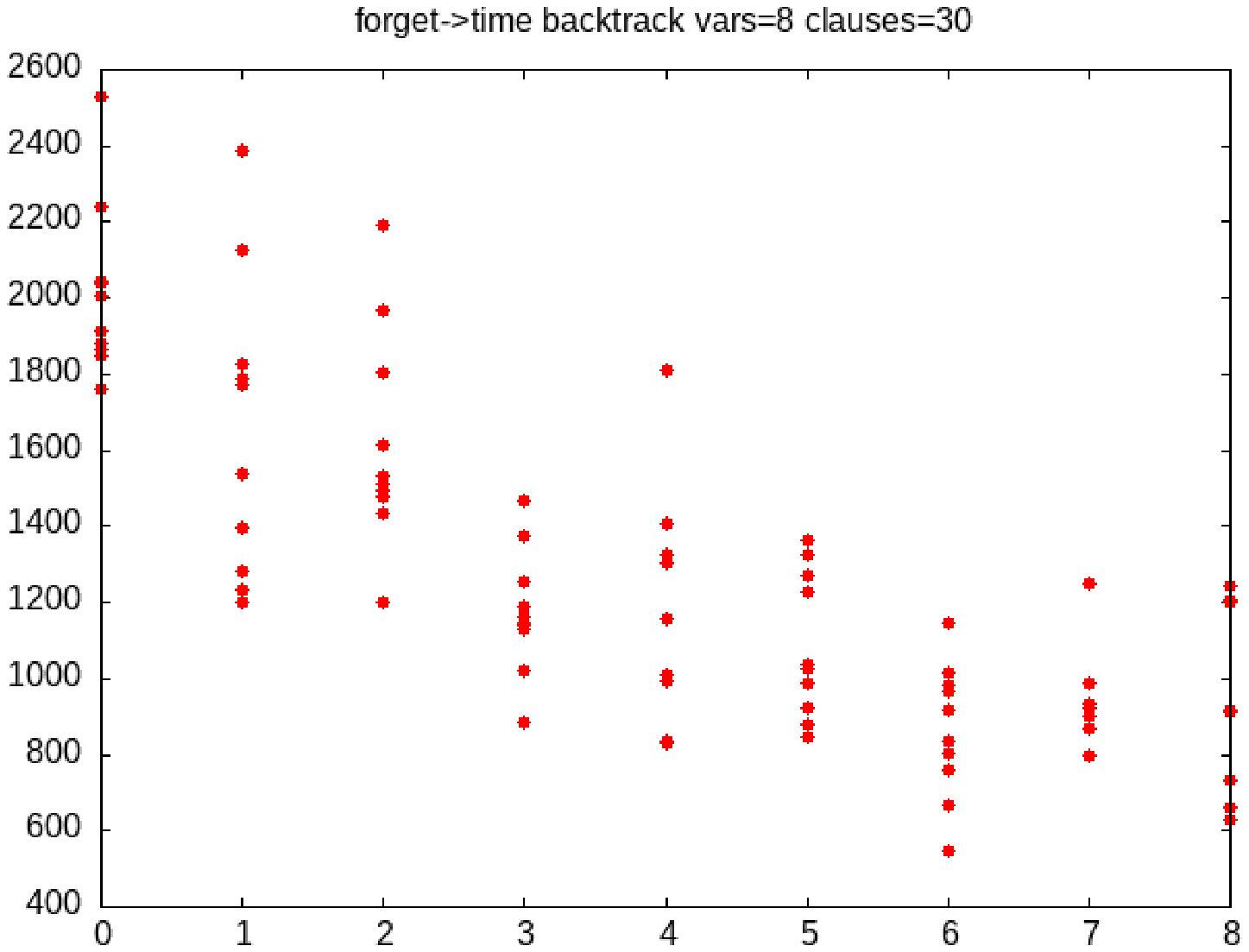}
\end{tabular}
}{
[forget->time]
}

\caption{Self-reported time, depending on the number of variables to forget}
\label{forget-time}
\end{center}
\end{figure}

Figure~\ref{forget-time} shows the self-reported running time depending on the
number of variables to forget when the total number of variables and the number
of clauses are respectively fixed at eight and eighteen. The close algorithm is
not affected by the number of variables to forget; this is expected, since the
slowest part of the algorithm is the generation of the resolution closure,
which does not even look at the variables to forget. The elimination algorithm
takes more time at the intermediate proportion of the variables to forget, but
the extent is not so marked as expected, and many difficult outliers emerge in
the left-to-middle of the diagram. A similar distribution is seen for the
linear resolution algorithm, with outliers in the right of the diagram. The
backtracking algorithm is unique: many variables to forget are slightly easier
than few, the contrary to the others; the dependency is small and is more
evident for many clauses; this is why the diagram is shown for thirty clauses
instead of eighteen.


\begin{figure}
\begin{center}

\ttytex{
\noindent
\begin{tabular}{cc}
\image{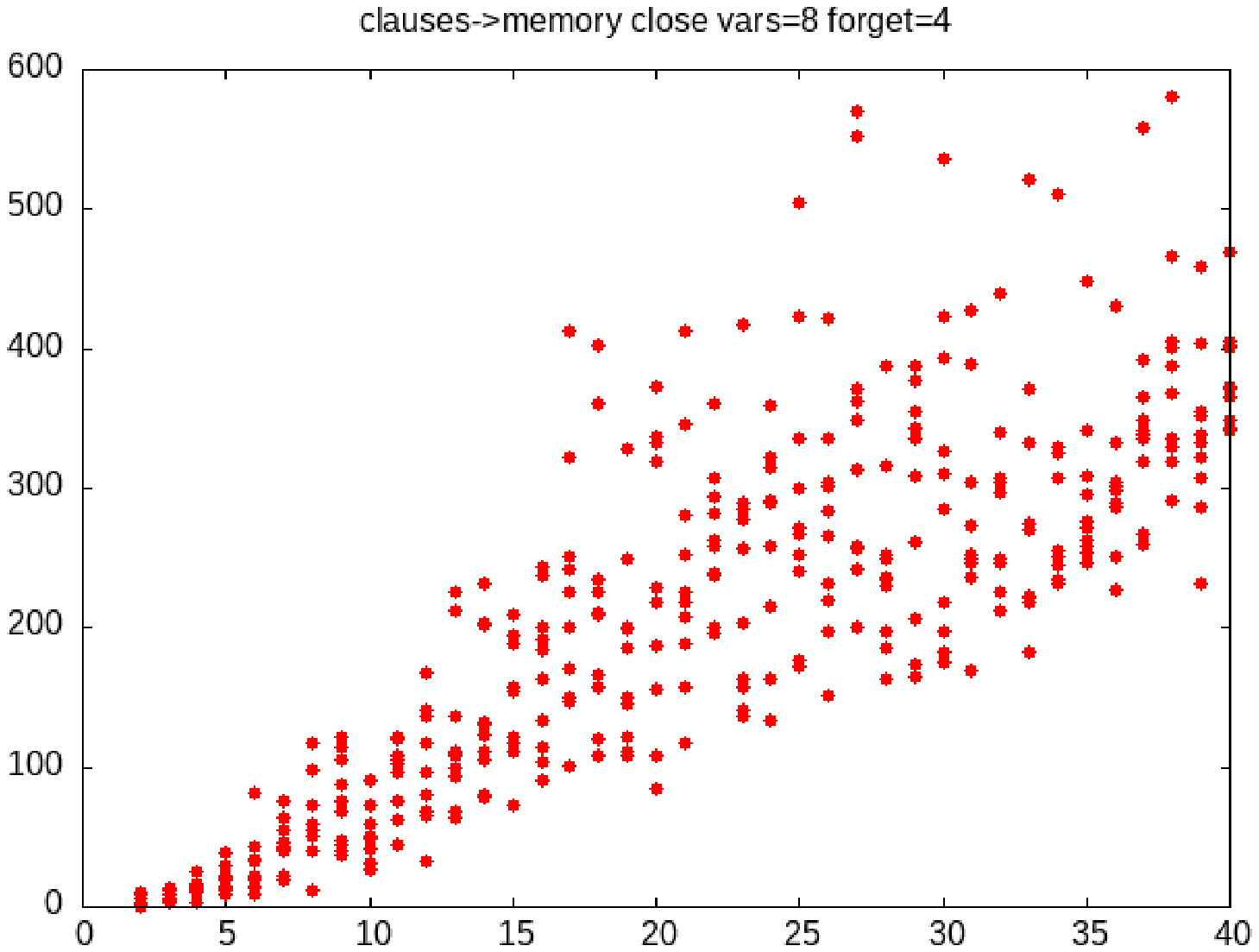}		&
\image{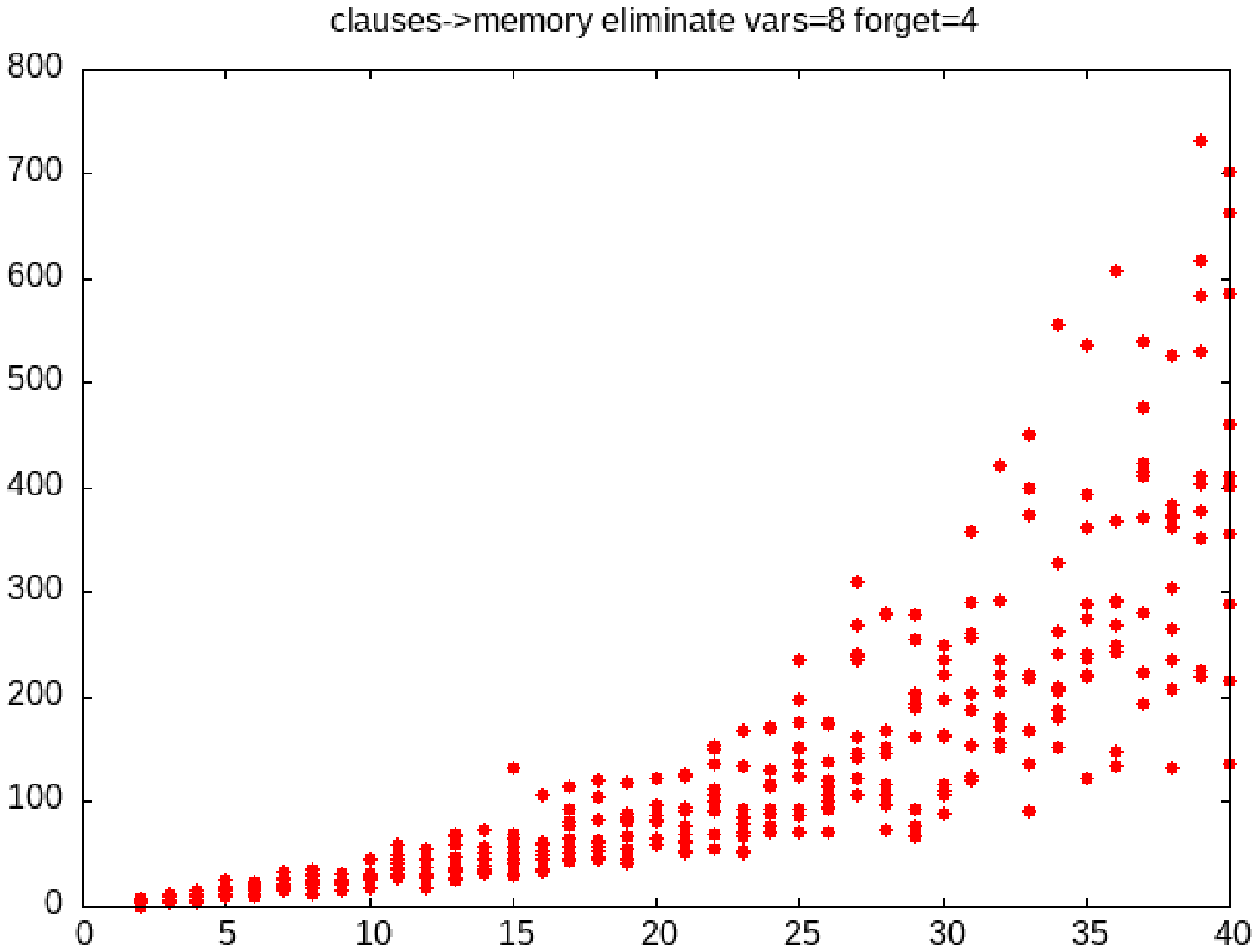}		\\
\image{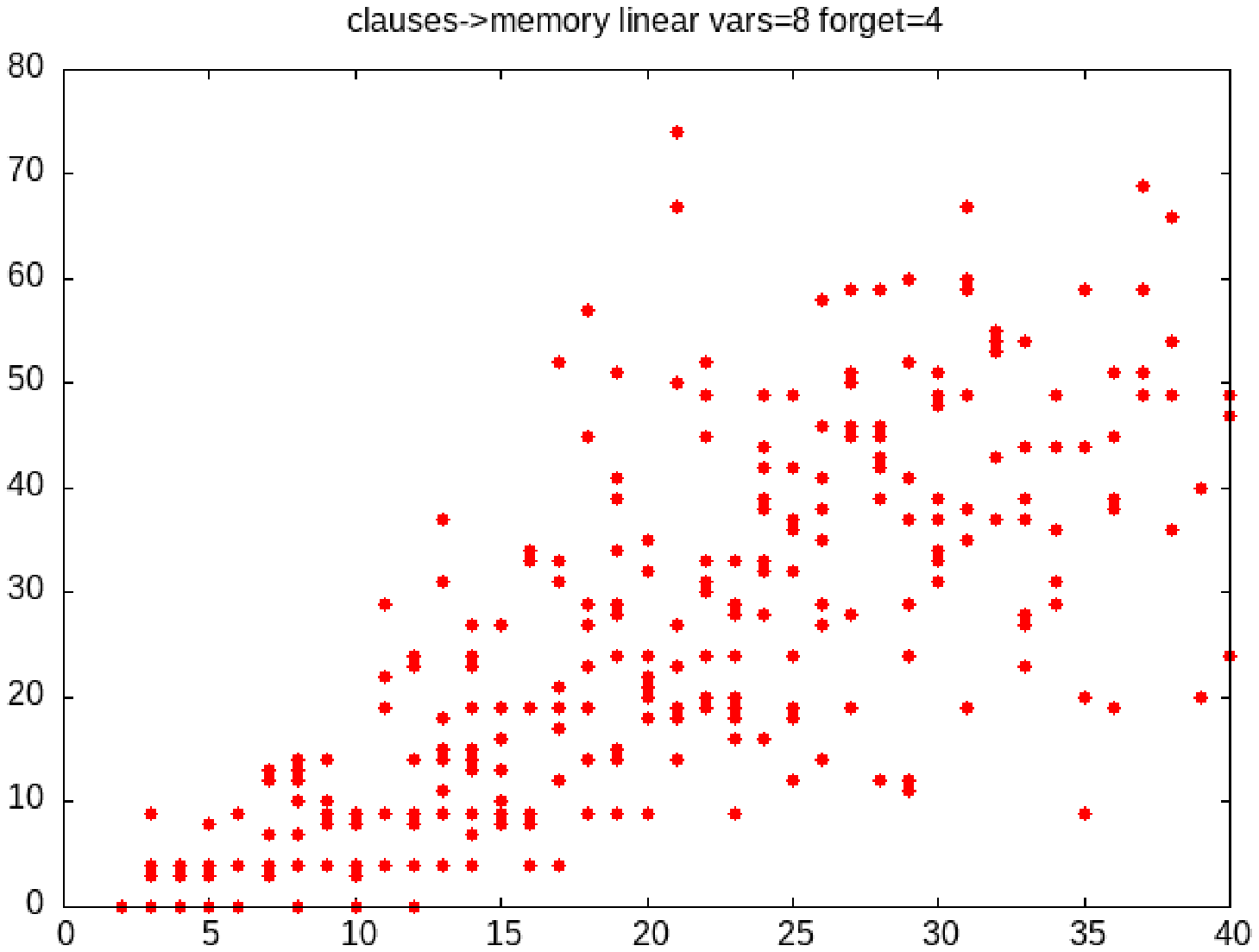}		&
\image{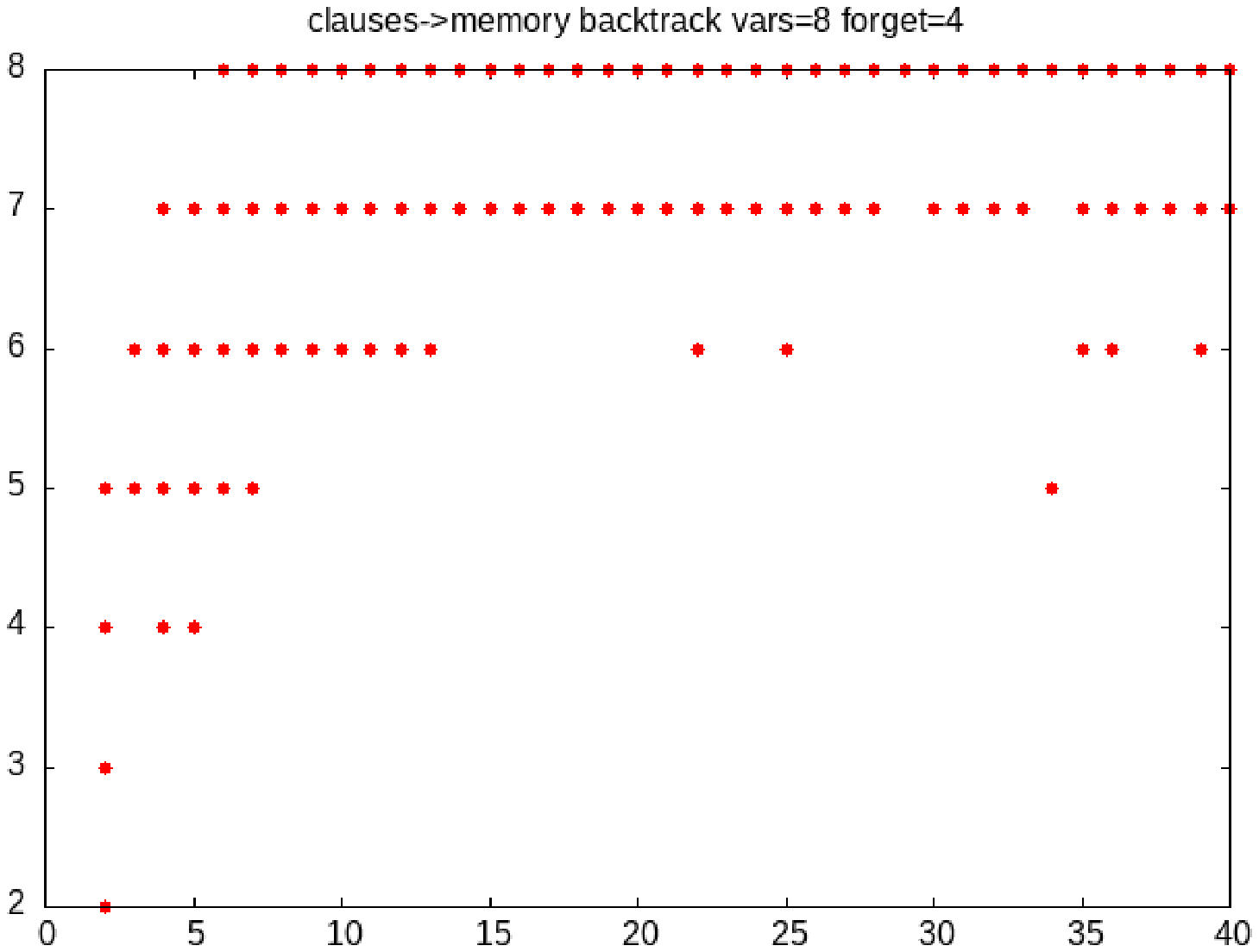}
\end{tabular}
}{
[clauses -> memory]
}

\caption{Self-reported memory usage, depending on the number of clauses}
\label{clauses-memory}
\end{center}
\end{figure}

The dependency of memory on the number of clauses is shown in
Figure~\ref{clauses-memory}. The curve of the close algorithm is similar to
that of time, with a steady increase. The elimination algorithm exhibits a
slightly superlinear increase in memory with the number of clauses; points are
distributed uniformly, and are more scattered when the clauses are many; none
is a real outlier. The linear resolution algorithm shows a linear increase with
a very uniform distribution of points. The diagram for the backtracking
algorithm looks weird at a first sight: its superior bound is a flat line, its
inferior bound is bell-shaped; the way the algorithm works explains the first:
the upper limit is due to the recursive subcalls being limited by the number of
variables; the lower limit is due to the hardest cases showing up for a middle
number of clauses, a phenomenon known as phase transition~\cite{xu-etal-12}.
Overall, the diagrams for memory are similar to the corresponding ones for time
except for the backtracking algorithm.


\begin{figure}
\begin{center}

\ttytex{
\noindent
\begin{tabular}{cc}
\image{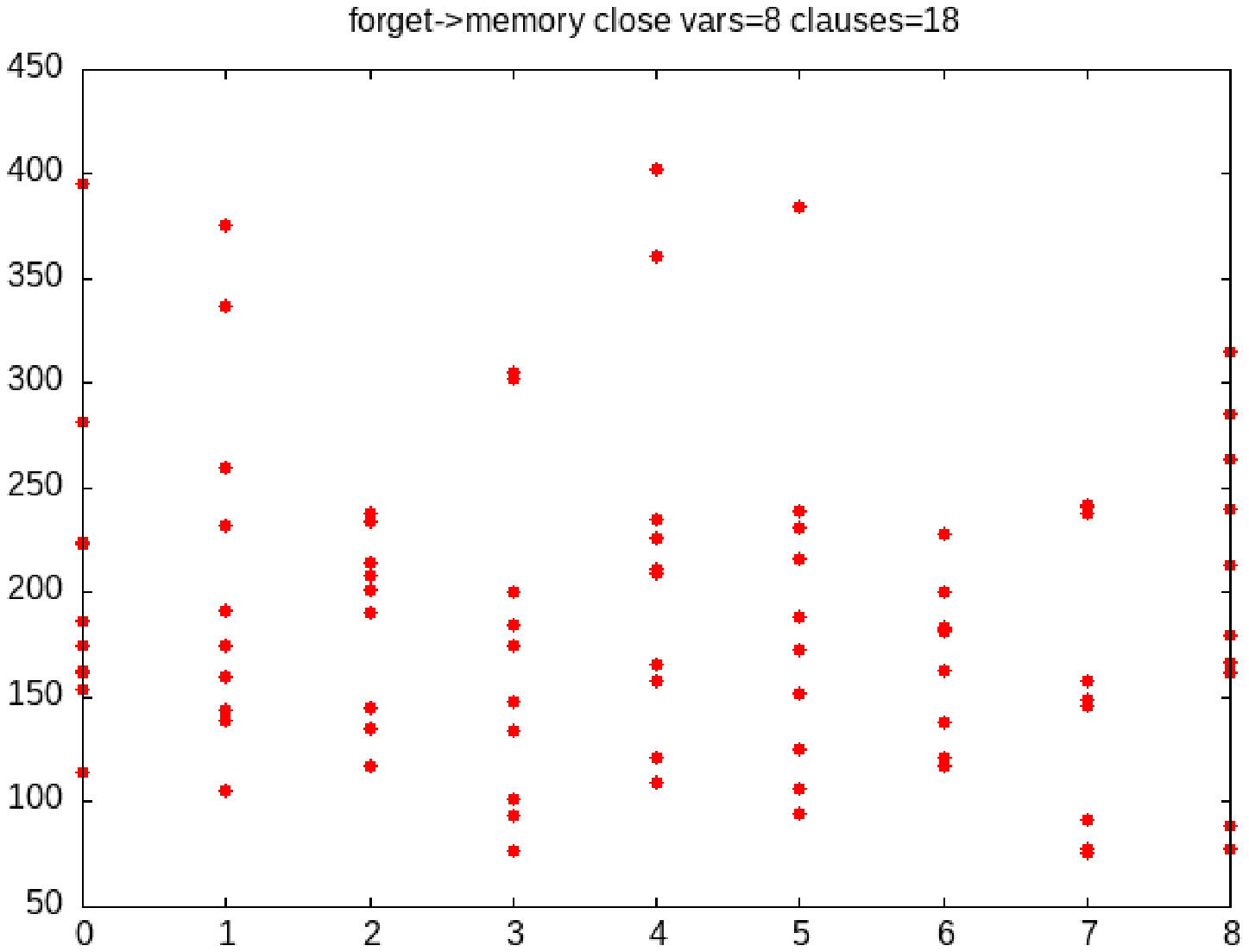}		&
\image{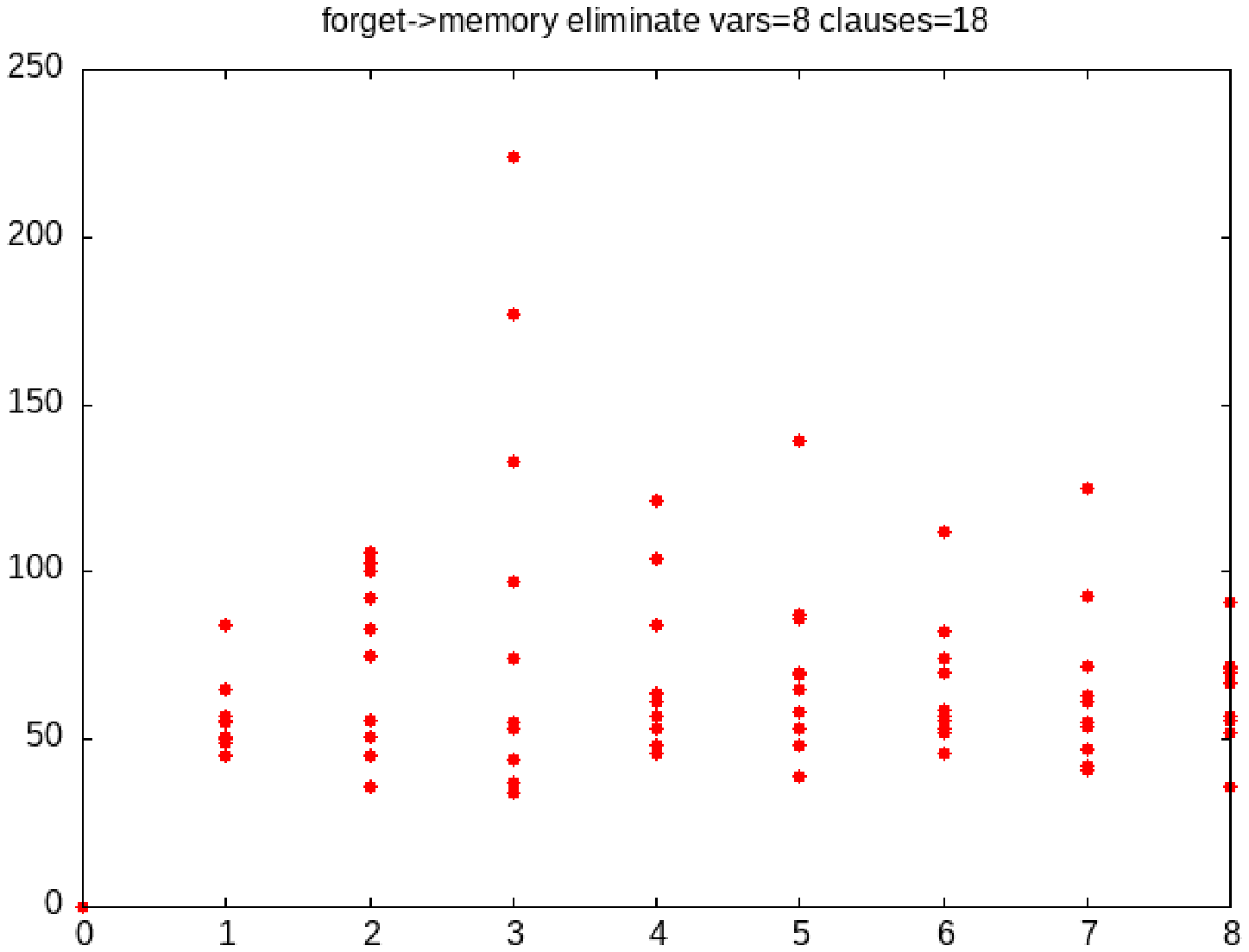}		\\
\image{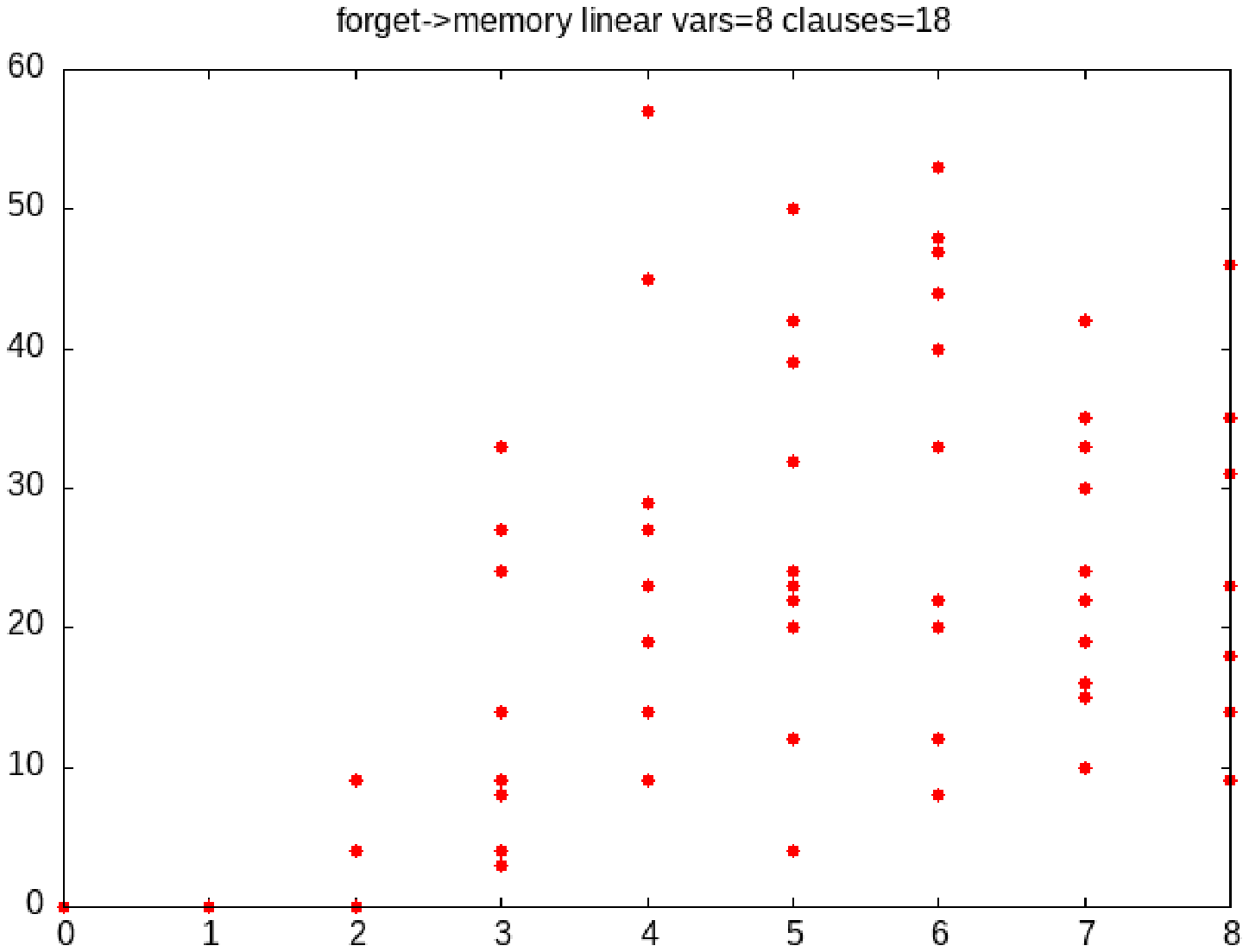}		&
\image{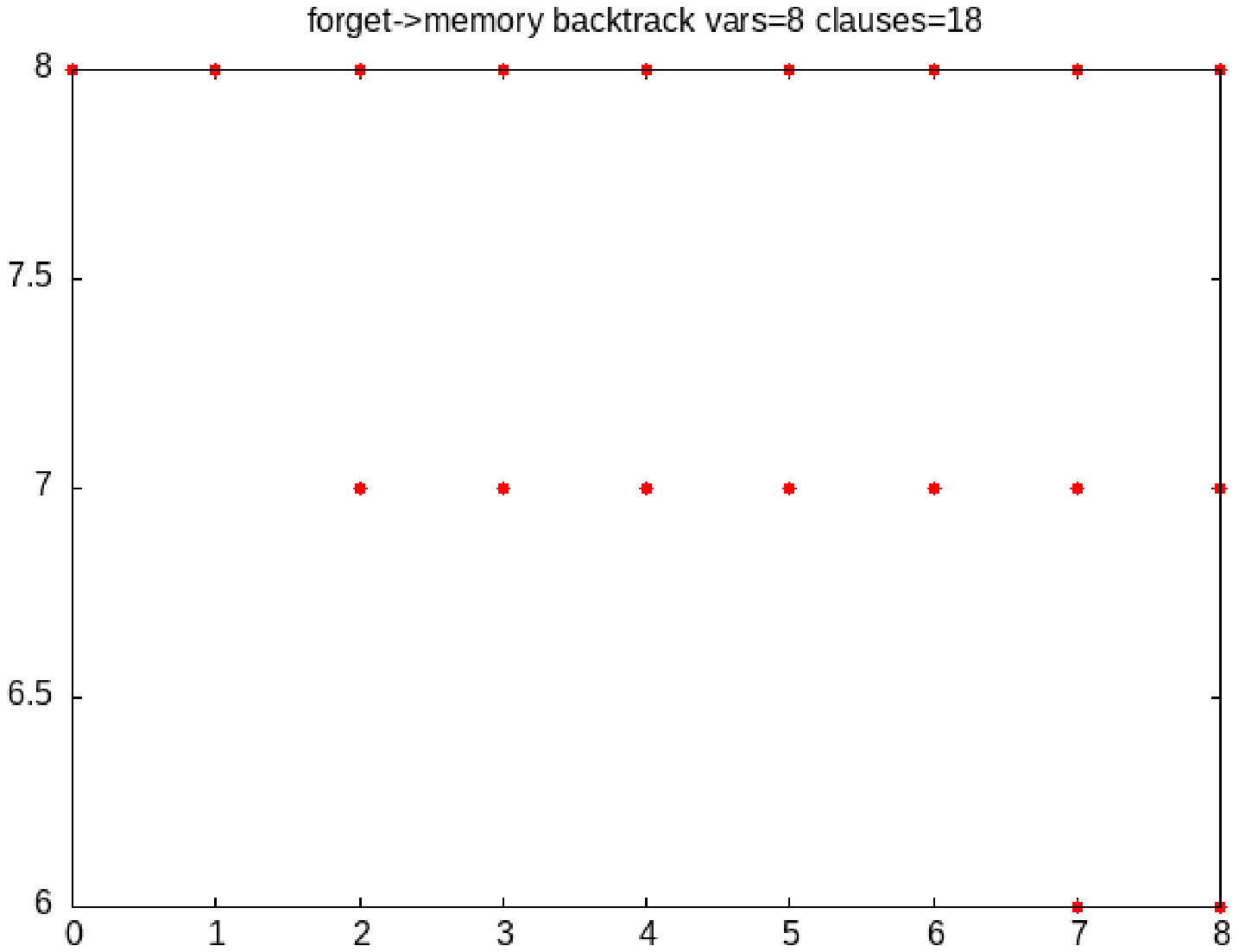}
\end{tabular}
}{
[forget -> memory]
}

\caption{Self-reported memory usage,
depending on the number of variables to forget}
\label{forget-memory}
\end{center}
\end{figure}

The memory used by the programs depending on the number of variables to forget
is shown in Figure~\ref{forget-memory}. The close algorithm does not depend on
the number of variables to forget, with points distributed uniformly. The
elimination algorithm is similar, with a slight increase in the middle. The
linear resolution algorithm is similar. The backtracking algorithm is bounded
in memory by the number of variables, which explains the flat upper limit; some
easy points are seen for a many variables to forget, but overall the difference
is very small.

\

This concludes the interpretation of the individual tests. Some algorithm is
faster than others, some algorithm uses less time, some algorithm generates the
shortest input. Yes, but which algorithm is the best?

Which algorithm is the best? Looking at actual time and memory, the winner is
the backtracking algorithm. It requires the least actual time and self-reported
memory, while being on the same scale as the close and elimination algorithms
on the size of output.

The close algorithm surprisingly produces the shortest outputs. Yet, the other
algorithms may still have a role if this measure is significant, as for example
the elimination algorithm is good for small inputs and the backtracking
algorithm is comparable for large inputs.

When the clarity of the resulting formula is important, the close and eliminate
algorithm win because they use the syntactic form of the input clauses. The
backtracking algorithm does not; it is semantical: it checks the effect of the
input clauses on the consistency of partial interpretations. For example, when
forgetting $b$ from
{} $\{a \rightarrow b, b \rightarrow c, c \rightarrow d, d \rightarrow a\}$,
it may produce
{} $\{a \rightarrow d, d \rightarrow c, c \rightarrow a\}$,
instead of
{} $\{a \rightarrow c, c \rightarrow d, d \rightarrow a\}$,
which is more similar to the input formula and is the result of the eliminate
algorithm.

\section{Any given algorithm}
\label{section-every}

Variable elimination and linear resolution are two forms of resolution, but
others exist, including many restrictions to linear resolution. This raises the
question: is forgetting possible in all of them? As an example, can s-linear
resolution be used for forgetting?

\proofonly{

\separator

{\bf Example: s-linear resolution could be used for forgetting.}

\

}

Variable elimination and linear resolution both forget variables by only
resolving on them, stopping when none is left. The same works for s-linear
resolution.

The resulting set of generated clauses expresses forgetting. This is proved by
showing that a subset of every full remembrance clause that is entailed by the
formula is generated. The full remembrance clauses are the ones that contain
all variables to forget.

Being entailed by the formula $F$, the negation of such a clause $C$ is
inconsistent with it: $F \cup \neg C \models \false$. Since s-linear resolution
is refutationally complete, $G \cup \neg C \vdash \bot$ follows: the empty
clause is generated from $G \cup \neg C$ by s-linear resolution. This
derivation is turned into $G \vdash C'$ where $C'$ is a subset of $C$ by the
transformation used for linear resolution, but this transformation does not
always maintain s-linearity. This condition allows a center clause $E$ to be
resolved with a previously generated clause $D$ only if the result $E'$ is a
subset of $E$.

\begin{center}
\setlength{\unitlength}{3947sp}%
\begingroup\makeatletter\ifx\SetFigFont\undefined%
\gdef\SetFigFont#1#2#3#4#5{%
  \reset@font\fontsize{#1}{#2pt}%
  \fontfamily{#3}\fontseries{#4}\fontshape{#5}%
  \selectfont}%
\fi\endgroup%
\begin{picture}(6852,1002)(2539,-3751)
\thinlines
{\color[rgb]{0,0,0}\put(5476,-3661){\line( 0, 1){300}}
}%
{\color[rgb]{0,0,0}\put(5251,-3661){\line( 1, 0){450}}
}%
{\color[rgb]{0,0,0}\put(6151,-3661){\line( 1, 0){450}}
}%
{\color[rgb]{0,0,0}\put(6376,-3661){\line( 0, 1){300}}
}%
{\color[rgb]{0,0,0}\put(7051,-3661){\line( 1, 0){450}}
}%
{\color[rgb]{0,0,0}\put(7276,-3661){\line( 0, 1){300}}
}%
{\color[rgb]{0,0,0}\put(7951,-3661){\line( 1, 0){450}}
}%
{\color[rgb]{0,0,0}\put(8176,-3661){\line( 0, 1){300}}
}%
{\color[rgb]{0,0,0}\put(8851,-3661){\line( 1, 0){450}}
}%
{\color[rgb]{0,0,0}\put(9076,-3661){\line( 0, 1){300}}
}%
{\color[rgb]{0,0,0}\put(2551,-3661){\line( 1, 0){450}}
}%
{\color[rgb]{0,0,0}\put(2776,-3661){\line( 0, 1){300}}
}%
{\color[rgb]{0,0,0}\put(3451,-3661){\line( 1, 0){450}}
}%
{\color[rgb]{0,0,0}\put(3676,-3661){\line( 0, 1){300}}
}%
{\color[rgb]{0,0,0}\put(3226,-3511){\line( 0, 1){750}}
\put(3226,-2761){\line( 1, 0){3150}}
\put(6376,-2761){\line( 0,-1){600}}
}%
{\color[rgb]{0,0,0}\put(4351,-3661){\line( 1, 0){450}}
}%
{\color[rgb]{0,0,0}\put(4576,-3361){\line( 0,-1){300}}
}%
\put(9376,-3736){\makebox(0,0)[lb]{\smash{{\SetFigFont{12}{14.4}
{\rmdefault}{\mddefault}{\updefault}{\color[rgb]{0,0,0}$\bot$}%
}}}}
\put(6826,-3736){\makebox(0,0)[b]{\smash{{\SetFigFont{12}{14.4}
{\rmdefault}{\mddefault}{\updefault}{\color[rgb]{0,0,0}$E'$}%
}}}}
\put(5926,-3736){\makebox(0,0)[b]{\smash{{\SetFigFont{12}{14.4}
{\rmdefault}{\mddefault}{\updefault}{\color[rgb]{0,0,0}$E$}%
}}}}
\put(3226,-3736){\makebox(0,0)[b]{\smash{{\SetFigFont{12}{14.4}
{\rmdefault}{\mddefault}{\updefault}{\color[rgb]{0,0,0}$D$}%
}}}}
\end{picture}%
\nop{
   |     |   |   |                |   |   |
  -+- D -+- -+- -+- E -+- E' c E -+- -+- -+-
      |                |
      +----------------+
}
\end{center}

The transformation used for linear resolution removes resolutions of center
clauses with the negation $\neg l$ of a literal of $C$. The effect is that $l$
is added to some of the following clauses. If neither $D$ nor $E$ change as
result, the condition $E' \subset E$ does not change. The same happens if both
are added $l$, or if $E$ is added $l$ and $D$ is not. The only case where a
violation occur is when $D$ is added $l$ and $E$ is not.

\begin{center}
\setlength{\unitlength}{3947sp}%
\begingroup\makeatletter\ifx\SetFigFont\undefined%
\gdef\SetFigFont#1#2#3#4#5{%
  \reset@font\fontsize{#1}{#2pt}%
  \fontfamily{#3}\fontseries{#4}\fontshape{#5}%
  \selectfont}%
\fi\endgroup%
\begin{picture}(6852,1002)(2539,-3751)
\thinlines
{\color[rgb]{0,0,0}\put(5476,-3661){\line( 0, 1){300}}
}%
{\color[rgb]{0,0,0}\put(5251,-3661){\line( 1, 0){450}}
}%
{\color[rgb]{0,0,0}\put(6151,-3661){\line( 1, 0){450}}
}%
{\color[rgb]{0,0,0}\put(6376,-3661){\line( 0, 1){300}}
}%
{\color[rgb]{0,0,0}\put(7051,-3661){\line( 1, 0){450}}
}%
{\color[rgb]{0,0,0}\put(7276,-3661){\line( 0, 1){300}}
}%
{\color[rgb]{0,0,0}\put(7951,-3661){\line( 1, 0){450}}
}%
{\color[rgb]{0,0,0}\put(8176,-3661){\line( 0, 1){300}}
}%
{\color[rgb]{0,0,0}\put(8851,-3661){\line( 1, 0){450}}
}%
{\color[rgb]{0,0,0}\put(9076,-3661){\line( 0, 1){300}}
}%
{\color[rgb]{0,0,0}\put(2551,-3661){\line( 1, 0){450}}
}%
{\color[rgb]{0,0,0}\put(2776,-3661){\line( 0, 1){300}}
}%
{\color[rgb]{0,0,0}\put(3451,-3661){\line( 1, 0){450}}
}%
{\color[rgb]{0,0,0}\put(3676,-3661){\line( 0, 1){300}}
}%
{\color[rgb]{0,0,0}\put(3226,-3511){\line( 0, 1){750}}
\put(3226,-2761){\line( 1, 0){3150}}
\put(6376,-2761){\line( 0,-1){600}}
}%
{\color[rgb]{0,0,0}\put(4351,-3661){\line( 1, 0){450}}
}%
{\color[rgb]{0,0,0}\put(4576,-3361){\line( 0,-1){300}}
}%
\put(5926,-3736){\makebox(0,0)[b]{\smash{{\SetFigFont{12}{14.4}
{\rmdefault}{\mddefault}{\updefault}{\color[rgb]{0,0,0}$E$}%
}}}}
\put(3226,-3736){\makebox(0,0)[b]{\smash{{\SetFigFont{12}{14.4}
{\rmdefault}{\mddefault}{\updefault}{\color[rgb]{0,0,0}$D \vee l$}%
}}}}
\put(6826,-3736){\makebox(0,0)[b]{\smash{{\SetFigFont{12}{14.4}
{\rmdefault}{\mddefault}{\updefault}{\color[rgb]{0,0,0}$E' \vee l$}%
}}}}
\end{picture}%
\nop{
   |         |   |   |                |   |   |
  -+- D v l -+- -+- -+- E -+- E' v l -+- -+- -+-
        |                  |
        +------------------+
}
\end{center}

The condition of s-linearity is violated since $E' \vee l$ contains $l$ while
$E$ does not. Yet, since $l$ is in $D \vee l$ but is not in $E$, it is resolved
with $\neg l$ in between.

\begin{center}
\setlength{\unitlength}{3947sp}%
\begingroup\makeatletter\ifx\SetFigFont\undefined%
\gdef\SetFigFont#1#2#3#4#5{%
  \reset@font\fontsize{#1}{#2pt}%
  \fontfamily{#3}\fontseries{#4}\fontshape{#5}%
  \selectfont}%
\fi\endgroup%
\begin{picture}(6852,1002)(2539,-3751)
\thinlines
{\color[rgb]{0,0,0}\put(5476,-3661){\line( 0, 1){300}}
}%
{\color[rgb]{0,0,0}\put(5251,-3661){\line( 1, 0){450}}
}%
{\color[rgb]{0,0,0}\put(6151,-3661){\line( 1, 0){450}}
}%
{\color[rgb]{0,0,0}\put(6376,-3661){\line( 0, 1){300}}
}%
{\color[rgb]{0,0,0}\put(7051,-3661){\line( 1, 0){450}}
}%
{\color[rgb]{0,0,0}\put(7276,-3661){\line( 0, 1){300}}
}%
{\color[rgb]{0,0,0}\put(7951,-3661){\line( 1, 0){450}}
}%
{\color[rgb]{0,0,0}\put(8176,-3661){\line( 0, 1){300}}
}%
{\color[rgb]{0,0,0}\put(8851,-3661){\line( 1, 0){450}}
}%
{\color[rgb]{0,0,0}\put(9076,-3661){\line( 0, 1){300}}
}%
{\color[rgb]{0,0,0}\put(2551,-3661){\line( 1, 0){450}}
}%
{\color[rgb]{0,0,0}\put(2776,-3661){\line( 0, 1){300}}
}%
{\color[rgb]{0,0,0}\put(3451,-3661){\line( 1, 0){450}}
}%
{\color[rgb]{0,0,0}\put(3676,-3661){\line( 0, 1){300}}
}%
{\color[rgb]{0,0,0}\put(3226,-3511){\line( 0, 1){750}}
\put(3226,-2761){\line( 1, 0){3150}}
\put(6376,-2761){\line( 0,-1){600}}
}%
{\color[rgb]{0,0,0}\put(4351,-3661){\line( 1, 0){450}}
}%
{\color[rgb]{0,0,0}\put(4576,-3361){\line( 0,-1){300}}
}%
\put(5926,-3736){\makebox(0,0)[b]{\smash{{\SetFigFont{12}{14.4}
{\rmdefault}{\mddefault}{\updefault}{\color[rgb]{0,0,0}$E$}%
}}}}
\put(3226,-3736){\makebox(0,0)[b]{\smash{{\SetFigFont{12}{14.4}
{\rmdefault}{\mddefault}{\updefault}{\color[rgb]{0,0,0}$D \vee l$}%
}}}}
\put(6826,-3736){\makebox(0,0)[b]{\smash{{\SetFigFont{12}{14.4}
{\rmdefault}{\mddefault}{\updefault}{\color[rgb]{0,0,0}$E' \vee l$}%
}}}}
\put(4576,-3286){\makebox(0,0)[b]{\smash{{\SetFigFont{12}{14.4}
{\rmdefault}{\mddefault}{\updefault}{\color[rgb]{0,0,0}$\neg l$}%
}}}}
\end{picture}%
\nop{
                -l
   |         |   |   |                |   |   |
  -+- D v l -+- -+- -+- E -+- E' v l -+- -+- -+-
        |                  |
        +------------------+
}
\end{center}

The transformation also removes this resolution and adds $l$ to all following
clauses. The result is that $E$ is added $l$.

\begin{center}
\setlength{\unitlength}{3947sp}%
\begingroup\makeatletter\ifx\SetFigFont\undefined%
\gdef\SetFigFont#1#2#3#4#5{%
  \reset@font\fontsize{#1}{#2pt}%
  \fontfamily{#3}\fontseries{#4}\fontshape{#5}%
  \selectfont}%
\fi\endgroup%
\begin{picture}(6852,1002)(2539,-3751)
\thinlines
{\color[rgb]{0,0,0}\put(5476,-3661){\line( 0, 1){300}}
}%
{\color[rgb]{0,0,0}\put(5251,-3661){\line( 1, 0){450}}
}%
{\color[rgb]{0,0,0}\put(6151,-3661){\line( 1, 0){450}}
}%
{\color[rgb]{0,0,0}\put(6376,-3661){\line( 0, 1){300}}
}%
{\color[rgb]{0,0,0}\put(7051,-3661){\line( 1, 0){450}}
}%
{\color[rgb]{0,0,0}\put(7276,-3661){\line( 0, 1){300}}
}%
{\color[rgb]{0,0,0}\put(7951,-3661){\line( 1, 0){450}}
}%
{\color[rgb]{0,0,0}\put(8176,-3661){\line( 0, 1){300}}
}%
{\color[rgb]{0,0,0}\put(8851,-3661){\line( 1, 0){450}}
}%
{\color[rgb]{0,0,0}\put(9076,-3661){\line( 0, 1){300}}
}%
{\color[rgb]{0,0,0}\put(2551,-3661){\line( 1, 0){450}}
}%
{\color[rgb]{0,0,0}\put(2776,-3661){\line( 0, 1){300}}
}%
{\color[rgb]{0,0,0}\put(3451,-3661){\line( 1, 0){450}}
}%
{\color[rgb]{0,0,0}\put(3676,-3661){\line( 0, 1){300}}
}%
{\color[rgb]{0,0,0}\put(3226,-3511){\line( 0, 1){750}}
\put(3226,-2761){\line( 1, 0){3150}}
\put(6376,-2761){\line( 0,-1){600}}
}%
{\color[rgb]{0,0,0}\put(4351,-3661){\line( 1, 0){900}}
}%
\put(5926,-3736){\makebox(0,0)[b]{\smash{{\SetFigFont{12}{14.4}
{\rmdefault}{\mddefault}{\updefault}{\color[rgb]{0,0,0}$E \vee l$}%
}}}}
\put(3226,-3736){\makebox(0,0)[b]{\smash{{\SetFigFont{12}{14.4}
{\rmdefault}{\mddefault}{\updefault}{\color[rgb]{0,0,0}$D \vee l$}%
}}}}
\put(6826,-3736){\makebox(0,0)[b]{\smash{{\SetFigFont{12}{14.4}
{\rmdefault}{\mddefault}{\updefault}{\color[rgb]{0,0,0}$E' \vee l$}%
}}}}
\end{picture}%
\nop{
   |         |       |                    |   |   |
  -+- D v l -+- --- -+- E v l -+- E' v l -+- -+- -+-
        |                      |
        +----------------------+
}
\end{center}

The condition $E' \vee l \subset E \vee l$ holds as a consequence of $E'
\subset E$. While removing a single resolution with $\neg l$ may violate
s-linearity, removing all of them restores it.

\proofonly{

\separator

{\bf Extension to an arbitrary resolution strategy.
Entailment completeness is not required nor desirable.}

\

}

Not only directional and linear resolution, but also A-order linear resolution
and s-linear resolution forget variables. Yet, each of these methods requires a
separate proof. A common point is that correctness is easy: resolution only
produces entailed clauses, and all entailed clauses comprising variables to
remember are correct when forgetting. Correctness is guaranteed by resolution.
Completeness is not.

A single proof of completeness for all forms of resolution would be ideal. The
second-best choice is reusing or adapting existing proofs. For example, many
resolution restrictions are proved entailment-complete: they generate a subset
of every clause entailed clause~\cite{slag-etal-69,inou-92,delv-99}.
Restricting to the clauses comprising only the variables to remember makes it a
complete forgetting algorithm.

A misleading solution emerges: forget by whichever form of resolution that is
entailment-complete. This solution is misleading because entailment
completeness is a.~not required and b.~not desirable.

Directional resolution is not entailment complete. Neither is A-ordering linear
resolution~\cite{amir-mcil-05}. Yet, they both forget variables.


The proof for linear resolution shows what is really required. The starting
point is an entailed full remembrance clause $C$, a clause comprising exactly
the variables to remember. The conclusion is that linear resolution produces a
subset of $C$. This is not entailment-completeness. It is
entailment-completeness on full remembrance clauses. Only clauses comprising
all variables to remember need to be generated, not all of them.

An example shows the difference: forgetting $b$ from
{} $\{a \vee b, \neg b \vee c, \neg c \vee d\}$.
Both variable elimination and A-ordering linear resolution only resolve the
first two clauses $a \vee b$ and $\neg b \vee c$ since their resolving variable
$b$ is the only one to forget. The result is $a \vee c$. Neither $a \vee d$ nor
any subset of its is generated, in spite of being entailed by the formula.

Completeness is not even required on prime implicates (the clauses that are
entailed while none of theirs subsets is). The previous example provides a
counterexample, since $a \vee d$ is such a prime implicate.

Entailment completeness is not required. Neither directional not s-linear
resolution meet it. Resolution by closure does. The least efficient method is
entailment-complete.

Resolution by closure resolves clauses in all possible ways and outputs the
ones comprising variables to remember. When forgetting $b$ from
{} $\{a \vee b, \neg b \vee c, \neg c \vee d\}$,
it produces $a \vee c$, but also $a \vee d$. The latter clause is redundant.
Forgetting does not require it, as it is entailed by $a \vee c$ and $\neg c
\vee d$. Entailment-completeness is not only unnecessary, it is not even
desirable. A shorter formula expressing forgetting is better than a larger one.

\proofonly{

The backtracking algorithm may generate formulae like
{} $\{\neg a \vee \neg c \vee d, a \vee c\}$.
The most efficient algorithm for forgetting does not seem to follow any
specific rule about the clauses it generates.

}

The proofs of correctness of linear resolution and its restrictions all start
from a full remembrance clause. When forgetting no variables, these are the
full clauses, the clauses that contain all variables. The generation of a
subset of all full clauses is necessary. The question is whether it is also
sufficient.

The proofs of completeness do not help. Their starting point is
refutational-completeness: if $F \cup \neg C \models \bot$ then $F \cup \neg C
\vdash \bot$. If a formula is contradictory, the empty clause is generated by
resolution. The proofs then proceed by turning $F \cup \neg C \vdash \bot$ into
$F \vdash C'$ with $C' \subseteq C$. This change may not maintain the
considered restriction to resolution, in general. It does for the specific case
of linear resolution, for A-ordering linear resolution and s-linear resolution,
but is not guaranteed in general. For some restrictions, removing the literals
of $\neg C$ from the derivation may violate the restriction.

\proofonly{

\separator

{\bf Extension to an arbitrary proof system.}

}

\

All four implemented algorithms are resolution, even if sometimes they do not
look so (backtracking). The question is whether every resolution restriction
can be adapted for forgetting. No general answer seems possible, as every
restriction requires a separate proof of completeness. Some general
observations are still possible: entailment completeness is not required nor
desirable.

The question is even more general: can every proof system be adapted to
forgetting?

A proof system is complete for refutation if it proves $F \vdash \bot$ whenever
$F$ is unsatisfiable. If $F \models C$, then $F \cup \neg C \models \bot$. By
refutational completeness, $F \cup \neg C \vdash \bot$ follows. This is not
enough, as only $F$ is given. Such a proof $F \cup \neg C \vdash \bot$ needs to
be converted into $F \vdash C$.

This is not necessary for all clauses $C$ that are entailed by $F$. The full
remembrance clauses are enough. If $C$ comprises all variables to remember and
is entailed by $F$, the refutation $F \cup \neg C \vdash \bot$ has to be
converted into $F \vdash C$. How to do this conversion, and whether it is even
possible, depends on the specific proof system.

\section{Conclusions}
\label{section-conclusions}


Four algorithms for propositional forgetting are compared. The first three are
based on resolution: unrestricted, ordered and linear. The fourth is based on
backtracking. They are implemented in Python and run on random formulae. The
time and memory usage and the size of the generated output are compared.


The backtracking algorithm wins on most measures. This is unsurprising given
the efficiency of backtracking-based satisfiability
algorithms~\cite{love-etal-16}. Still, forgetting is not the same as
satisfiability. It requires not just to prove the existence of a model but to
derive a formula that expresses forgetting.

Variable elimination is a good competitor. Linear resolution seems unusable as
far as the experiments show. This is unexpected, given the amount of work done
on this form of resolution. Yet, only the basic version of linear resolution
was implemented, with inverse alphabetic ordering of the variables. Many
variants exist. Some of them may prove efficient for forgetting. At the same
time, the other algorithms were also implemented in their basic version:
variable elimination with alphabetic removing order, backtracking with basic
unoptimized unit propagation and choice of the branching variable, resolution
closure with no additional data structures to optimize the resolutions. In
spite of this, these algorithms did not perform as poorly as linear resolution.

Variable elimination and resolution closure appear to work well on some
measures and some specific cases. For example, resolution closure produces the
shortest formulae when the input formula is not very small, when it is
surpassed by variable elimination. The difference is however not large. It may
be due to the way the algorithms are implemented.


Further optimizing them is a possible direction for future work. Others include
the extension to other logics, especially Answer Set Programming, which is
relatively similar to propositional logics and forgetting was thoroughly
investigated~\cite{wang-etal-14,gonc-etal-16}. Another direction is the
adaptation of other algorithms for satisfiability to forgetting.

Resolution and backtracking can be used for forgetting, but they both where
designed for propositional satisfiability. Do all algorithms for satisfiability
work for forgetting?

Since forgetting preserves all consequences of the original formula on the
variables to remember, it is expressed by a formula comprising only clauses
that are entailed by the original formula. If an algorithm produces all
consequences of the original formula, it can be adapted to forgetting by
selecting only the clauses that do not contain variables to forget.

Being able to generate a subset of every clause entailed by a formula is called
entailment completeness~\cite{slag-etal-69,inou-92,delv-99,amir-mcil-05}.
Unrestricted resolution possesses this property, for example. The algorithm
based on resolution closure works this way.

A first possible requirement for an algorithm to be adapted to forgetting is
its entailment completeness: if a clause is entailed by the formula, a subset
of it is found by the algorithm.

This condition is sufficient for expressing forgetting, but is not necessary.
It is too stringent. It does not hold for A-ordering linear resolution, which
still expresses forgetting. While A-ordering linear resolution does not
generate all minimal clauses entailed by the formula, it generates enough of
them. Enough for forgetting. Enough to entail all consequences of the original
formula on the variables to remember.

To express forgetting, an algorithm needs not to generate all consequences of
the formula. As an example, an algorithm may not generate $\neg a \vee d$ when
forgetting $b$ from
{} $\{a \vee b, \neg b \vee c, \neg c \vee d\}$,
even though $\neg a \vee d$ is a prime implicate. It may still express
forgetting as
{} $\{\neg a \vee c, \neg c \vee d, \neg d \vee a\}$.
Neither variable elimination nor linear resolution output $\neg a \vee d$, for
example. What matters is not that $\neg a \vee d$ is in the output. What
matters is that $\neg a \vee d$ is entailed by the output. Generating all
consequences of the original formula or even just all prime implicates is not
required. It may even be a drawback, when it results in oversized outputs.

Another metric for evaluating different proof mechanisms is their proof size.
For example, general resolution proofs are always comparable in size to the
backtracking proofs for the same formula, but the converse does not hold for
all formulas~\cite{beam-pita-01}.
The minimal length of the proofs is a lower bound on the time used by an
algorithm. It is not on memory usage. The backtracking algorithm is a
counterexample: a proof is a tree of recursive calls; when a branch is over,
the memory it requires can be recycled. Even when measuring time, the size of a
proof is only a lower bound. To find a proof of a length $n$, no less than $n$
steps are necessary. Necessary, not sufficient. Finding the proof may be hard,
requiring more than $n$ steps~\cite{buss-12}.

Proof size is still an indicator of the memory efficiency of the algorithms
when memory cannot be reused. This is the case for directional resolution
(variable elimination): viewing a proof as a tree of clauses generated by
resolution, a whole level has to be maintained at every time, and the width of
a level is related to the overall size of the proof. Other resolution
strategies are able to generate smaller proofs. Yet, the experiments on linear
resolution excludes it as a practical alternative in spite of its theoretical
properties.

A further direction of study is approximate forgetting: instead of maintaining
exactly the consequences on the variables to remember, just keeping most of
them may be enough. Or the opposite: maintain all of them at the cost of adding
some new consequences. Incomplete satisfiability methods such as local
search~\cite{kaut-selm-94-b,yolc-pocz-19} may help.


\proofonly{

TODO:
the efficiency of the algorithms depends on their choices: clauses to resolve
first, sequence of variables to forget, branching variables; these choices are
currently not optimized

}

\bibliographystyle{alpha}
\newcommand{\etalchar}[1]{$^{#1}$}

\end{document}